\documentclass[twocolumn,english,prb]{revtex4-2}
\usepackage{color}
\usepackage[utf8]{inputenc}
\usepackage{graphicx}
\usepackage[T1]{fontenc}
\usepackage{float}
\usepackage{textcomp}
\usepackage{amsmath}
\usepackage{amssymb}
\usepackage{graphicx}
\usepackage{esint}
\usepackage{xcolor}
\usepackage{multirow}
\usepackage{dsfont}
\usepackage{physics}
\usepackage{bbold}

\begin{document}

%\title{Discrete bound states at electric field inversion loops in buckled silicene:
%topologically and trivially confined states}

\title{Snakelike trajectories of electrons released from quantum dots driven by the spin Hall effect}

\author{B. Szafran}
\email{bszafran@agh.edu.pl}
\affiliation{AGH University of Krakow, Faculty of Physics and Applied Computer Science, Al. Mickiewicza 30, 30-059 Krakow, Poland}

\author{P.  W\'ojcik}
\email{pawelwojcik@agh.edu.pl}
\affiliation{AGH University of Krakow, Faculty of Physics and Applied Computer Science, Al. Mickiewicza 30, 30-059 Krakow, Poland}

\begin{abstract}
Using time-dependent simulations, we analyze the trajectories of electrons released from a quantum dot in a waveguide made of a spin-orbit-coupled material (InSb). An electron released from the quantum dot, when driven by an electric field follows a trajectory that is deflected by spin-orbit interaction and undergoes spin precession that results in a spin-dependent, snake-like trajectory. The trajectory strongly depends on the initial state of the electron, enabling detection of the electron quantum state in the dot when connected to the T-junction. Notably, we show that the snake-like trajectory persists even under a small external magnetic field with low, incomplete initial electron spin polarization. Our findings are supported by semiclassical calculations of the electron trajectory, which show good agreement with full quantum mechanical simulations. 
\end{abstract}

\maketitle

\section{introduction}

The manipulation and detection of electron spin in low-dimensional systems form the cornerstone of spintronics~\cite{Fabian2004} and quantum information processing~\cite{Ladd2010}. Quantum dots (QDs) provide a highly controllable platform for initializing, manipulating, and studying single-electron spin states.  Consequently, spin qubits in semiconductor  QDs~\cite{Burkard2023,Philips2022,DiVincenzo1998} have emerged as promising building blocks for the development of scalable quantum computers, owing to their long coherence times, full electrical tunability, and compatibility with well-established semiconductor fabrication technologies~\cite{Hanson2007,Zwanenburg2013,Wang2024,Philips2022, Takeda2024,Morello2010}.

The spin state of electrons confined in quantum dots has been probed typically using techniques such as spin-to-charge conversion, a method that infers the electron's spin state by correlating it with a detectable charge signal~\cite{Koppens2006,Koppens2005,Petta2005,Nowack2007,Barthel2009,Petta2022,Nadj-Perge2010,Nadj-Perge2012,Koppens2008}. Another widely employed techniques include energy-selective tunneling between a quantum dot and a nearby electron reservoir — known as the Elzerman method~\cite{Elzerman2004}, RF reflectometry~\cite{Chittock-Wood2026,Colless2013,Liu2021,Crippa2019}, gate-based sensing~\cite{Urdampilleta2019}, dispersive readout using resonators~\cite{KamHu2024}, and the Pauli spin blockade mechanisms~\cite{Ono2002,Johnson2005,Petta2005,Hanson2007,Nadj-Perge2010,Nadj-Perge2012,Nowak2012}. 
Regardless of the method, once the spin state is mapped onto a charge state, its detection is performed using sensitive electrical measurement techniques, such as those employing quantum point contacts (QPCs) or single-electron transistors~\cite{Elzerman2004,Hanson2007,Barthel2009}, which involve complex experimental setups or are constrained by limited temporal resolution.

Recent advances in spin-orbit-coupled materials have opened new pathways for controlling/detecting electron spin via purely electrical means. In particular, the spin Hall effect (SHE) has emerged as a powerful mechanism to generate spin-dependent forces on electrons, leading to transverse spin currents without the need for magnetic fields~\cite{ruskie,sinova2004,sinova2015}. In particular, when the electron with a well-defined spin is injected into a spin-orbit-active waveguide, their trajectories are no longer dictated solely by classical drift. Instead, the spin–orbit interaction induces spin-dependent deflections and precessions, giving rise to characteristic oscillatory motion known as Zitterbewegung~\cite{Schliemann2006,Shermans}. These trajectories encode information about the electron’s initial spin state, offering a novel route for spin detection. Recently, this mechanism has been proposed as a means to detect the helical state in T-junctions~\cite{Wojcik_2021}.

In this work, we investigate the dynamics of electrons released from quantum dots in spin-orbit-coupled waveguides, focusing on the spin-orbital dynamic driven by the SHE. Specifically, we demonstrate that the electronic state released from the quantum dot, when subjected to an effective spin–orbit field, undergoes spin precession that results in a spin-dependent path following a snake-like trajectory. This trajectory can ultimately be resolved through nonlocal measurements performed on the T-terminal junction coupled to the quantum dot. Interestingly, we demonstrate that the spin-dependent trajectories remain robust even under weak magnetic fields and incomplete spin polarization of electrons confined in the QD.

The organization of this paper is as follows: In Sec. II, we present the theoretical model along with a schematic illustration of the proposed setup. Sec. III provides an analysis of the electronic spectra of a quantum dot and a detailed investigation of the spin trajectory and spin state detection applied to InSb. Finally, the summary and conclusions are given in Sec. IV.

\section{Theoretical model}

\begin{figure}[!t]
\includegraphics[width=1.0\columnwidth]{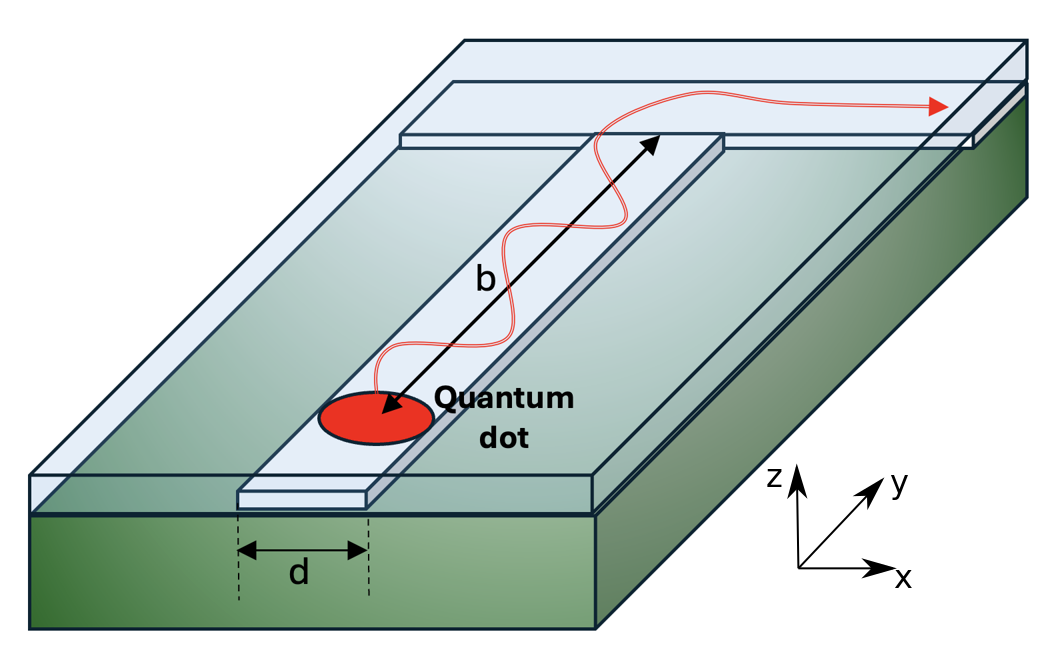}
\caption{Schematics of the QD embedded within the chanel of width $d$. The electrostatic confinement of the quantum dot is switched off at $t=0$ and replaced by a homogenous electric field oriented in the $y$ direction. The split in the channel is located at a distance $b$ from the QD. The red arrow schematically indicates the trajectory of the electron released from the QD.} \label{s0}
\end{figure}

We consider a T-junction embedded in the 2DEG, with the quantum dot located within a channel of width $d$ (see Fig.~\ref{s0}). Due to the strong confinement in the direction perpendicular to the plane, we assume that the vertical degree of freedom is frozen, which allows us to reduce the problem to two dimensions. The lateral confinement potential is given by $V(x,y;t<0)=V_{\infty}(x,y)+V_{QD}(x,y)$, where $V_{\infty}(x,y)$ describes the hard-wall confinement of the channel — zero potential inside the channel and infinite potential barriers at its edges — and $V_{QD}$ is the electrostatic potential that defines the quantum dot, applied for $t<0$.
At $t=0$, we switch off the quantum dot confinement potential ($V_{QD}=0$) and introduce a weak homogeneous electric field along the channel, such that the total potential becomes $V(x,y,t>0)=V_{\infty}(x,y)-eF y$. Taking the initial wave function as the electronic state confined in the quantum dot, we analyze its time-dependent evolution by solving the Schrödinger equation.
As we demonstrate, due to the spin Hall effect (SHE), the lateral motion of the wave packet depends on its initial state. This property enables the distinction between the two components of the Kramers doublet by splitting the channel into two drain leads, positioned a distance $b$ from the center of the quantum dot. 

\section{Results and Discussion}
Let us consider a simple 2D Hamiltonian for an electron confined in InSb quantum well with a  Rashba type spin-orbit coupling~\cite{Rashba2003}. Hamiltonian of the system is given by
\begin{equation}
H=-\frac{\hbar^2}{2m^*}\nabla^2+V(x,y)+H_{SO}+\frac{g\mu_B}{2}{\vec{ B}}\cdot{\bf \vec{\sigma}},
\end{equation}
with the spin-orbit coupling term 
\begin{equation} H_{SO}=\alpha ( k_x\sigma_y -k_y\sigma_x)\label{hso}.
\end{equation} 
The calculations are carried out for the following values of the parameters corresponding to InSb: the electron effective mass $m^*=0.014m_0$, the Rashba SO coupling constant of $\alpha=50$~meVnm and the Land\'e factor $g=-49$.

\subsection{Transport of the electronic state released from QD in semiconductor channel}

\begin{figure}[htbp]
 \begin{tabular}{l}
\includegraphics[width=0.7\columnwidth]{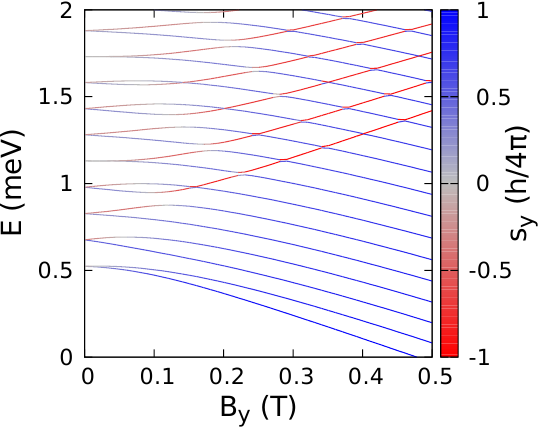} \put(-140,30){(a)} \\
\includegraphics[width=0.6\columnwidth]{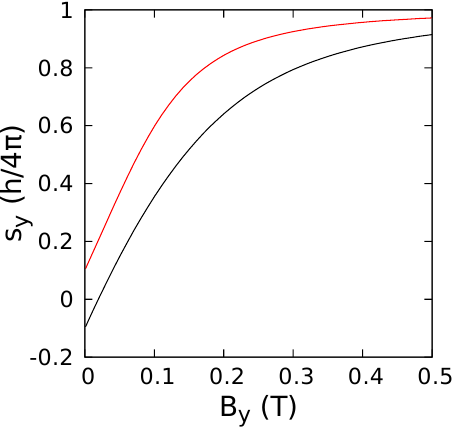} \put(-115,30){(b)} \\
\includegraphics[width=0.7\columnwidth]{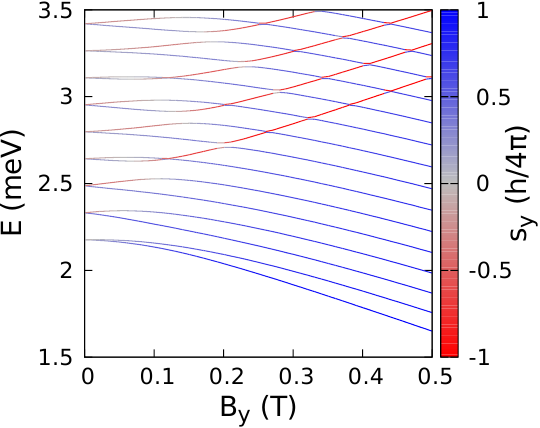} \put(-140,30){(c)} \\
\includegraphics[width=0.6\columnwidth]{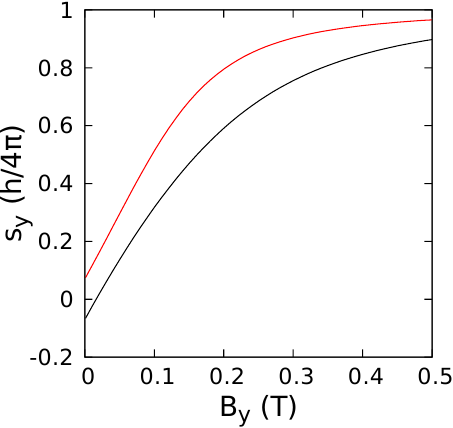} \put(-115,30){(d)} \\
 \end{tabular}
\caption{(a) The energy spectrum for the electron confined in a harmonic oscillator quantum dot potential
with $\hbar\omega=0.15625$~meV, the width of the InSb channel $d=175$~nm and the magnetic field oriented 
parallel to the $y$ axis. The color of the lines indicates the spin $y$ component of the eigenstates.
In (b) we show the average $y$ component of the spin for the ground-state (red line) and the first-excited state (black line).  (c,d) same as (a,b) only for $d=100$ nm. 
} \label{spectrum}
\end{figure}

For the system under consideration, the quantum dot is defined by an electrostatic confinement potential, which we model as a harmonic oscillator $V_{QD}(x,y)=\frac{m^*\omega^2}{2}(x^2+y^2)$ with $\hbar \omega=0.15625$~meV. The quantum dot energy spectrum for the channel width $d=175$~nm is shown in Fig. \ref{spectrum}
as a function of the in-plane magnetic field oriented along the $y$ axis for which the orbital effects of the in-plane magnetic field are absent. 
We use a finite difference approach with a square mesh of step $\Delta x=1.56$~nm. The computational box in the $y$ direction is very long (2.1$~\mu$m for this calculation).
The finite but very large size in $y$ direction does not affect the spectrum, in contrast to the lateral confinement in the $x$ direction which does influence the results due to its limited width -  the oscillator radius $R_o=\sqrt{\frac{\hbar}{m^*\omega}}$ for our parameters is $93.32$ nm and is larger than the halfwidth of the channel. The $E(B_y)$ dependence is dominated by the Zeeman effect due to the large value of the $g$ factor. Nevertheless, a complete spin polarization is obtained only at large magnetic field
since in InSb the Rashba constant is significant and the mixing of the spin states by the SO coupling prevents a complete spin-polarization. In the following procedure, to distinguish between the initial states of the Kramers doublet, we apply a very small in-plane magnetic field $B_y=10$~$\mu$T, which is just sufficient to slightly lift the degeneracy of the ground-state doublet.
The results for $d=100$ nm are shown in Fig. \ref{spectrum}(c,d). The energy spectrum for this value of $d$ is shifted up due to a stronger lateral confinement and the average values of $s_y$ at $B=0$ are closer to zero.

\begin{figure}[!t]
 \begin{tabular}{ll}
\includegraphics[width=0.35\columnwidth]{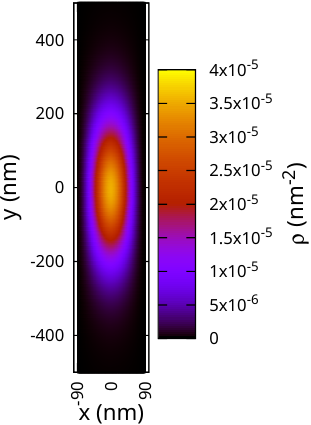} (a) &  \includegraphics[width=0.35\columnwidth]{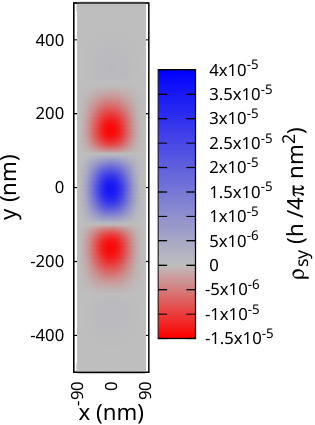} (b)\\
\includegraphics[width=0.3\columnwidth]{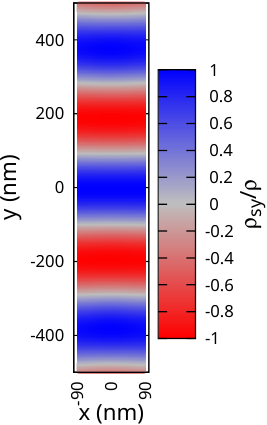} (c) & \includegraphics[width=0.3\columnwidth]{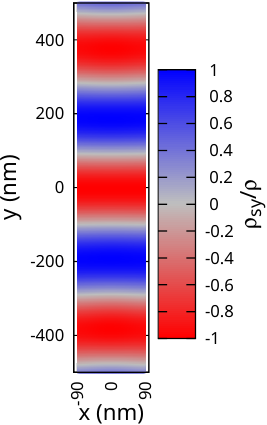}(d)
\end{tabular}
\caption{
(a) The probability density for ground state of the Kramers doublet, together with (b) the spin density $s_y$ - the spin density for the second state of the Kramers pair is opposite. (c) and (d) the spin density $s_y$ divided by the probability density $\rho$, determined for both states of the Kramers doublet. The results correspond to the QD confinement energy $\hbar\omega=0.15625$~meV and the channel width $d=175$~nm.
\label{wf}
}
\end{figure}

\begin{figure}[!t]
 \begin{tabular}{lllllll}
\includegraphics[trim= 10cm 0 10cm 0,clip, width=0.15\columnwidth]{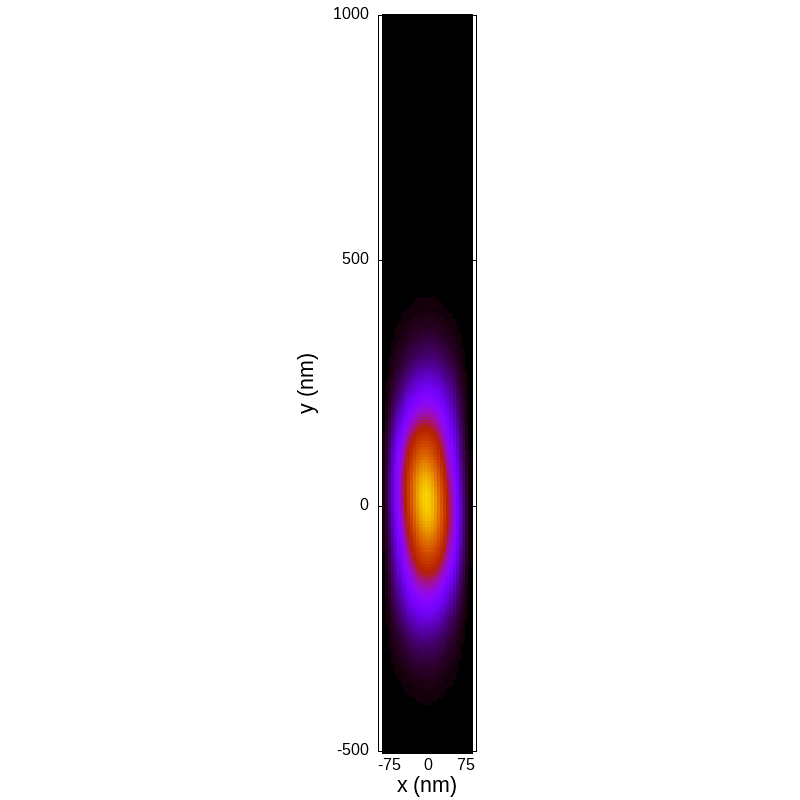} &
\includegraphics[trim= 10cm 0 10cm 0,clip, width=0.15\columnwidth]{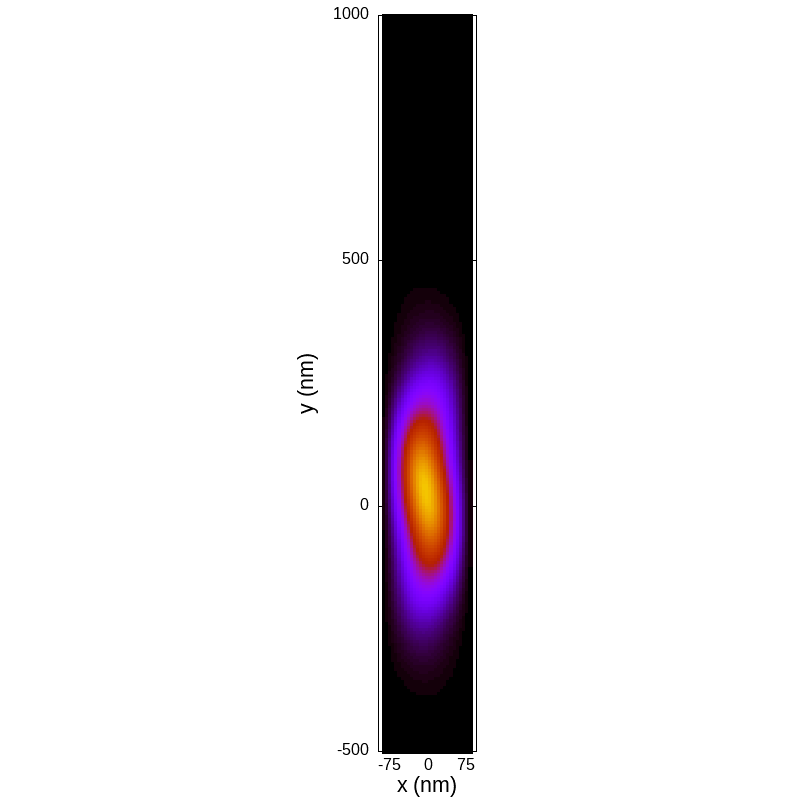} &
\includegraphics[trim= 10cm 0 10cm 0,clip, width=0.15\columnwidth]{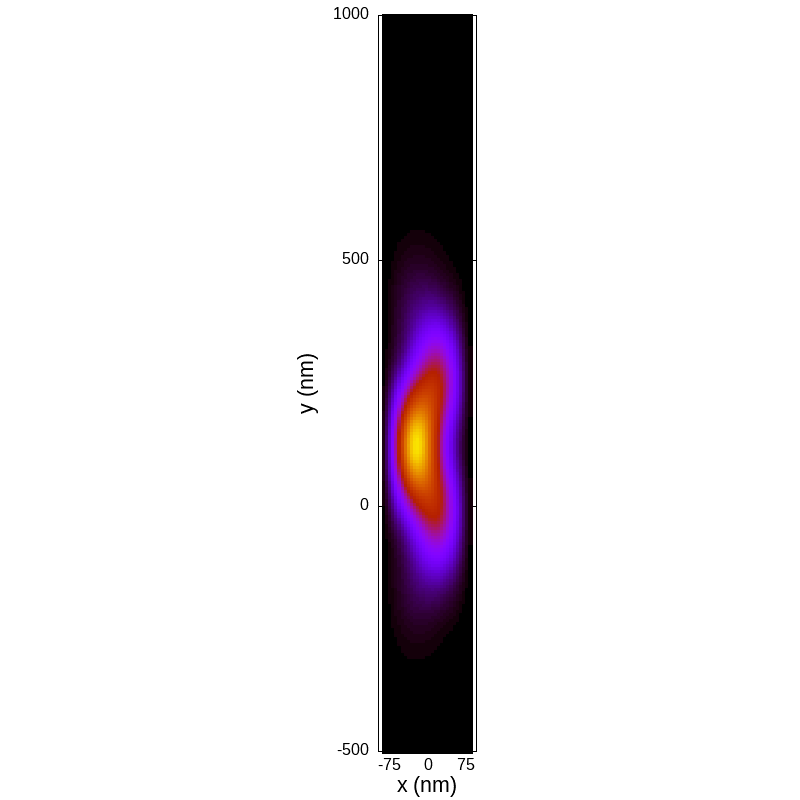} &
\includegraphics[trim= 10cm 0 10cm 0,clip, width=0.15\columnwidth]{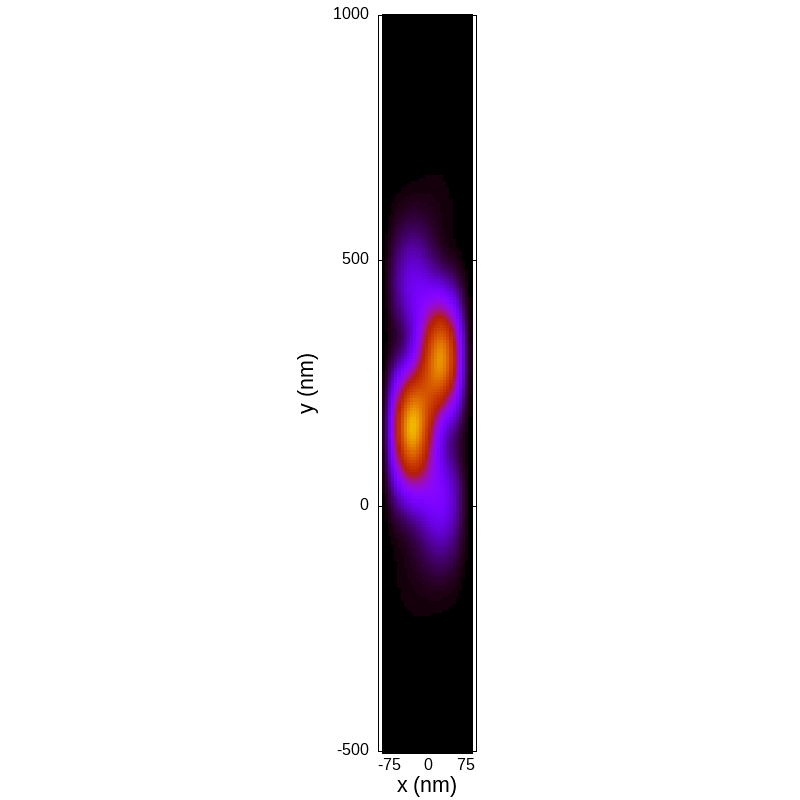} &
\includegraphics[trim= 10cm 0 10cm 0,clip, width=0.15\columnwidth]{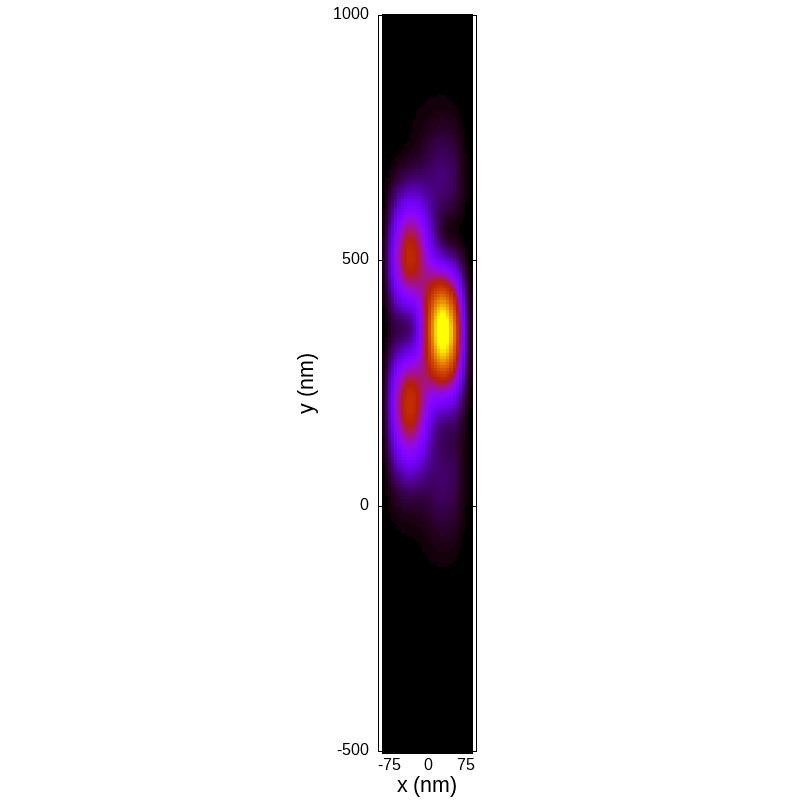} &
\includegraphics[trim= 10cm 0 10cm 0,clip, width=0.15\columnwidth]{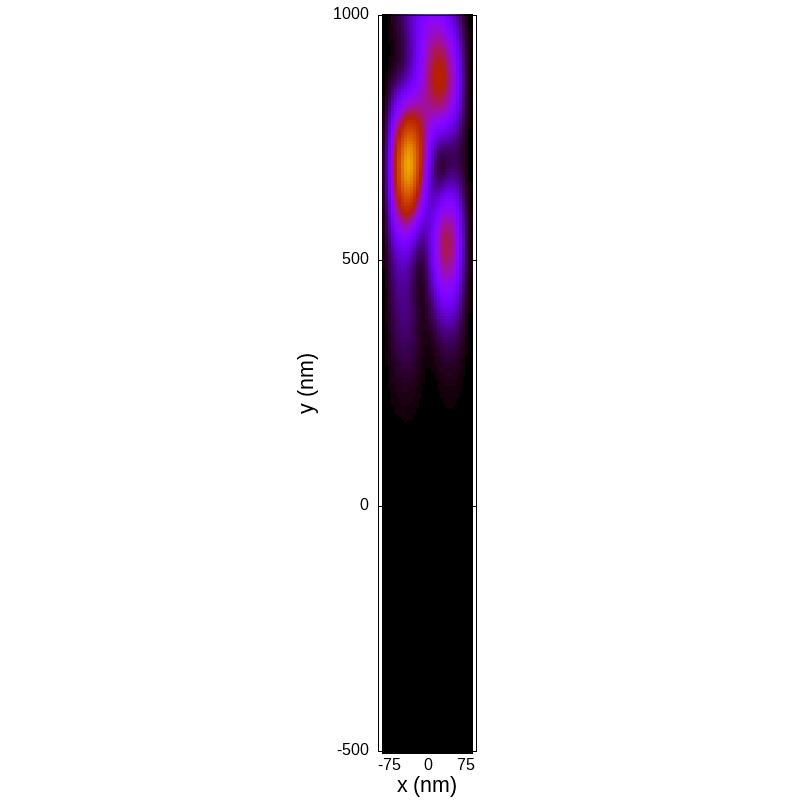}  \put(-6,100){(a)}\\
\includegraphics[trim= 10cm 0 10cm 0,clip, width=0.15\columnwidth]{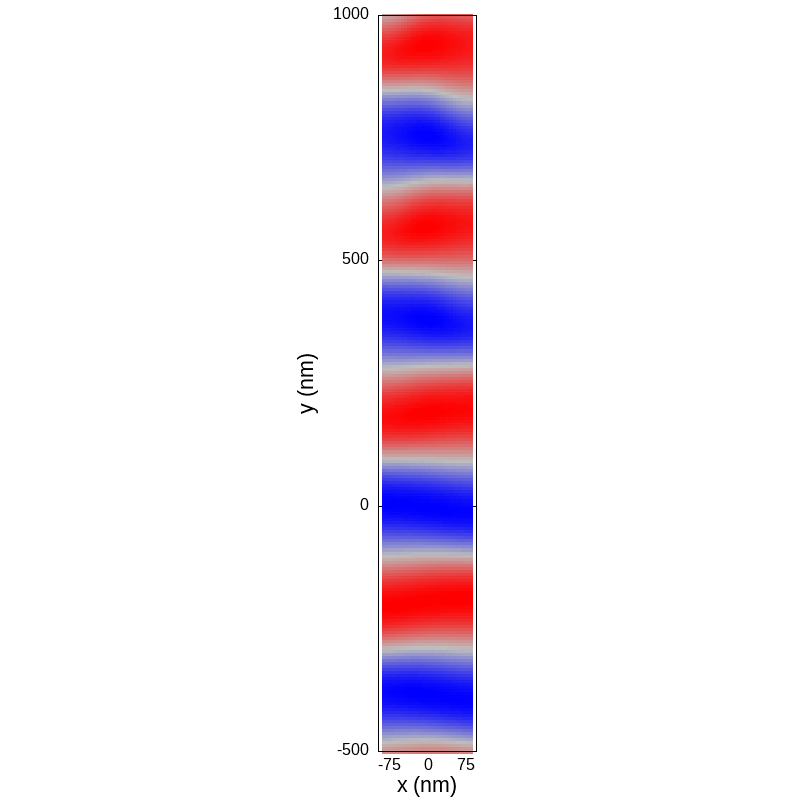} &
\includegraphics[trim= 10cm 0 10cm 0,clip, width=0.15\columnwidth]{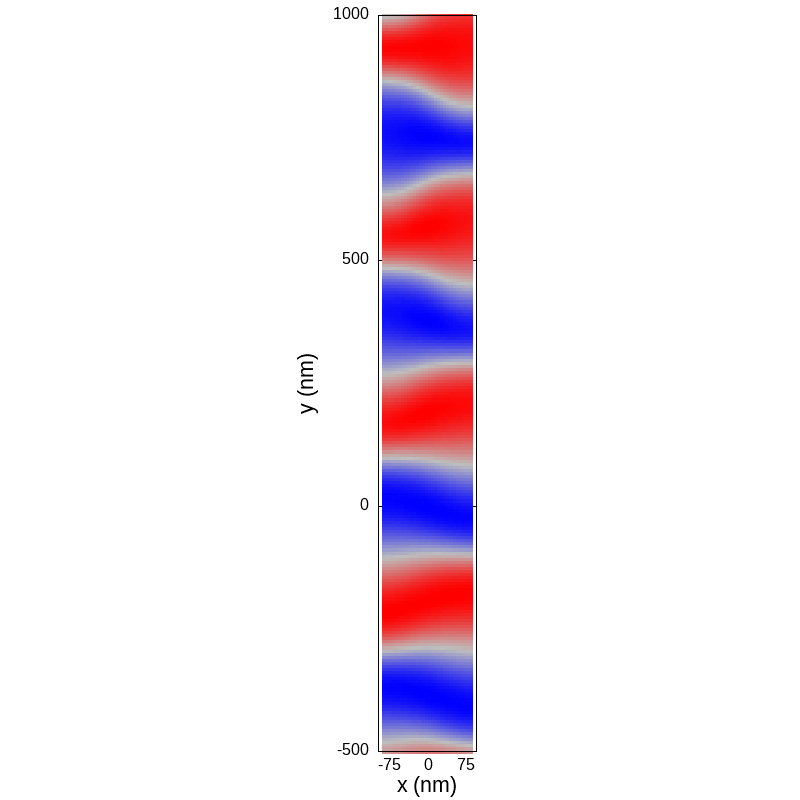}  &
\includegraphics[trim= 10cm 0 10cm 0,clip, width=0.15\columnwidth]{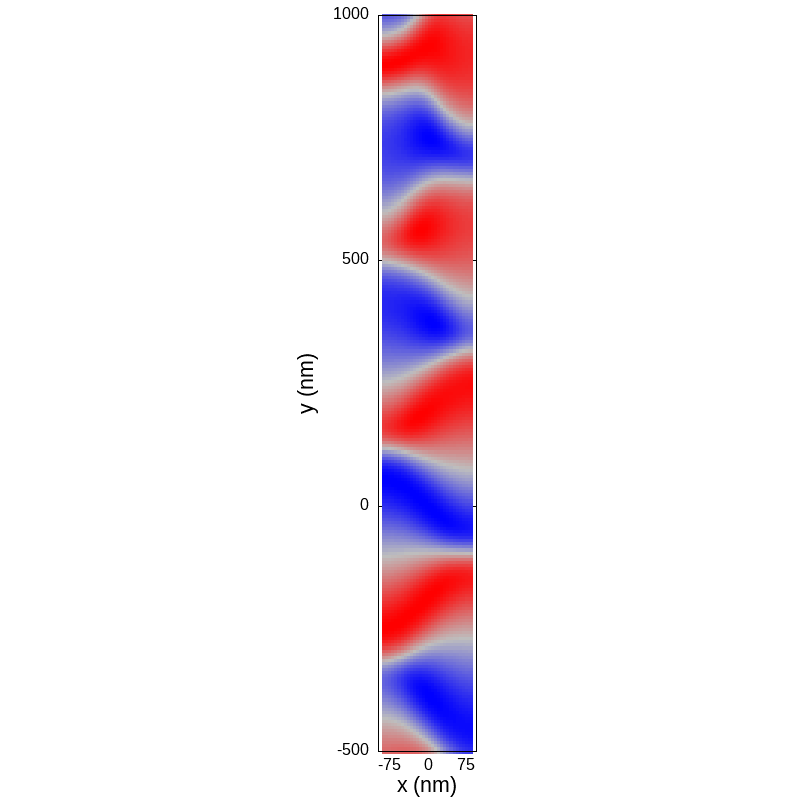}  &
\includegraphics[trim= 10cm 0 10cm 0,clip, width=0.15\columnwidth]{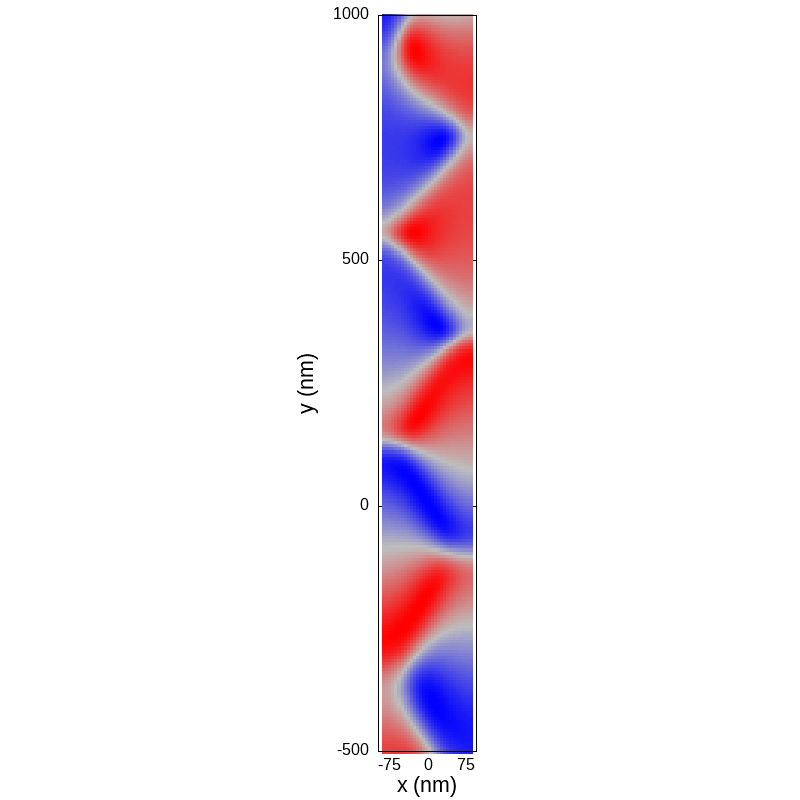}  &
\includegraphics[trim= 10cm 0 10cm 0,clip, width=0.15\columnwidth]{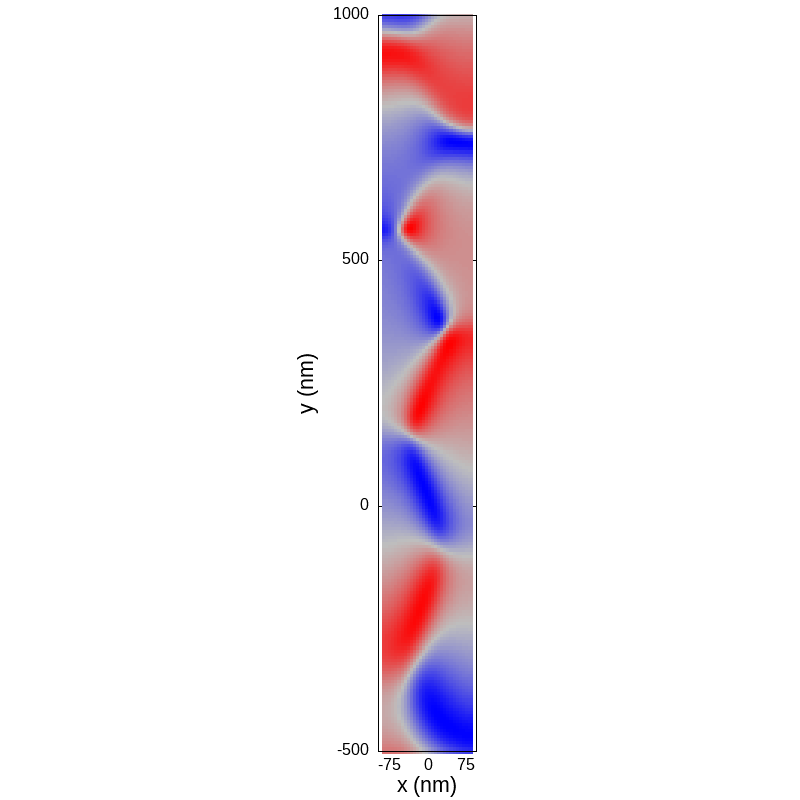}  &
\includegraphics[trim= 10cm 0 5cm 0,clip, width=0.24\columnwidth]{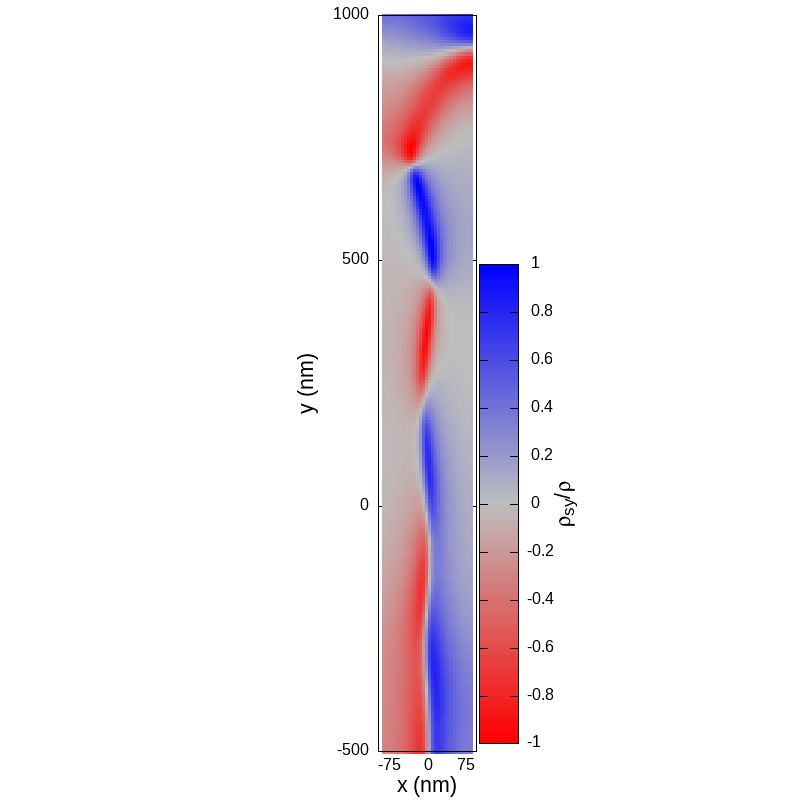}  \put(-25,100){(b)} \\
\includegraphics[trim= 10cm 0 10cm 0,clip, width=0.15\columnwidth]{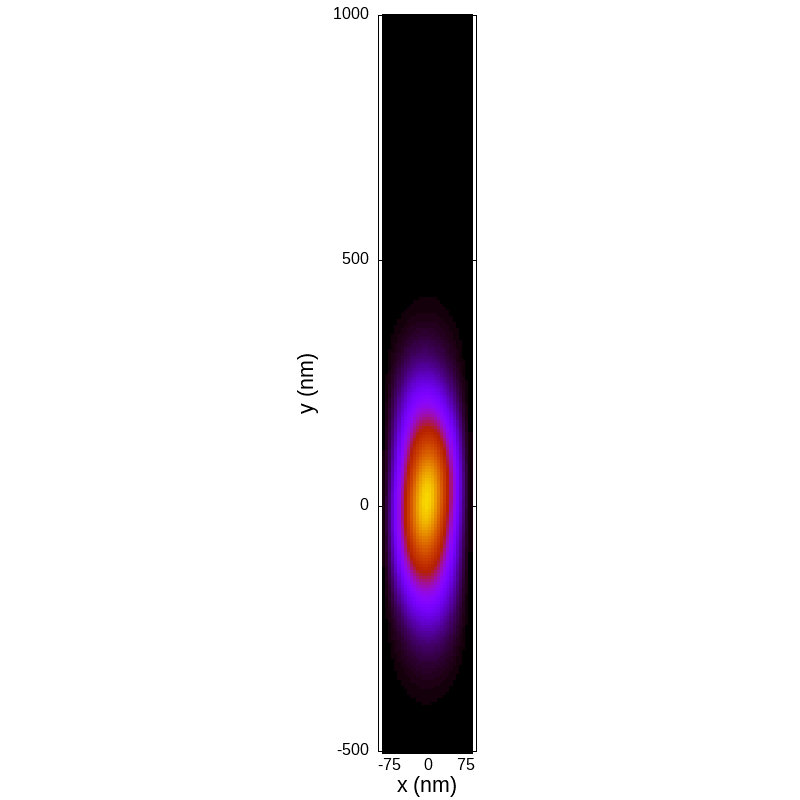}\put(-24,132){\tiny 0.48ps} &
\includegraphics[trim= 10cm 0 10cm 0,clip, width=0.15\columnwidth]{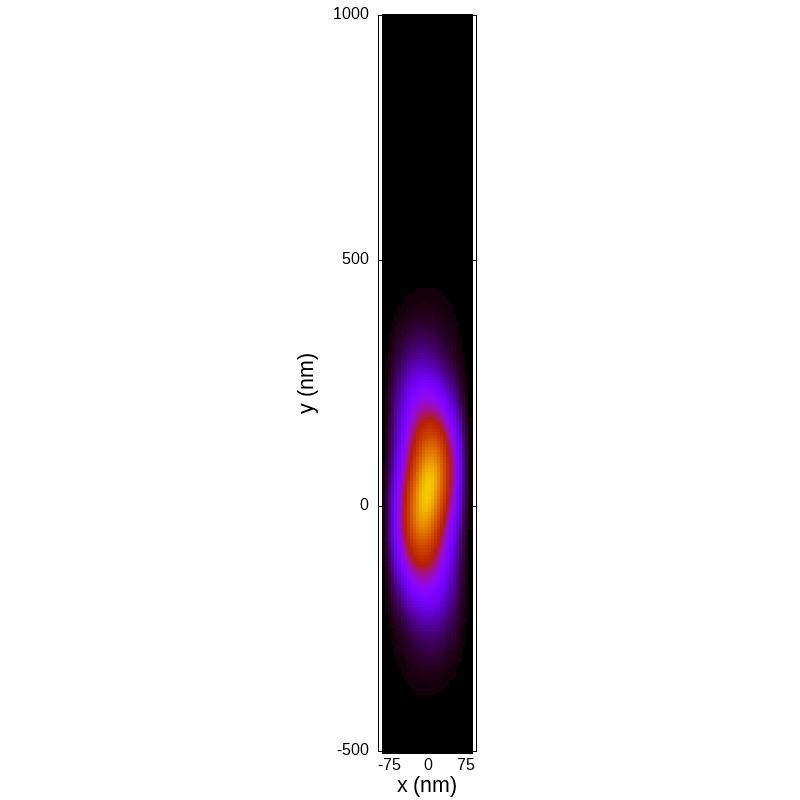} \put(-24,132){\tiny 0.73ps}&
\includegraphics[trim= 10cm 0 10cm 0,clip, width=0.15\columnwidth]{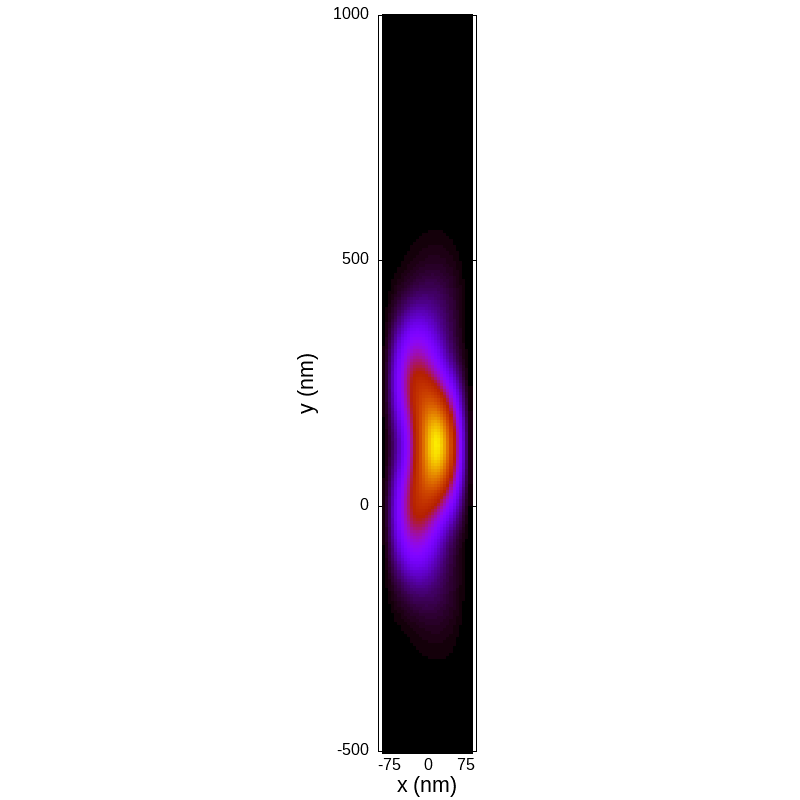} \put(-24,132){\tiny 1.48ps}&
\includegraphics[trim= 10cm 0 10cm 0,clip, width=0.15\columnwidth]{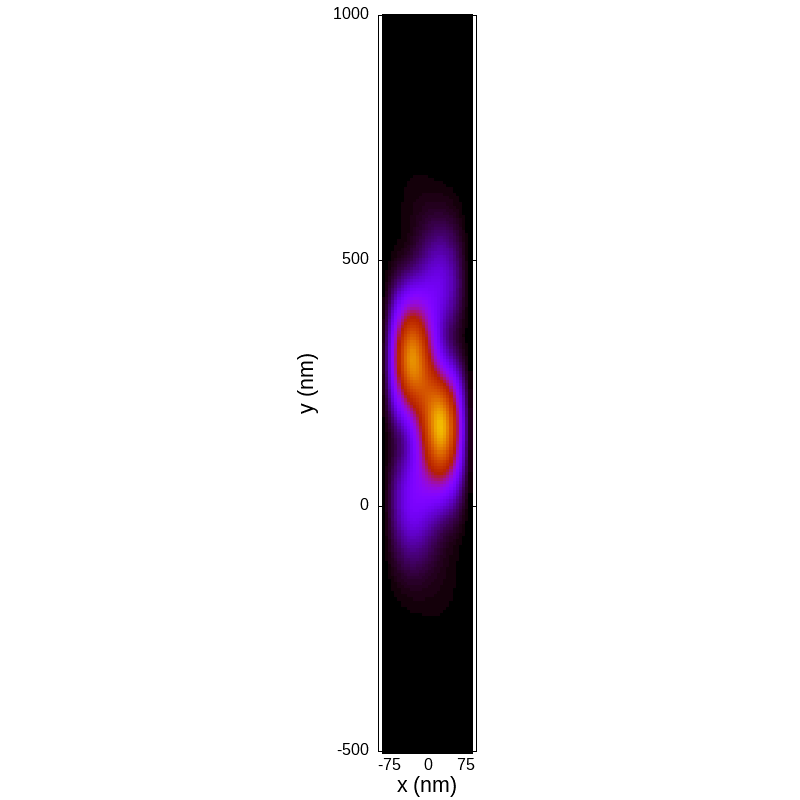} \put(-24,132){\tiny 1.98ps}&
\includegraphics[trim= 10cm 0 10cm 0,clip, width=0.15\columnwidth]{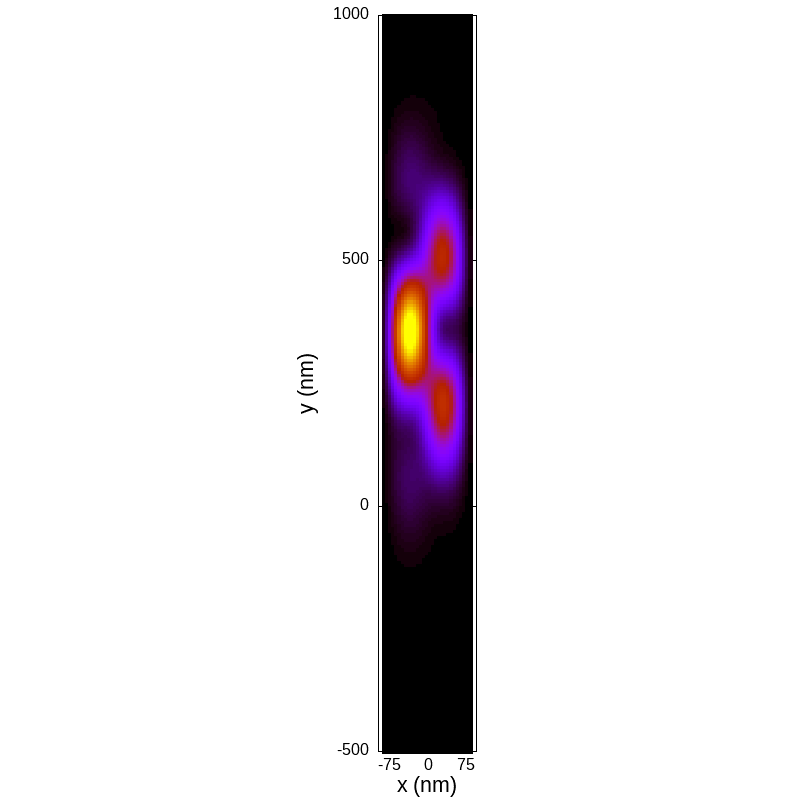} \put(-24,132){\tiny 2.53ps}&
\includegraphics[trim= 10cm 0 5cm 0,clip, width=0.24\columnwidth]{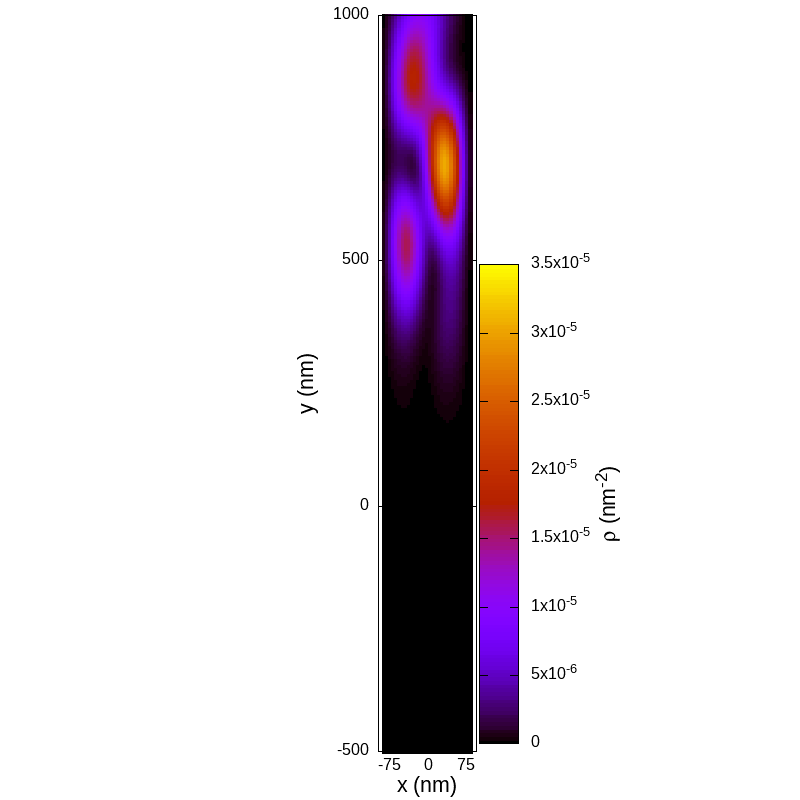} \put(-44,132){\tiny 3.72ps}\put(-25,120){(c)} \\
\includegraphics[trim= 10cm 0 10cm 0,clip, width=0.15\columnwidth]{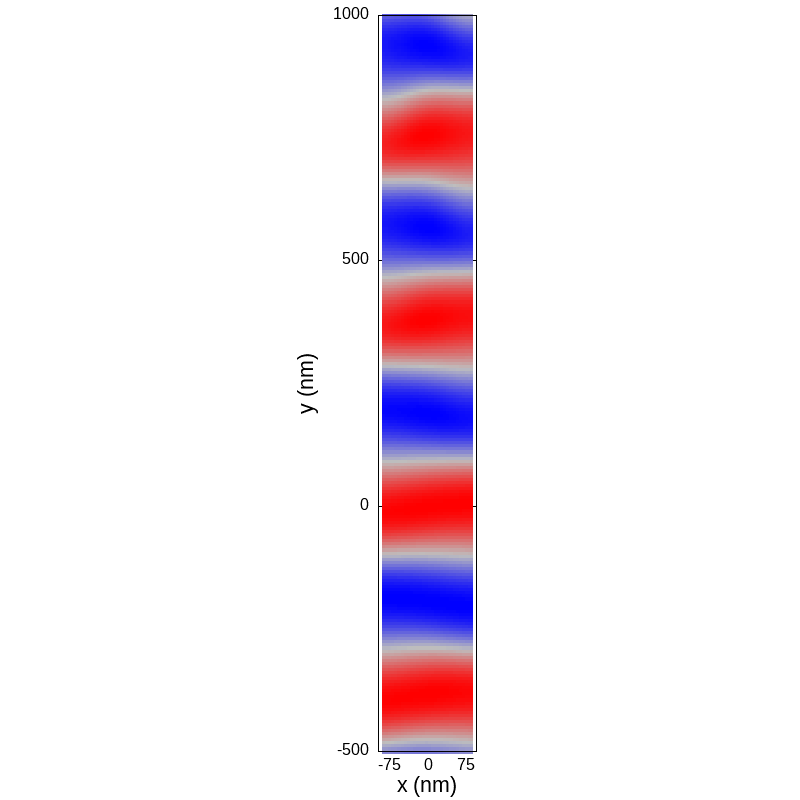} &
\includegraphics[trim= 10cm 0 10cm 0,clip, width=0.15\columnwidth]{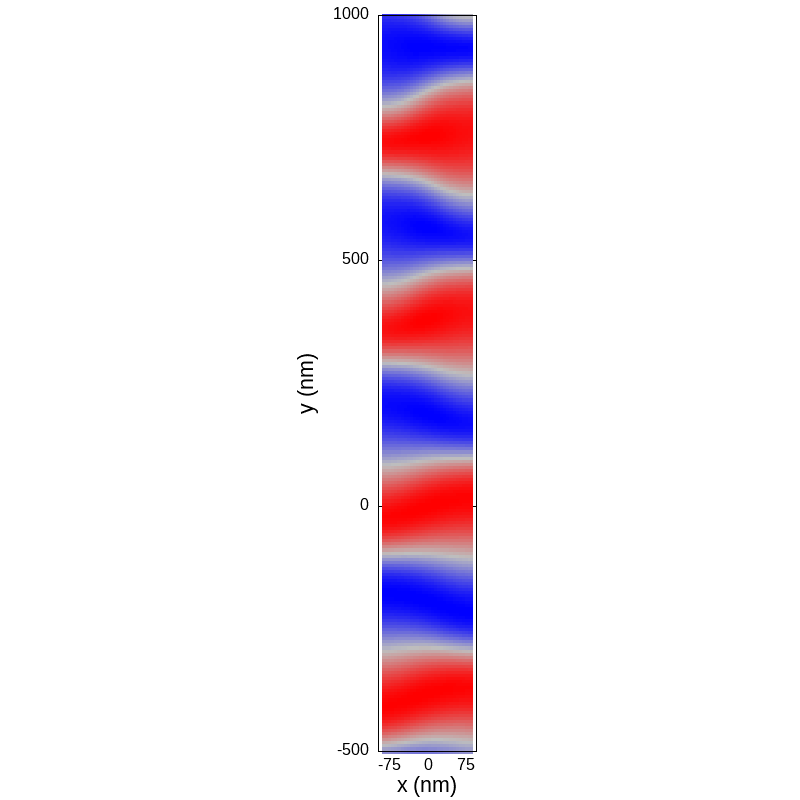}  &
\includegraphics[trim= 10cm 0 10cm 0,clip, width=0.15\columnwidth]{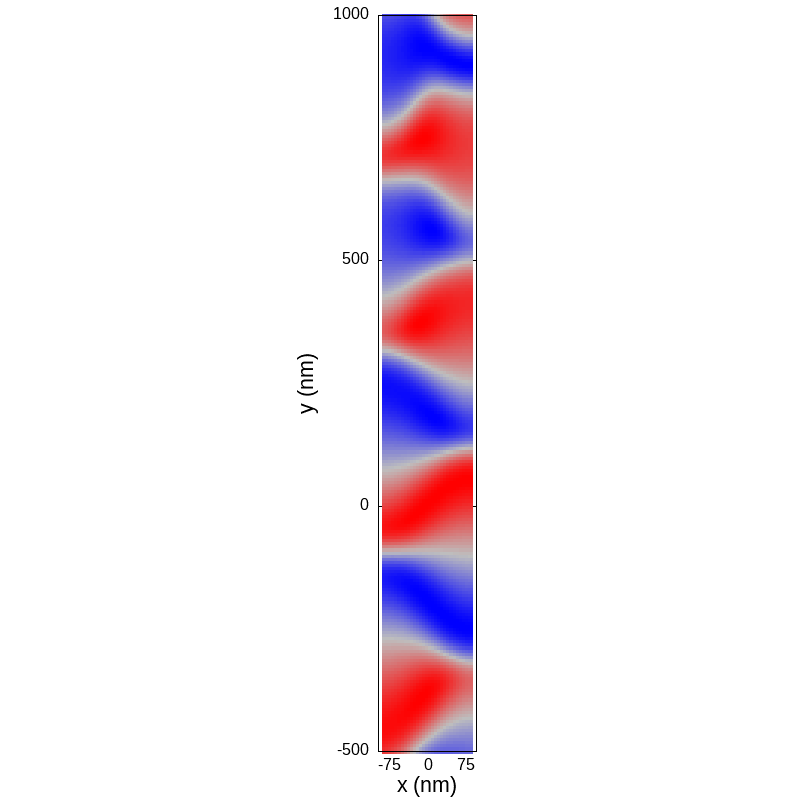}  &
\includegraphics[trim= 10cm 0 10cm 0,clip, width=0.15\columnwidth]{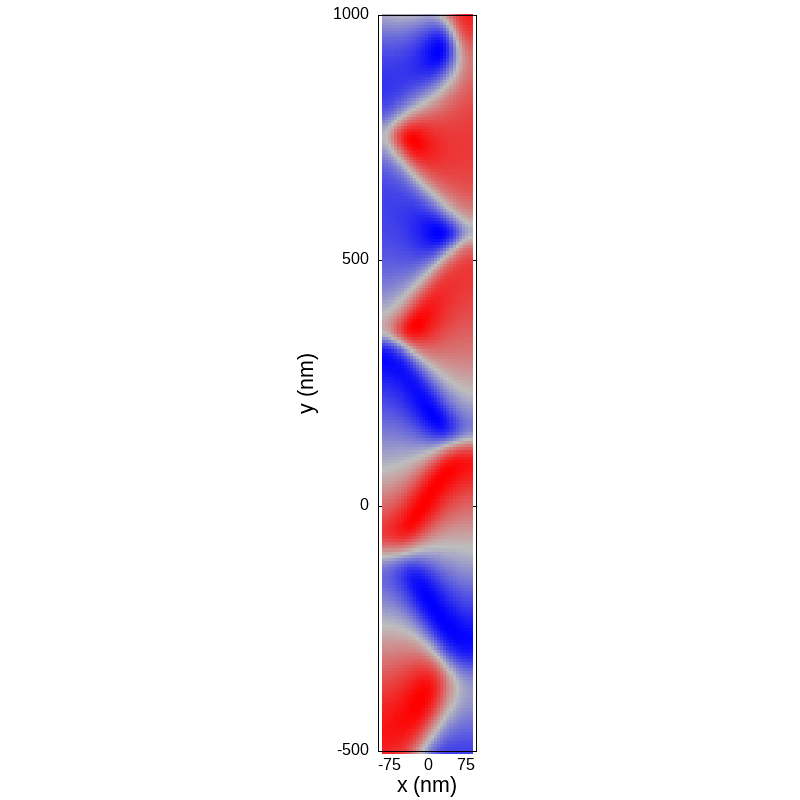}  &
\includegraphics[trim= 10cm 0 10cm 0,clip, width=0.15\columnwidth]{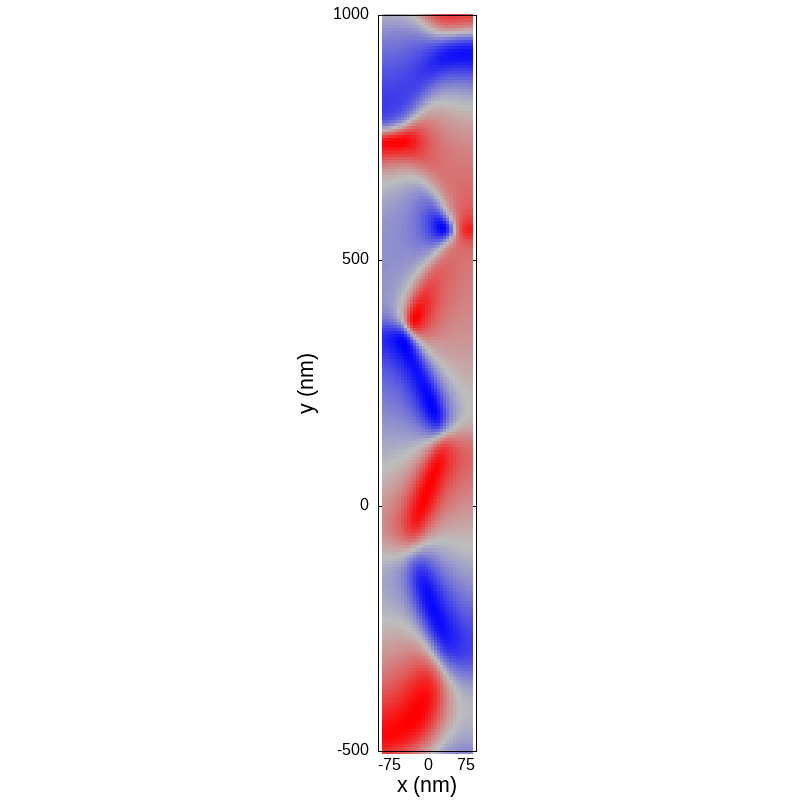}  &
\includegraphics[trim= 10cm 0 10cm 0,clip, width=0.15\columnwidth]{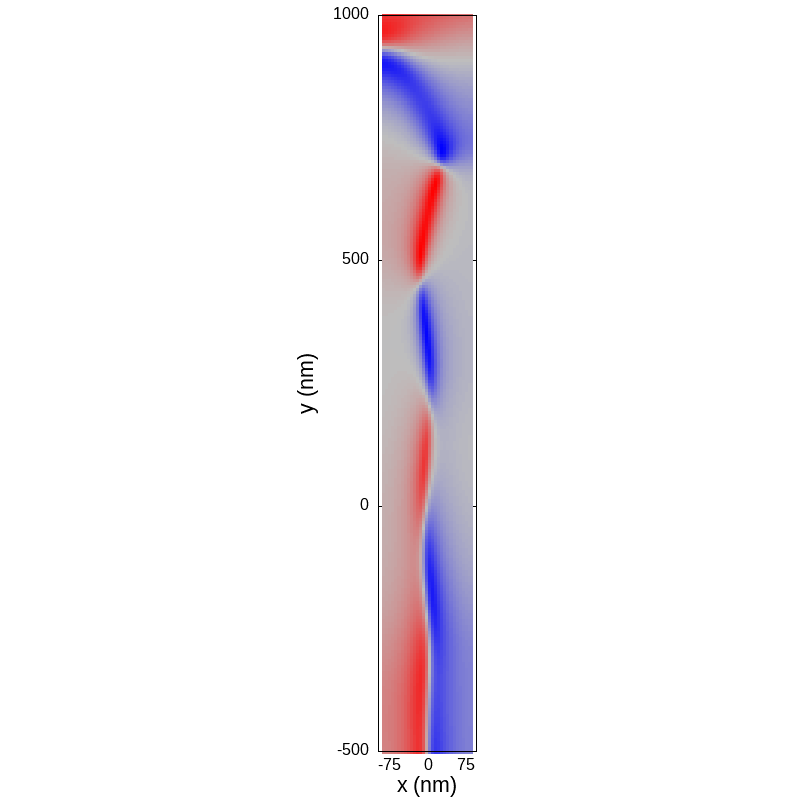}  \put(-6,100){(d)} \\
\end{tabular}
\caption{ Snaphots of the probability density (a,c) and spin-y component (b,d), calculated as a ratio of the $s_y$ component density to the probability density, for selected moments in time listed on top of panel (a). The initial condition set in one of the states of the Kramers doublet in the QD; (a,b) for the lower
energy state and (c,d) for the higher energy state. The results for the InSb channel of width $d=175$~nm and the magnetic field $B_y=10~\mu$T. The color scale for (a,b) as well as (c,d) is the same. } \label{evolution}
\end{figure}

Figure \ref{wf}(a) shows the probability density which is identical for both the states of the ground-state Kramers doublet and Fig. \ref{wf}(b) shows the spin $s_y$ density for the lower-energy state.  The corresponding spin density for the higher energy state of the doublet is opposite (not shown).
 Figure \ref{wf}(c) and \ref{wf}(d) display
the spin $y$ density divided by the probability density with a regular oscillation along the $y$ axis.
The sign of the spin texture is opposite for both the states of the Kramers doublet. 
We show below that the spin polarization along the channel -- even incomplete -- produces a distinctly different electron trajectory for the initial state set from one of the doublet states.

At $t=0$ we switch off the quantum dot confinement potential ($V_{QD}=0$) and introduce a weak homogenous electric field along the channel $V(x,y,t>0)=V_{\infty}(x,y)-eF y$ with $eF=0.01$ meV/nm. 
Snapshots of the time evolution, i.e., the charge density, and the spin texture for the initial condition set in one of the states of the doublet, is given  in Fig. \ref{evolution} (a) and (c) for the  lower energy state and the higher energy state, respectively. The spin-density to probability density ratio is displayed in Fig. \ref{evolution}~(b) and (d). 
Upon its release from the quantum dot the electron moves up along $y$ axis in the direction of the electric field.
We notice that at $t=0.73$ ps the packet no longer moves exactly parallel to the channel axis but acquires a wiggle that is opposite for the ground [Fig. \ref{evolution}(a)]  and the excited state [Fig. \ref{evolution}(c)]. 

In terms of spin evolution, the effect of Rashba spin–orbit coupling on the spin can be interpreted as arising \cite{Meier2007} from an effective spin–orbit-induced magnetic field  ${\bf B}_{SO}=\frac{2\alpha}{g\mu_B}\left(-k_y, +k_x, 0\right)^T$. 
The effective field induces a precession of the spin in the direction perpendicular to the direction of motion, 
within the plane of confinement - see in Fig. \ref{evolution}(b,d). At the beginning ${\bf B}_{SO}$ is oriented exactly 
antiparallel to the $x$ axis and the spin  initially oriented along the $y$ direction precesses with respect to this axis. As the trajectory of the electron 
is deflected a component of ${\bf B}_{SO}$ oriented along the axis of the channel appears, hence the sign of $y$ component of the spin changes along the lines that are no longer perpendicular to the $y$ axis. 

 \begin{figure}[!t]
\begin{tabular}{l}
\includegraphics[width=0.8\columnwidth]{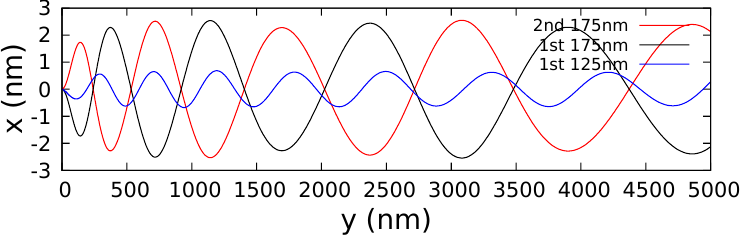}\put(-25,25){(a)}\\
\includegraphics[width=0.8\columnwidth]{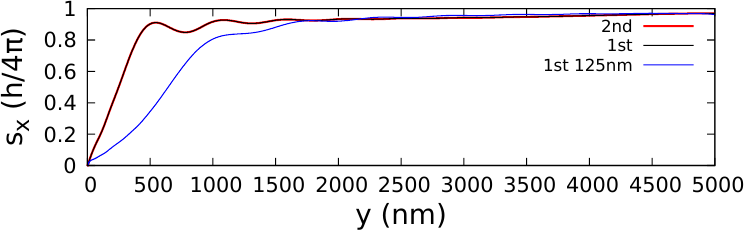}\put(-25,25){(b)}\\
\includegraphics[width=0.8\columnwidth]{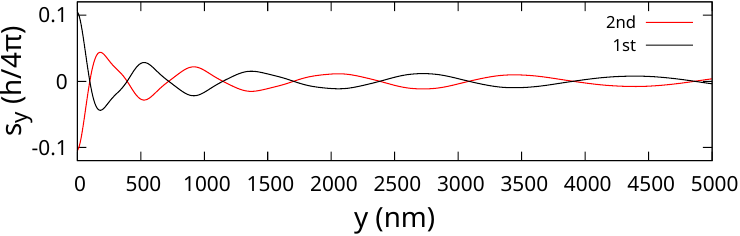}\put(-25,25){(c)}\\
\includegraphics[width=0.8\columnwidth]{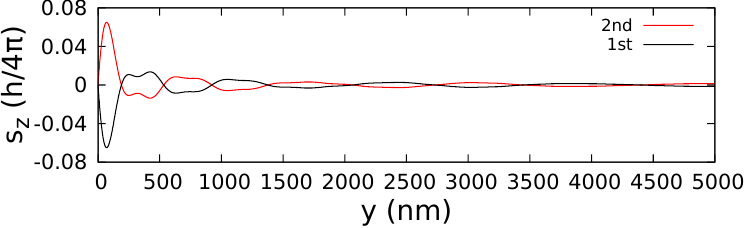}\put(-25,25){(d)}
\end{tabular}
\caption{(a) Trajectory and (b–d) average spin components of the Kramers doublet states flowing throughout the channel. Black and red lines correspond to the first and second states of the Kramers doublet for $d=175$~nm. In (a) and (b) additionally blue line markes results for the lowest state and  a thinner channel $d=125$~nm. Results for for $B_y=10$~$\mu$T.
 } \label{insb14}
\end{figure}

In Fig. \ref{insb14}(a) we plotted the average electron position for the  initial state set in the ground-state (black line) and the first excited state (red line). The period of mean $x$-position oscillations increases with $y$, as the electron packet is accelerated along the $y$-axis by the external field.
Importantly, the average $x$ positions for the two initial conditions are exactly opposite. The amplitude of the $x(y)$ oscillations is small in spite of the fact that the 
deflection of the wave packet in the $x$ direction [cf. Fig. \ref{evolution}(a,c)] is quite pronounced. The packet, however,  is not deflected as a whole and its parts are smeared over the opposite ends of the channel.

Figure \ref{insb14}(b-d) shows the evolution of the average spin components. As we can see, the polarization is initially present only for the $y$ component of the spin [Fig. \ref{insb14}(c)].
Initially, the packet moves parallel to the $y$ axis, which induces precession of the spin along the $x$ axis that promotes a polarization of the $z$-component of the spin [Fig. \ref{insb14}(d)]
which is opposite for both initial conditions.  Note that the average values of $s_y$ and $s_z$ are opposite for the initial states of the Kramers doublet during the entire time evolution. The absolute amplitude of their oscillations decreases along the channel, as the packet spreads over time and traverses regions with varying local spin texture [cf. Fig. \ref{evolution}(b,d)].
Remarkably, the electron packets in both initial conditions acquire the same component of the spin in $x$ direction. The wiggling of the electron path off the axis of the channel induces spin precession that leads to the accumulation of the spin in the $x$ direction. A semi-classical interpretation of this process is provided below.
In Fig. \ref{insb14} (a) and (b) we additionally plotted the results for the width of the channel decreased 
from $175$~nm to $125$~nm (blue line). As we can see, the period of the $x(y)$ oscillations decreases for a thinner channel, while the length over which the $s_x$ component accumulates increases. In particular, for a strictly one-dimensional (1D) channel, one would expect spin precession in the $(y,z)$ plane with no generation of the $s_x$ component. \\

Explanation of the charge and the spin dynamics for an electron released from QD ground state in the channel can be provided within a semi-classical approach, in which one treats the electron as a classical particle carrying a quantum spin. 
Let us consider Hamiltonian (1) as a classical Hamilton function. We neglect the Zeeman term, since the external magnetic field is very weak and applied solely to distinguish between the Kramers pairs.  The set of Hamilton equations for the particle position $x(t),y(t)$ and momentum $p_x(t),p_y(t)$ is given by
\begin{eqnarray}
\frac{dp_x}{dt}&=&-\frac{\partial H}{\partial x}=-\frac{\partial V}{\partial x}, \label{zaba}\\ 
\frac{dp_y}{dt}&=&-\frac{\partial H}{\partial y}=-\frac{\partial V}{\partial y}, \label{zaba2}\\
\frac{dx}{dt}&=&\frac{\partial H}{\partial p_x}=\frac{p_x}{m}+\alpha \sigma_y, \label{xt} \\
\frac{dy}{dt}&=&\frac{\partial H}{\partial p_y}=\frac{p_y}{m}-\alpha \sigma_x.\label{yt} 
\end{eqnarray} 
 We solve the set of equations (\ref{zaba} -- \ref{yt}) simultaneously with the Schr\"odinger equation for the spinor $S(t)$  \cite{sherman}
\begin{equation}
i\hbar \frac{\partial S}{\partial t}= H_{SO} S. \label{shso}
\end{equation} 
In Eq. (\ref{shso}) we use the Hamiltonian (\ref{hso}) but with classical momentum of the electron.
In our semi-classical model we will plug the average spin components $\sigma_{x/y}=\langle S|\sigma_{x/y}|S \rangle$  in the velocity 
equations Eq. (\ref{xt}) and (\ref{yt}).
For the purpose of the present paper we will refer to the calculation integrating Eqs. (\ref{zaba}--\ref{shso}) as a 'Hamilton model'.

For the classical simulation, we adopted a smooth lateral confinement instead of the  potential well to avoid hard collisions from the edges of the channel. We took a potential in the form $V(x)=V_0\exp(-x^4/h^4)$ with $2h=180$ nm.
Note, however, that the semi-classical approach implies separation of the spin and position degrees of freedom. The separated spinor is a pure state that lies on the Bloch sphere. On the other hand, in the quantum calculation  the spin orientation  depends on the electron position in space and the average spin values for the initial condition $\langle \sigma_x\rangle =\langle\sigma_z\rangle =0$ and $\langle\sigma_y\rangle \simeq \pm 0.1$ - see Fig. \ref{insb14} - do not lie on the Bloch sphere.
Therefore, within the semi-classical model we cannot
reproduce the quantum initial conditions for the spin. Instead we take the eigenstates of $\sigma_y$ operator, e.g. the orientation of the spin in the maximum of the initial wave packet 
in the center of the quantum dot [see Fig. \ref{wf}(b-d) and Fig. \ref{evolution}].\\
 \begin{figure}[!t]
\begin{tabular}{l}
\includegraphics[width=0.8\columnwidth]{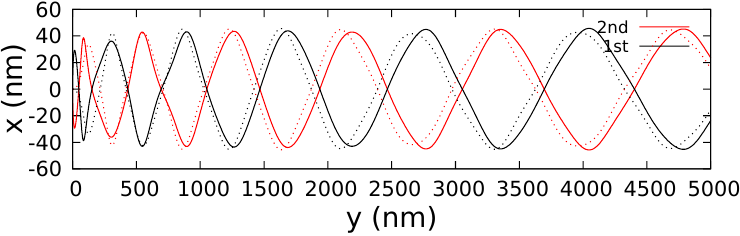}\put(-25,25){(a)}\\
\includegraphics[width=0.8\columnwidth]{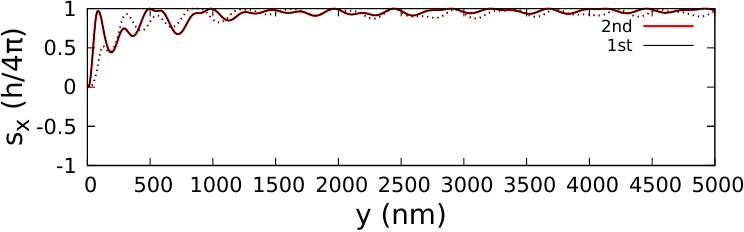}\put(-25,25){(b)}\\
\includegraphics[width=0.8\columnwidth]{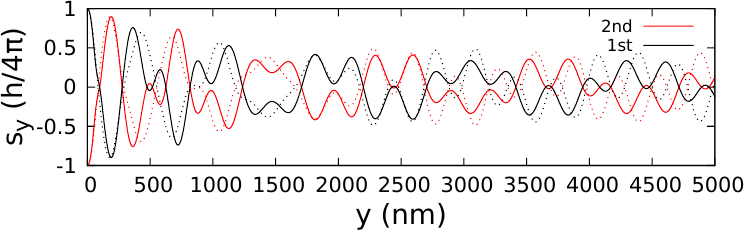}\put(-25,25){(c)}\\
\includegraphics[width=0.8\columnwidth]{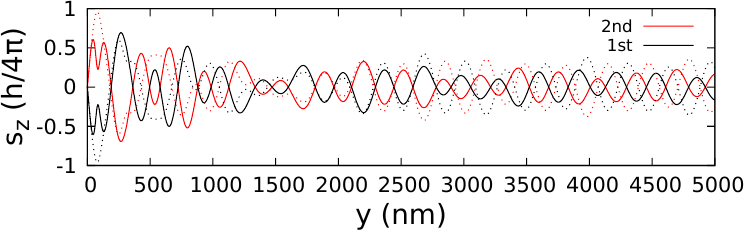}\put(-25,25){(d)}
\end{tabular}
\caption{As Fig. \ref{insb14} but for a classical model: solid -- Hamilton equations, dashed -- Heisenberg-like model.} 
\label{kinsb14}
\end{figure}
The results for the position and the spin as calculated via this semi-classical 'Hamilton model' are given in Fig.~\ref{kinsb14} with the solid lines. The $x(y)$ oscillation in  Fig. \ref{kinsb14}(a) is qualitatively similar to the quantum results [cf.  Fig. \ref{insb14}(a)].
 We note that the scale of the oscillations in the semi-classical and the quantum models
are very different. The classical point-charge electron can reach the edges of the channel while the quantum wave function occupies the entire channel and even if it shifts to the edge it is never exactly localized there. Note that the applied electron effective mass is very small even for III-V materials which prevents a stronger localization of low-energy wave functions.  
 The accumulation of the $x$ component of the spin for both the initial conditions [Fig. \ref{insb14}(b)] is reproduced by the semi-classical model, albeit the values in Fig. \ref{kinsb14}(b) (solid line) do not saturate as in the quantum calculation and weak oscillations are still observed at large $y$. Also, the oscillations of the $s_x$ and $s_y$ average values do not fall to zero at later stages of the semi-classical calculation in contrast to the quantum calculation. In the semi-classical model, the average values of the spin states lie on the Bloch sphere all along the time evolution. 

A less naive semi-classical model can be produced with the evaluation of the 
''force'' with the Heisenberg equation ${\bf F^{SO}}=m\frac{d^2{\bf r}}{dt^2}=-\frac{m}{\hbar^2}\left[H,[H,r]\right]$. 
For the evaluation of the commutators, we take the Hamiltonian (1) without the Zeeman term that is neglected
since the applied external magnetic field is very small. 
The calculation produces 
\begin{equation} 
(F^{SO}_x,F^{SO}_y)=\frac{2\alpha^2m}{\hbar^2} (k_y,-k_x)\sigma_z\label{force} \end{equation}
which is the spin-dependent force that leads to the redistribution of electrons with opposite spins toward opposite edges of the sample, known as the SHE~\cite{sinova2004}.
We solve the Newton equations including the SO force (Eq.~\ref{force}) with the spin evolution integrated with the Schr\"odinger equation as in the previous model and use the average $z$ component of the spin $\sigma_z=\langle S|\sigma_z |S \rangle$ for Eq. (\ref{force}). We will refer to the calculation with the SO force as the Heisenberg-like model. The results of the Heisenberg-like model are given in Fig. \ref{kinsb14} with the  dotted lines.

Although a different spin component enters the equation of motion in the Hamilton [Eq. (\ref{xt}) and (\ref{yt})] and Heisenberg [Eq. (\ref{force})] models, both calculations provide qualitatively similar results for the position and the spin -- see Fig. \ref{kinsb14}.  
\begin{figure}[!t]
\begin{tabular}{l}
\includegraphics[width=0.8\columnwidth]{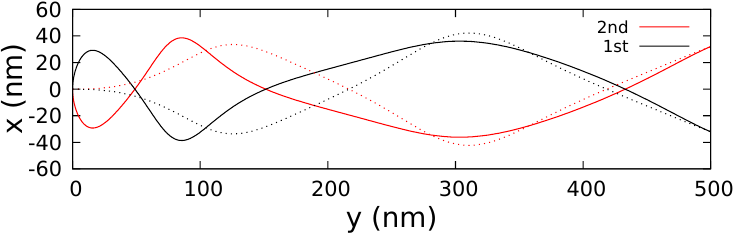}\put(-25,25){(a)}\\
\includegraphics[width=0.8\columnwidth]{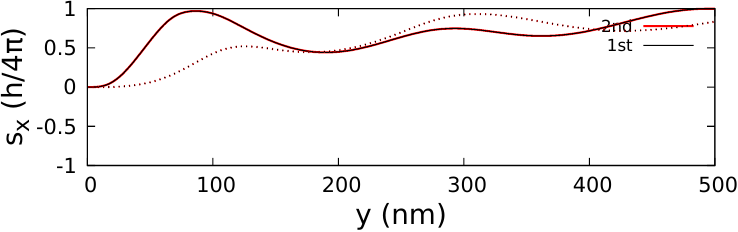}\put(-25,25){(b)}\\
\includegraphics[width=0.8\columnwidth]{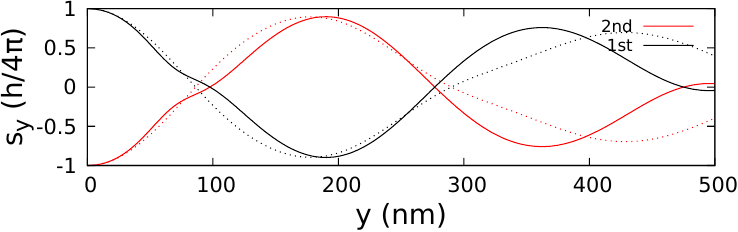}\put(-25,25){(c)}\\
\includegraphics[width=0.8\columnwidth]{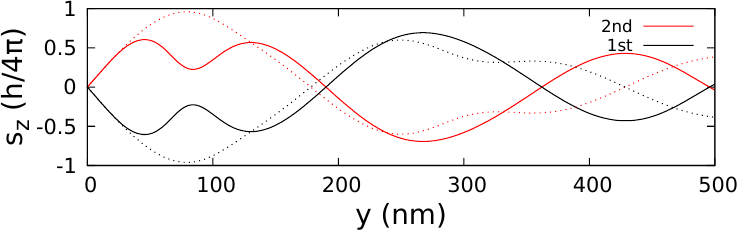}\put(-25,25){(d)} \\
\includegraphics[width=0.8\columnwidth]{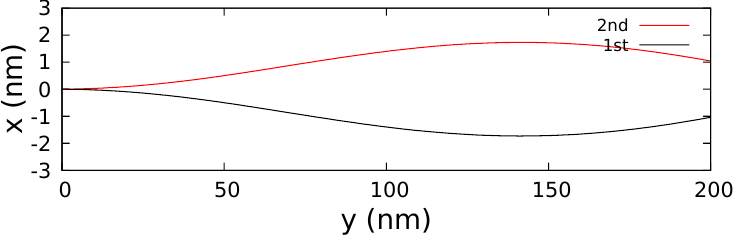}\put(-25,25){(e)}
\end{tabular}
\caption{(a-d) Zoom of Fig. \ref{kinsb14} and (e) zoom of Fig. \ref{insb14}(a)} \label{zoom}
\end{figure}
A qualitative difference is only visible at the start of the simulation that is depicted in an  enlarged fragment in Fig. \ref{zoom}(a-d). In the Heisenberg-like model, the displacement off the $x$ axis appears only  after the spin initially oriented in the $\pm y$ direction precesses to $z$ direction
along the $x$ axis perpendicular to the electron velocity. On the other hand, in the  Hamilton model, the  shift in the $x$ direction appears right after the start since the $v_x$ component due to spin in Eq. (\ref{xt}) is nonzero at $t=0$. 
In the Heisenberg-like model the spin acquires the $x$ component with a delay, since the deflection of the electron path appears only after $s_z$ component is accumulated by the precession of the spin initially oriented in the  $\pm y$ direction. 
For comparison Fig. \ref{zoom}(e) shows the zoom on the quantum mechanical average position. The path of the average values of the quantum calculation resembles not the Hamilton model but the Heisenberg-like model, which is therefore more adequate.
 
In both the models the $x(y)$ oscillation appears after the release of the electron from the quantum dot  and the oscillatory motion at a later stages of the motion pertains by the energy conservation 
since both $s_z$ and $s_y$ values are rather small for $y>2000$ nm.
Note that in both the quantum calculation [Fig.\ref{insb14}(c,d)] and semi-classical models [Fig.\ref{kinsb14}(c,d)] the oscillations amplitude is smaller for $s_z$ than for $s_y$. The period of $x(y)$
oscillations increases in $y$ as in the quantum calculation due to the motion in the electric field applied for $t>0$.

The classical models can also shed some light on the evolution of the electron spin observed in the quantum  calculations. For a classical particle, the behavior of its spin can be understood as a precession in the effective SO magnetic field ${\bf B}_{SO}$. 

The precession can be described in terms of the Bloch equations $\dot {\bf s}=\frac{g\mu_B}{2} {{\bf B}_{SO}}\times{\bf s}$  with \begin{eqnarray}
\dot s_x&=&\alpha s_z k_x, \label{dsx} \\ \dot s_y&=&\alpha s_z k_y, \label{dsy} \\ \dot s_z&=&-\alpha (k_ys_y+k_xs_x). \label{dsz}\end{eqnarray}
The electron spin is initially oriented in the $\pm y$ direction [Figs. 
\ref{insb14}(b-d) and \ref{zoom}(b-d)]. The electron starts to move in $+y$ direction
by the homogeneous electric field applied for $t>0$. By Eq. (\ref{dsz}) with nonzero $s_y$ and $k_y$ the electron acquires $s_z$ component of the opposite sign
of the initial sign of $s_y$. In the Heisenberg model
the force in the $x$ direction has the orientation parallel to $s_z$ [Eq. (\ref{force})] and thus  after the start of the motion the sign of $k_x$ is opposite to the initial $s_y$ orientation  [cf. Fig. \ref{kinsb14}(a) and (c)]. 
The sign of the product of $s_z$ and $k_x$ in Eq. (\ref{dsx}) is then positive for both initial $s_y$ orientations and 
$s_x$ starts to grow for both initial conditions, i.e. when the electron starts to move to positive $x$ with positive $s_z$ {\it or}
when the electron starts to move to negative $x$ with negative $s_z$ [see the dotted lines in Fig. \ref{zoom}(a-d)].
Hence, for both initial $s_y$ polarizations, we find the same effect on $s_x$ similarly as in the quantum calculation
(Fig. \ref{insb14}). 

For a strictly 1D channel the electron will move parallel to $y$ axis ($k_x=0$)
only and a precession of the spin would involve only its $y$ and $z$ components. However, when the electron path is deflected in the direction perpendicular
to the channel axis non-zero values of $s_x$ can appear. 
According to Eq. (\ref{dsx}) the value of $s_x$ changes when $\sigma_z$ and $k_x$ are simultaneously non-zero. 
The electron spin rotates from $z$ to $x$ direction when it moves along the $x$ axis. 
For the Hamilton model (solid lines in Fig. \ref{zoom}) we see that $s_x$ reaches temporarily 1 for $y=90$ nm -- when the electron acquires a local extremum of $x(y)$ or $k_x$ changes sign in agreement with Eq. (\ref{dsx}).
In the Heisenberg-like model (dotted lines in Fig. \ref{zoom}) $s_x$ starts to grow only when $s_z$ becomes large enough to produce the lateral force. In this model, $s_x$ stops to grow for $y\simeq 120$ nm  when the lateral motion of the electron is reversed, also in agreement with Eq. (\ref{dsx}).  
In the quantum and both semi-classical approaches for large $y$ the spin becomes nearly polarized in the $x$ direction [cf. Fig. \ref{insb14}(b) and Fig. \ref{kinsb14}(b)]. The electron is accelerated by the electric field, resulting in an increase in its momentum in the $y$-direction. In the limit of large $k_y$ the magnetic field ${\bf B}_{SO}$ becomes parallel to the $x$ direction, which stabilizes the spin polarization in this direction.

Summarizing the above findings, for an electron released into a straight channel from one of the two quantum dot lowest-energy states initially spin-polarized in the direction of the channel,
the spin-orbit coupling declines its path in opposite directions for both the states of the Kramers ground-state doublet. The effect
is pronounced also in a small external magnetic field in spite of an incomplete initial spin polarization and in spite of the fact
that the spin polarization soon after the electron release turns similar for both the states and is oriented perpendicular to the axis of the channel.
\begin{figure}[!t]
\begin{tabular}{ll}
\includegraphics[trim= 0 0 0 1cm,clip, width=0.5\columnwidth]{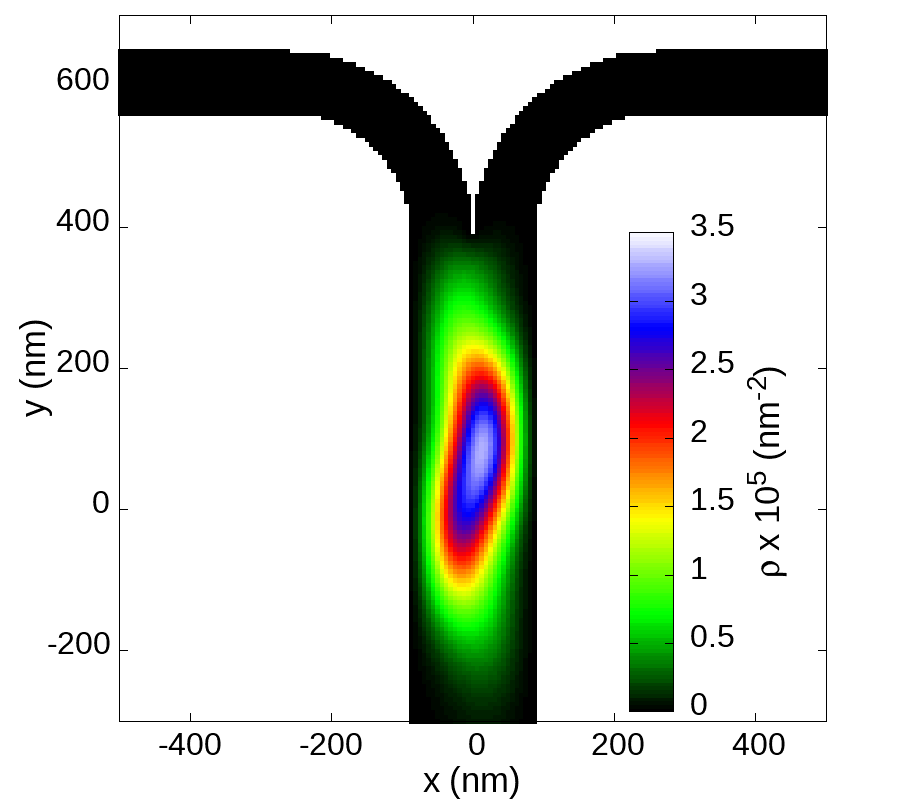}\put(-84,110){\tiny 1.06ps} \put(-99,20){(a)} &
\includegraphics[trim= 0 0 0 1cm,clip, width=0.5\columnwidth]{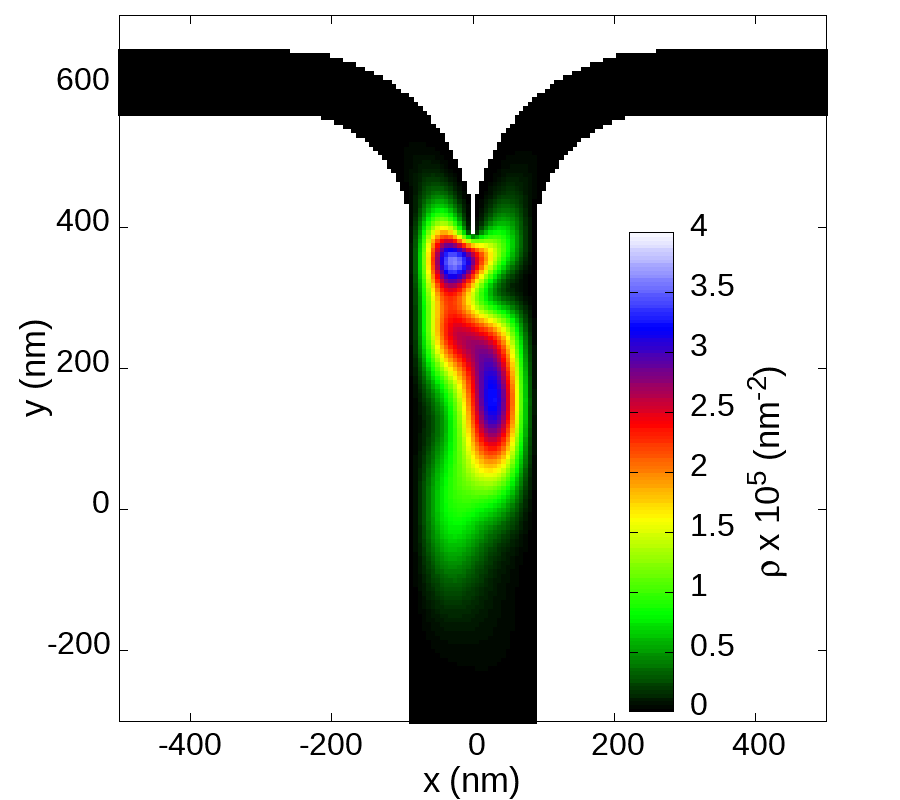}\put(-84,110){\tiny 1.93ps} \put(-99,20){(b)} \\
\includegraphics[trim= 0 0 0 1cm,clip, width=0.5\columnwidth]{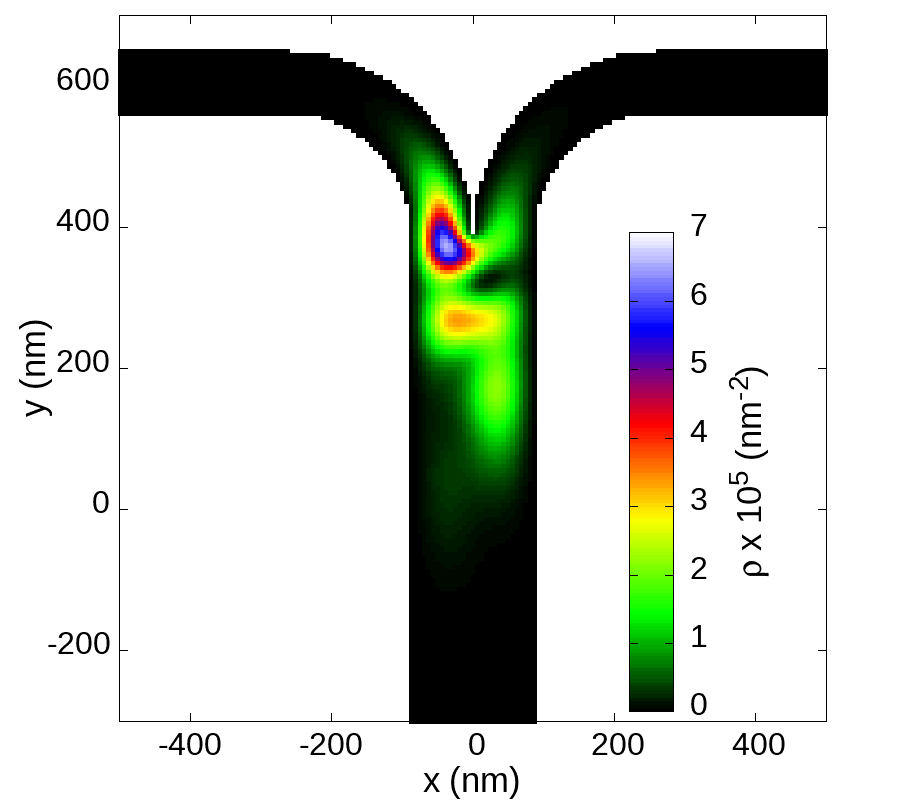}\put(-84,110){\tiny 2.42ps} \put(-99,20){(c)} &
\includegraphics[trim= 0 0 0 1cm,clip, width=0.5\columnwidth]{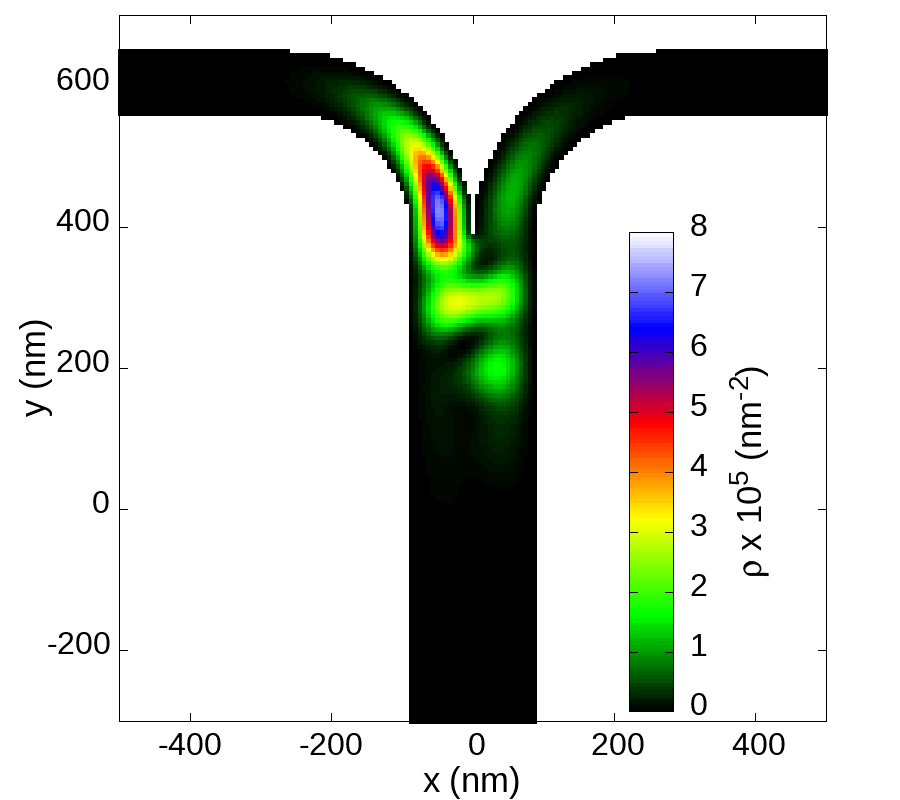}\put(-84,110){\tiny 2.90ps} \put(-99,20){(d)}  \\
\includegraphics[trim= 0 0 0 1cm,clip, width=0.5\columnwidth]{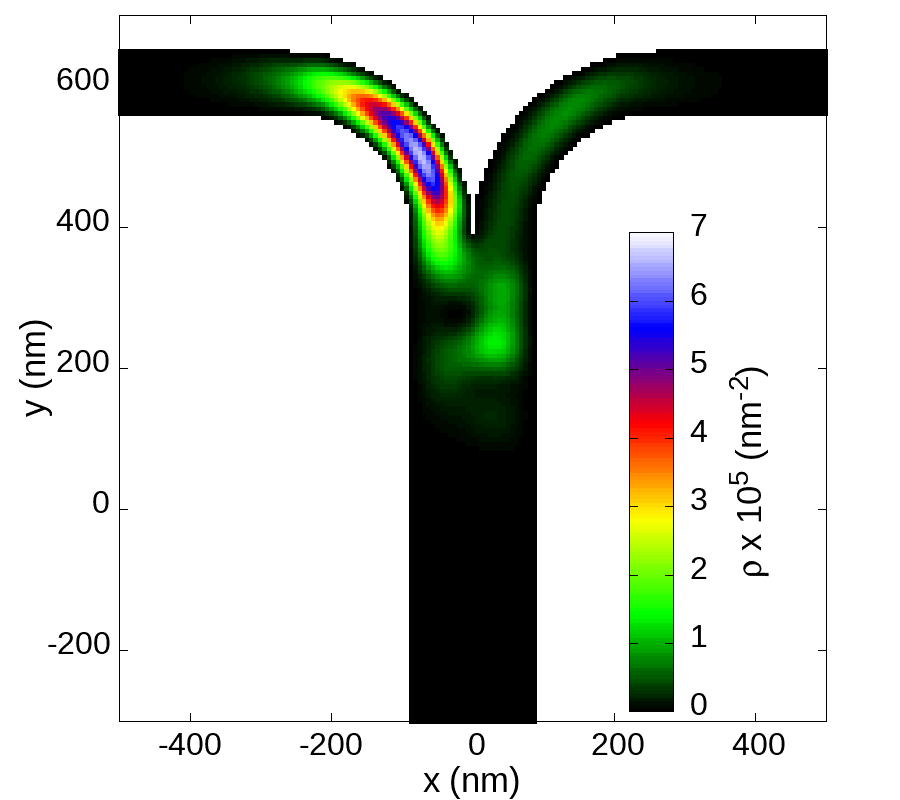}\put(-84,110){\tiny 3.38ps} \put(-99,20){(e)} &
\includegraphics[trim= 0 0 0 1cm,clip, width=0.5\columnwidth]{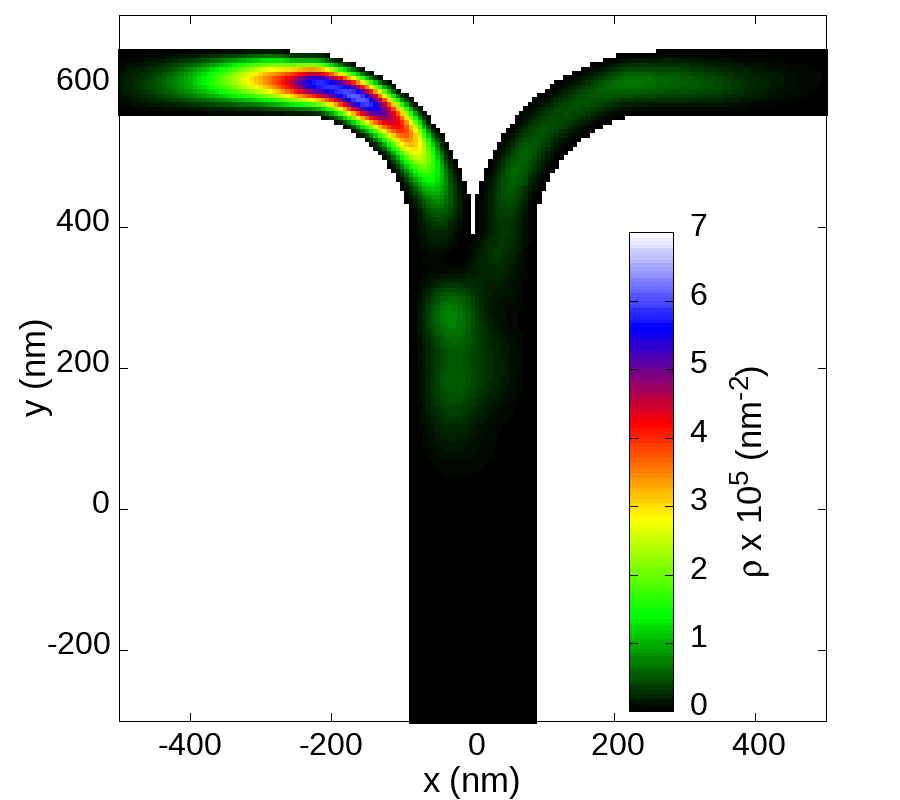}\put(-84,110){\tiny 3.87ps}\put(-99,20){(f)}  \\
\includegraphics[trim= 0 0 0 1cm,clip, width=0.5\columnwidth]{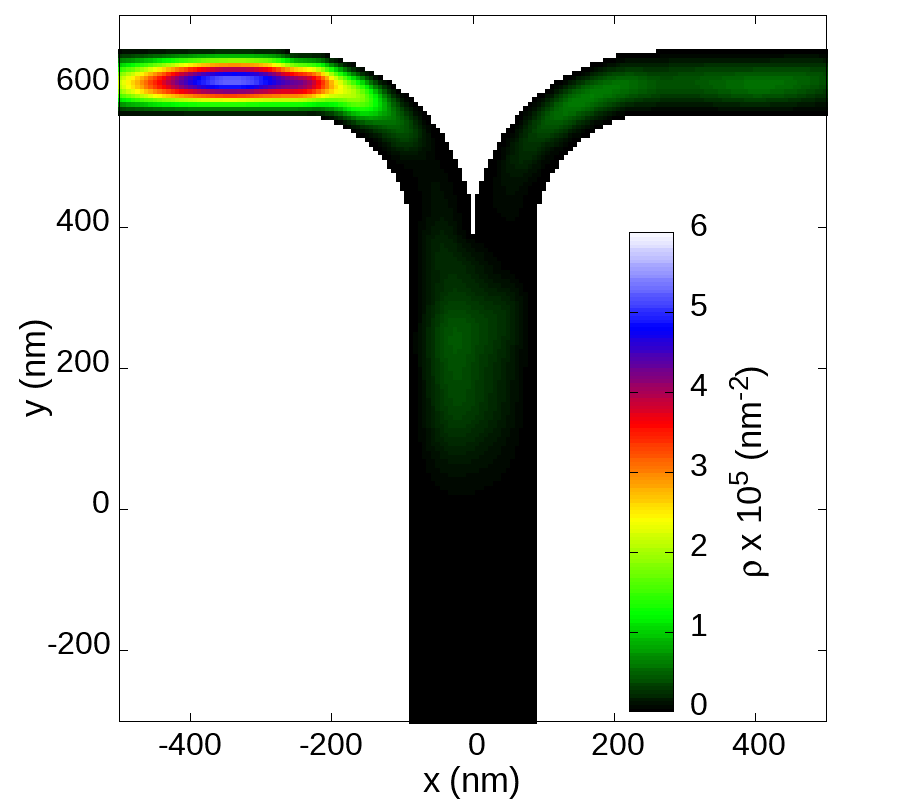}\put(-84,110){\tiny 4.35ps} \put(-99,20){(g)} &
\includegraphics[trim= 0 0 0 1cm,clip, width=0.5\columnwidth]{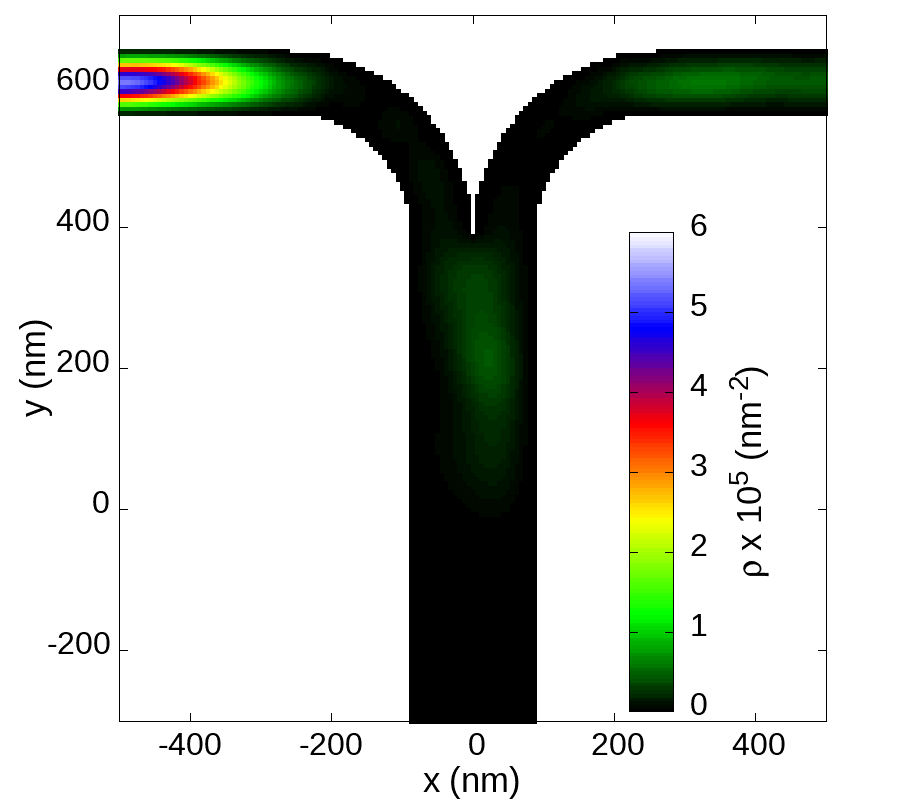}\put(-84,110){\tiny 4.84ps} \put(-99,20){(h)} \\
\end{tabular}
\caption{
Snapshots of the probability density packet for the initial state prepared as in Fig. \ref{evolution}(c)
i.e. in the state with a partial spin polarization antiparallel to the $y$ axis. Results for the channel that is split [see Fig.~1] at a distance of $b=387$ nm from the center of
the quantum dot. } 
\label{movies}
\end{figure}

\subsection{The initial state detection}

The lateral motion of the packet that depends on the initial condition can be used to distinguish between the two states of the Kramers doublet with the channel splitted into two drain leads [Fig.~1] at a distance $b$ from the quantum dot center - see Fig.~\ref{s0}. The leads have width $t=d/2$ and the split of the input channel is modeled by arcs centered at $P=(\pm 3t,b)$ with inner and outer radii of $2t$ and $3t$ respectively. The ratio of $t/d$ and the shape of the split were chosen for their relatively high selectivity in the detection of the initial condition and for the relatively short time that the electron packet takes to leave the input channel to the drain outputs. A similar experimental setup has been recently used in Ref.~\cite{Strambini2009}.
 \begin{figure}[!h]
 \begin{tabular}{l}
\includegraphics[width=0.65\columnwidth]{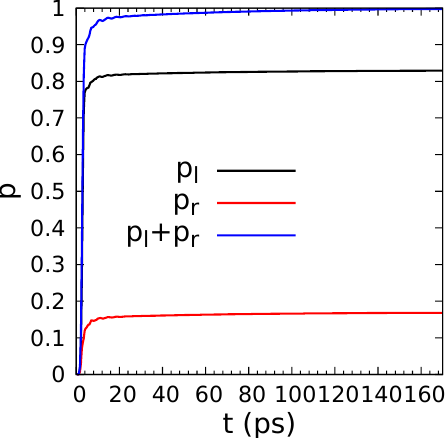} %\put(-29,35){(a)} 
\\
\end{tabular}	
\caption{The parts of the electron packet in the left ($p_l$, black line) and right output ($p_r$, red line) channels and their sum as a function of time for the input channel of width $d=175$~nm and the split from the center of the dot at a distance of $b=387$~nm. The QD state with a partial spin polarization antiparallel to the $y$ axis is taken as the initial condition. The results correspond to the time evolution shown in Fig. \ref{movies} and to the leftmost extrema in $p_{l/r}(b)$ dependence of Fig. \ref{wfub}(a), respectively. 99.7\% of overall transfer reached after 155ps.  99.8\% after $t=173$~ps.
} 
\label{tmovies}
\end{figure}

In Fig. \ref{movies} we show the snaphots for the evolution of the packet  for the segment length of $b=387$ nm and the initial condition set at the ground-state with positive $s_y$. The parts of the packet flowing out through the left and right output leads are plotted in Fig. \ref{tmovies}.
In the considered case, the packet enters the split near $1.93$~ps [Fig. \ref{movies}(b)] and most of the packet goes into the left output lead [Fig. \ref{movies}(e-f)] but with non-zero transfer to the right output. The input channel is nearly emptied for $t>4$ ps [Fig. \ref{movies}(g,h)], however, for $t=4.83$ps [Fig. \ref{movies}(h) and Fig. \ref{tmovies}] still only about 91.5\% of the packet
left to the drain electrodes from the input channel. The residual parts of the packet eventually leave the channel thanks
to the electric potential slope [Fig. \ref{tmovies}]. We integrate the Schrödinger equation until at least 99.8\% of the packet has exited the input channel into one of the drain leads. For $b=387$ nm the overall transfer level of 99.8\% is reached for $t=173$ps. 
%The time to be covered by this calculation depends on $b$ and is usually the longest for shortest $b$ where the drop of the electric potential due to the homogeneous field along the channel is small.
Fig. \ref{movies} shows only the crucial but small part of the computational box, which covers the input channel of length 10~$\mu$m, and output channels of length 39~$\mu$m each to avoid any effects of packet backscattering from the edges of the channel for a simulation that needs to last long.

 \begin{figure}[!t]
\begin{tabular}{l}
\includegraphics[width=0.8\columnwidth]{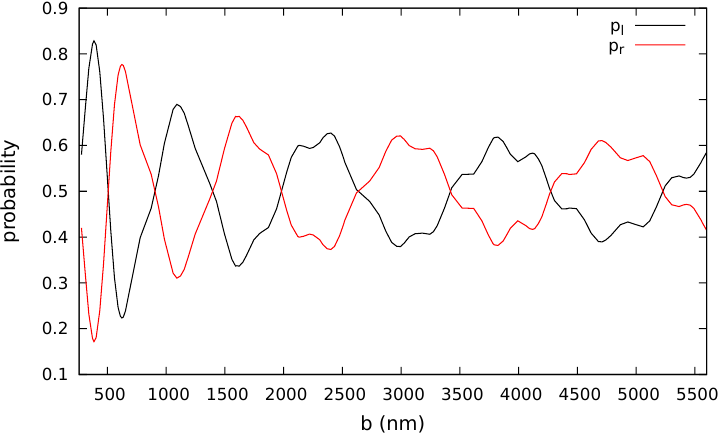}\put(-25,25){(a)}\\
\includegraphics[width=0.8\columnwidth]{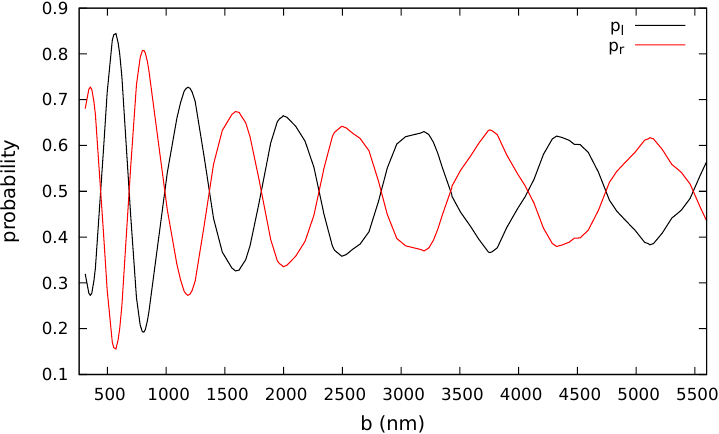}\put(-25,25){(b)}\\
\includegraphics[width=0.8\columnwidth]{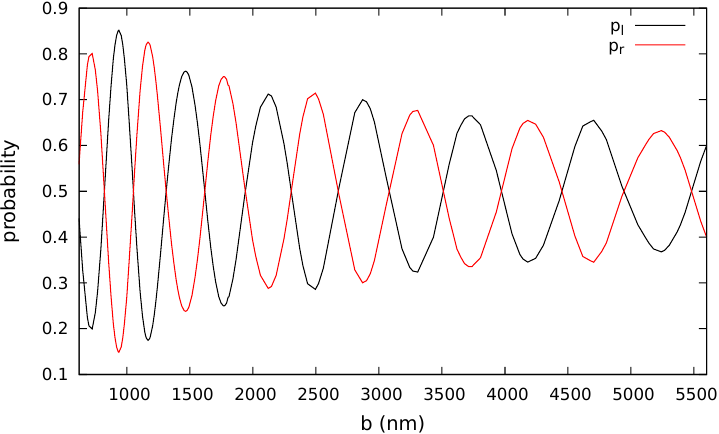}\put(-25,25){(c)}\\
\includegraphics[width=0.8\columnwidth]{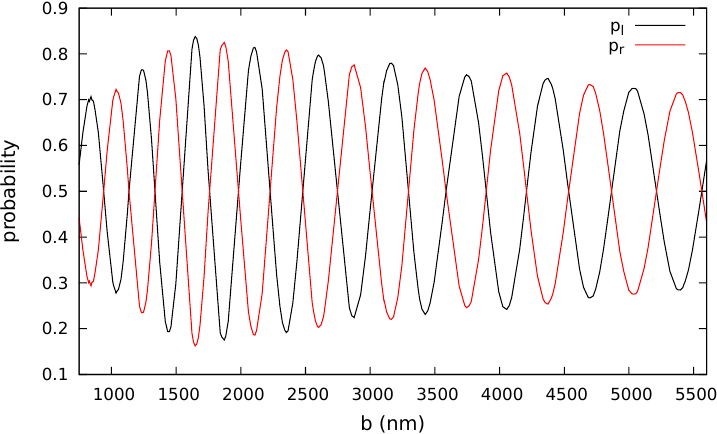}\put(-25,25){(d)}
\end{tabular}
\caption{The transfer probability of the wave packet released from the QD to the left and right output channels as a function of $b$. Results are shown for channel widths: (a) $d=175$~nm, (b) $d=150$~nm (c) $d=125$~nm (a) $d=100$~nm and the initial state set as  the higher state of the Kramers doublet, with partial spin polarization along the $-y$-direction.}
\label{wfub}
\end{figure}

Figure \ref{wfub}(a) shows the transfer probability to the left and right output channels as a function of  $b$ -- the distance between the quantum dot and the channel split.
The maximal value of $|p_l-p_r|$ is obtained for $b=387$ nm, where the transfer probability of the ground state, i.e. lower of the Kramers doublet state with a partial spin polarization along the $y$ axis, to the left output lead is about $p_l=0.82$ with the transfer probability to the right drain lead of about $p_r=0.18$. The probabilities for the higher-energy state of the  Kramers doublet with the partial spin polarization along $-y$ direction are exactly opposite. 
The probability of the correct identification of the state by the charge detection is then about 82\%. 
The second extremum is found for $b=624$ nm [see Fig. \ref{wfub}(a)] with $p_r=77.1\%$ and $p_l=22.9\%$. 
Note that the optimal $b$ depends on the width of the channel.  Figure \ref{wfub}~(b),(c) and (d) show the results for $d=150$ nm, $125$ nm and $100$ nm, respectively. The optimal value of $b$ is larger for lower $d$.  
For $d=150$~nm the optimal $b$ is 564 nm with $p_l\simeq 84.5\%$ and $p_r\simeq 15.5\%$. The corresponding values for $d=125$ nm and $d=100$ nm are: 973 nm with $p_l\simeq 85.1\%$ and $p_r\simeq 14.9\%$
and 1641 nm with  $p_l\simeq 83.8\%$ and $p_r\simeq 16.2\%$. Note, on the other hand, that the maximal spread of $p_l$ and $p_r$ values
weakly depends on $d$. For thinner $d$ the period of $p_{l/r}$ oscillations as functions of $b$ increases due to a larger number of scattering from the  edges of the thinner channel per unit channel length.

Note, that the results given above indicate that the proposed method works even for very low initial spin polarization, e.g. it produces very different electron trajectories for the states 
of the Kramers doublet which differ more in the spatial distribution of the spin density than the average spin. 

As a final remark, it is important to note that the results presented above correspond to spins initially oriented in the $y$-direction, i.e., along the channel. For the magnetic field oriented perpendicular to the channel, in the $x$ direction, the Kramers doublet quantum-dot-confined states used as the  initial conditions are polarized in the $x$ direction. The charge density for both the initial conditions evolves in time in exactly the same manner producing $p_l=p_r$ that prohibits detection of the initial state from the orbital motion of the electron.

\subsection{Channel potential and lateral oscillations of the wave packet}
\subsubsection{harmonic lateral confinement}
Electron motion in quantum wires with harmonic oscillator lateral potential $V_h(x)=\frac{1}{2}m^* \omega_x^2 x^2$ without and in-plane electric field was studied 
in Ref. \cite{Schliemann2006} in a subspace spanned by four lowest-energy eigenstates of the
lateral harmonic confinement: two lowest-energy states of the spatial quantization and both spin $z$ component eigenstates. The model \cite{Schliemann2006} used as an initial condition a Gaussian wave packet in the lateral direction $x$ with spin polarized along the $z$ direction and a definite wave vector in the $y$ directon $k_{y}$. A maximal amplitude of $\langle x\rangle(t)$ oscillations was found for a resonant condition
$2\alpha k_y=\hbar \omega_x$. 
The present approach differs in a few aspectes with the model \cite{Schliemann2006}: i) the lateral confinement in the present approach is defined by an infinite quantum well potential; ii) the initial condition is taken as the ground-state of the quantum dot confinement, i.e. the initial state is localized also in the $y$ direction;  iii) due to the in-plane orientation of the applied magnetic field and the spin-orbit coupling the initial spin is not in an eigenstate of $\sigma_z$ operator; iv) the momentum along the $y$ component is acquired from the in-plane electric field and not assumed a priori, iv) the wave vector is non-definite and its average value grows in time. It is nevertheless interesting to relate the present model with the results of Ref. \cite{Schliemann2006}. 

For the purpose of the comparison we set the channel potential to harmonic $V_\infty(x,y)=V_h(x)$, removed the external magnetic field and set the initial condition in form of a Gaussian wave packet
\begin{equation} \Psi(x,y,t=0)=N\exp(-\frac{m\omega_x}{2\hbar}(x^2+y^2)+ik_y y)\left(\begin{array}{l}1 \\ 0 \end{array}\right), \label{gauszian} \end{equation} where $N$ is the normalization factor and the spin is polarized along the $z$ axis.  Note, that for the wave vector introduced in this way $\hbar k_y$ is an average value of the initial momentum in the $y$ direction and not a definite quantity. We used $\hbar\omega_x=0.15625$ meV as above for the quantum dot potential. The width of the computational box was increased to 1.4 $\mu$m to allow for a free lateral motion of the wave packet.

 \begin{figure}[!t]
\begin{tabular}{l}
\includegraphics[width=0.6\columnwidth]{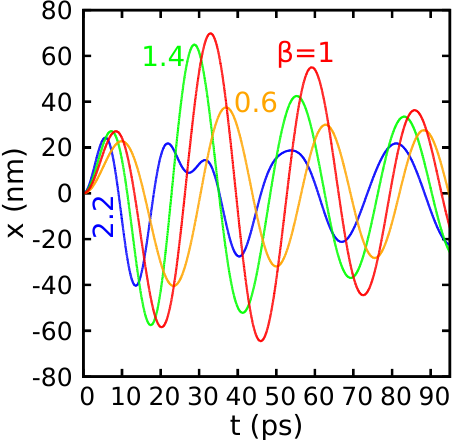}\put(-20,25){(a)}\\
\includegraphics[width=0.7\columnwidth]{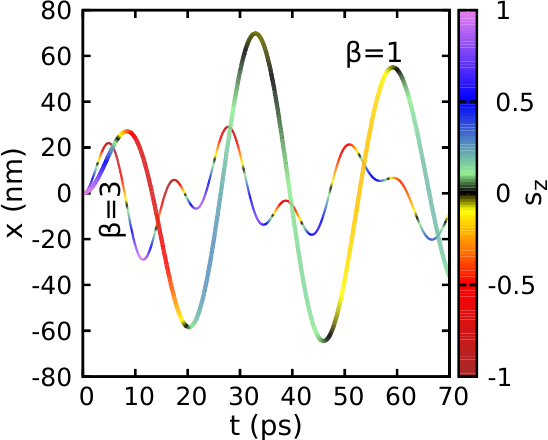}\put(-45,25){(b)}
\end{tabular}
\caption{ Average $x$ position of the wave packet for the channel potential given by harmonic oscillator  confinement $\frac{m^*\omega_x^2 x^2}{2}$ with $\omega_x=0.15625$ meV, the initial condition set as a Gaussian with average momentum equal to $\hbar k_y=\frac{\hbar\omega_x}{2\alpha}\times \beta$
and the electron spin initially polarized in the $z$ direction (see text) and $\alpha=5$meV nm. In (a) results for various values of $\beta$ are given by different colors. 
In (b) the color stands for the average $z$-component of the spin. }
\label{gat5}
\end{figure}

 \begin{figure}[!t]
\begin{tabular}{l}
\includegraphics[width=0.6\columnwidth]{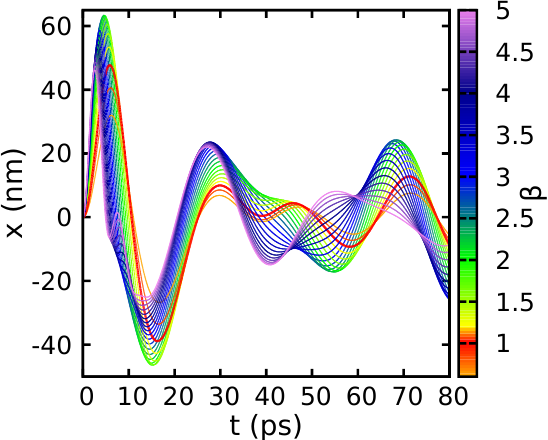}\put(-45,25){(a)}\\
\includegraphics[width=0.6\columnwidth]{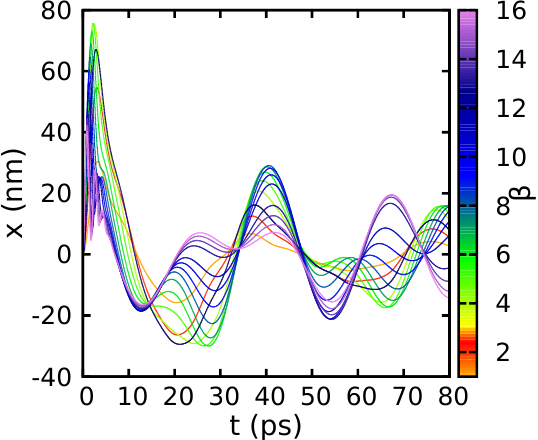}\put(-45,25){(b)}\\
\includegraphics[width=0.6\columnwidth]{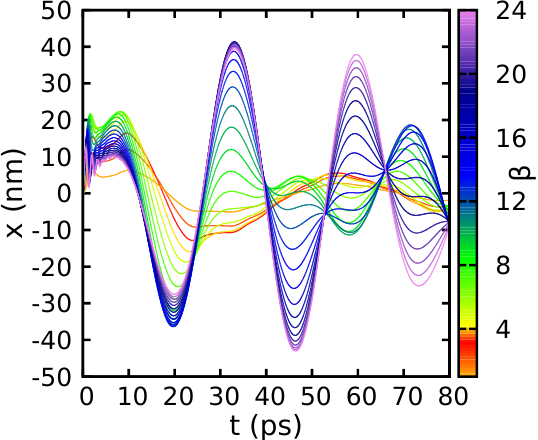}\put(-45,25){(c)} \\
\includegraphics[width=0.6\columnwidth]{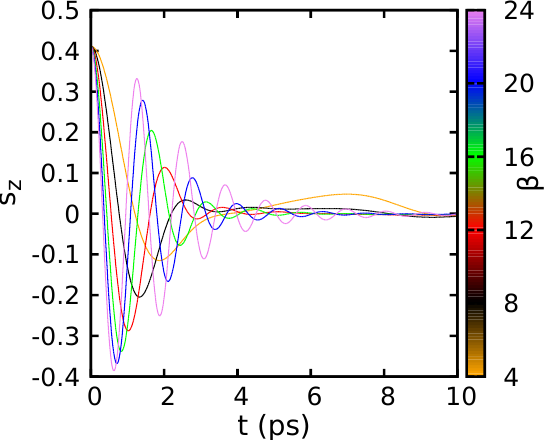}\put(-45,25){(d)}
\end{tabular}
\caption{Same as Fig. \ref{gat5}(a) but for $\alpha=20$ meV nm (a) and  $\alpha=50$ meV nm (b).
In (c) the QD ground state with $B_z=0.1$ T was applied as the initial condition. (d) shows 
the average spin $z$ component in $\hbar/2$ units for the setup of panel (c).}
\label{gat20}
\end{figure}

The $\langle x\rangle(t)$ dependence for spin-orbit coupling constant reduced from to $\alpha=50$ meV nm applied above to $\alpha=5$ meV nm are given in Fig. \ref{gat5}(a) for the average wave vector in the initial condition scaled by $\beta$ from the resonant \cite{Schliemann2006}  value 
$k_y=\frac{\hbar\omega_x}{2\alpha}\times \beta$ for $\beta=0.6, 1$ and 1.4. 
Initially the average packet goes to the right side of the channel the faster the larger is the wave vector -- due to the spin-orbit coupling deflection (see Eq. (\ref{force})). However, the amplitude of the oscillations is not a monotonic function of $\beta$ and the largest amplitude is found near $\beta=1$ in agreement with findings of Ref. \cite{Schliemann2006}. In Figure \ref{gat5}(b) we marked the orientation of the $z$ component of the spin on the $\langle x \rangle (t)$ curve for $\beta=1$ and $\beta=3$. For $\beta=1$ 
the z-component of the spin changes sign for $t=6.65$ ps [Fig. \ref{gat5}(b)]. Upon the inversion of the spin the ''force'' [Eq. \ref{force}] due to the spin-orbit coupling is directed to $-x$ direction.
Another inversion occurs for $t\simeq 20$ ps and at $t\simeq 32$ ps the $x$ obtains its maximal value of nearly 70 nm. At least up to this moment the spin Hall ''force'' (Eq. (\ref{force})) superposes with the force due to the harmonic oscillator potential.
On the other hand for $\beta=3$ the inversions of the spin occur due to the precession appear at a shorter time intervals than the motion of the wave packet. The direction of the motion of the average position changes with a larger frequency and in consequence
the absolute value of $\langle x \rangle$ does not reach the values found for $\beta=1$.  

The resonant condition \cite{Schliemann2006} was obtained in the approximation using four subbands 
for which the spin-orbit interaction was introduced as an off-diagonal term. The approximation is justified
when the scale of the Rashba interaction $\epsilon_R=\frac{m^*\alpha^2}{\hbar^2}$ is small compared
to the energy of the lateral quantization $\Delta E=\hbar\omega$. For the values applied in Fig. \ref{gat5},
$\epsilon_R=0.00459$ meV or $\epsilon_R=0.0294 \Delta E$. 
Figure \ref{gat20} shows the results for $\alpha=20$ meV nm (\ref{gat20}(a)) and $\alpha=50$ meV nm
(\ref{gat20}(b)), i.e. for $\epsilon_R=0.47\Delta E$ and $\epsilon_R=2.94\Delta E$, when
the spin-orbit energy is comparable and exceeds the lateral quantization, respectively.
In both cases the maximal amplitudes are obtained for $\beta>1$. For $\alpha=20$ meV nm 
 (Fig. \ref{gat20}(a)) the first two extrema of $\langle x \rangle$ (for $t=3$ ps and 15 ps) are obtained the largest for $\beta\simeq 2$,  the third ($t=28$ ps) is the largest for $\beta\simeq 4$ and the fourth is maximal ($t=40$ ps for $\beta\simeq 5$.
For $\alpha=50$ meV nm (\ref{gat20}(b)) the first maxima are obtained for $\beta \simeq 4$ 
and the value of $\beta$ corresponding to maximal amplitude grows with $t$. 

Concluding, the simulations of the wave packets reproduce the resonant condition between the spin-orbit force and the lateral confinement potential  for the
amplitude of the lateral oscillations
between the wave vector, the confinement energy and the Rashba coupling constant when the spin-orbit
coupling energy is much smaller than the quantization energy. For stronger spin-orbit coupling
the largest is the first maximum of the lateral quantization which is found for a wave vector
larger than the resonant one. The subsequent oscillations have smaller amplitudes and get larger
for even larger initial $k_y$.

In order to draw nearer the system to our main calculation in Figure \ref{gat20}(c) we plotted the results for the quantum dot ground-state  instead 
a Gaussian packet with polarized spin in the initial condition with  $\alpha=50$ meV nm and $B_z=0.1$T. 
The ground-state wave function is multplied by a plane wave $\exp(ik_y y)$ for setting 
the initial condition.
In the calculations we introduced $B_z$ only to the spin Zeeman interaction and neglected the orbital
effects for consistence of the comparison with the previous results. 
The average value of the initial spin is now $\langle s_z\rangle= 0.469 \frac{\hbar}{2}$ instead of  $1\times \hbar/2$ units.
Since the initial spin is only weakly polarized, the first extremum of $\langle x \rangle(t)$ is now much lower than for the polarized spin in Fig. \ref{gat20}(b). The subsequent extrema are larger with the value of $\beta$ corresponding to the largest
absolute value of the average $x$ growing in time. 

In Fig. \ref{gat20}(c) we see an appearance of a very rapid oscillations of $\langle x \rangle(t)$ within first few ps of the motion for largest $\beta$ which correspond
to the fastest precession of the spin and the corresponding flips of the spin-Hall force orientation. The average $s_z$ was plotted in Fig. \ref{gat20}(d) for the
set up of Fig. \ref{gat20}(c). A similar rapid oscillations with even larger amplitude  was found for the spin initially polarized in the z-direction in Fig. \ref{gat20}(b).

For $\alpha=5$ meV nm the results obtained for the QD ground-state (not shown) are very similar
to the Gaussian wave packet as the initial condition, i.e. to the results of Fig. \ref{gat5}
since for this value of $\alpha$ the spin is already nearly polarized for $B_z=1$ T, i.e.  $\langle s_z \rangle=0.9926  \frac{\hbar}{2}$.

\subsubsection{Quantum well confinement}
\begin{figure}[!t]
\begin{tabular}{l}
\includegraphics[width=0.6\columnwidth]{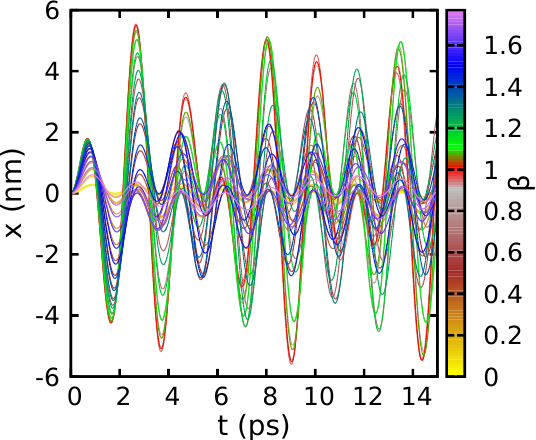}\put(-45,25){(a)}\\
\includegraphics[width=0.6\columnwidth]{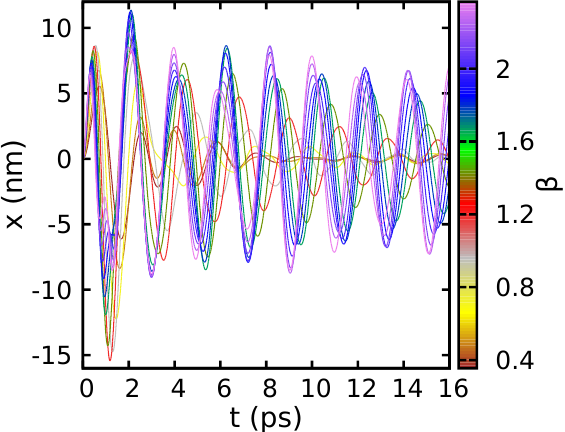}\put(-45,25){(b)}\\
\end{tabular}
\caption{Average value of $x$ for the initial condition set by the ground-state wave function
multiplied by a plane wave $\exp(ik_yy)$ for $\alpha=5$ meV nm (a) and $\alpha=50$ meV nm (b).
The channel is taken in form of a quantum wire of length $d=175$ nm with an infinite quantum well
lateral potential. External perpendicular magnetic field $B_z=0.1$ T is applied and its orbital
effects are neglected. 
The initial average momentum is set us $\hbar k_y=\frac{\Delta E}{2\alpha}\times \beta$,
and the values of $\beta$ are given by the colorscale.}
\label{qw20}
\end{figure}

Let us now reconsider the results presented above in this subsection but for the lateral quantum wire 
confinement set as an infinite quantum well as in the main results of the present model (Fig. 2-11).
The initial condition is taken as the ground-state wave function of the QD with the magnetic field oriented
in the $z$ direction multiplied by a plane wave as above. The orbital effects of the magnetic field are still neglected.   We set the width of the wire $d=175$ nm. The energy spacing between the lowest and second
subband of the wire is $\Delta E=\frac{3}{2} \frac{\pi^2\hbar^2}{m^*d^2}=2.64$ meV.
The resonant condition of Ref. \cite{Schliemann2006} can be translated replacing $\hbar \omega_x$ by $\Delta E$,
$k_y=\frac{\Delta E}{2\alpha}$. We study  $\langle x \rangle (t)$
as a function of $\beta$ where the initial wave vector is set as $k_y=\beta \frac{\Delta E}{2\alpha}$.
For the small value of the spin-orbit constant $\alpha=5$ meV nm (Fig. \ref{qw20}(a)) the maximal amplitude is obtained for $\beta\simeq 1$  -- similarly as for the harmonic confinement.
For $\alpha=50$ meV nm (Fig. \ref{qw20}(b)) the first extrema of the oscillations are obtained for $\beta\simeq 1.2$
and subsequent oscillations get larger for larger values of $\beta$, i.e. qualitatively similar to the harmonic lateral confinement studied in Fig. \ref{gat20}(c,d).

\begin{figure}[!t]
\begin{tabular}{l}
\includegraphics[width=0.6\columnwidth]{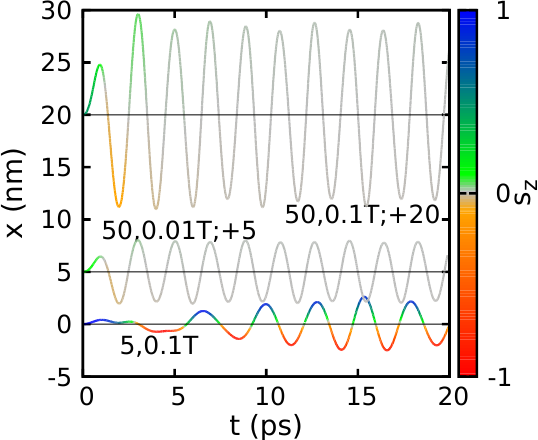}
\end{tabular}
\caption{Average value of $x$ for the initial condition set by the ground-state wave function
and in-plane electric field of $eF=0.01$meV/nm. We consider the values of $\alpha=5$  and $\alpha=50$ meV nm (b) and 
The channel is taken in form of a quantum wire of length $d=175$ nm with an infinite quantum well
lateral potential. External perpendicular magnetic field $B_z=0.1$ T (for $\alpha=5$ and 50 meV nm)
and $B_z=0.01T$  is applied (for $\alpha=50$ meV nm). The colorscale gives the average $z$ component of the spin in $\hbar/2$ units. The curves for $\alpha=50$ meVnm are shifted up for clarity by +5 and +20
for $B_z=0.01$T and $B_z=0.1$T, respectively.
}
\label{lastguwno}
\end{figure}

Finally, we introduce the in-plane electric field $eF=0.01$ meV/nm along the axis of the channel and we no longer impose artifically the momentum by multiplication of the QD ground-state wave function by a plane wave. 
The results for the amplitude of the oscillations are given in Fig. \ref{lastguwno}.The results of Fig. \ref{lastguwno} correspond to the model used for Fig. \ref{insb14}(a) only with the magnetic field oriented along the $z$ direction and not $y$. 
For $\alpha=50$ meVnm we set $B_z$ to 0.01T and 0.1T.
The initial ground-state values of $\langle s_z\rangle$ are 0.126 and 0.4692 respectively.
The non-zero values of $\langle s_z\rangle$ are observed up to the third extremum. 
The amplitude of the oscillations is much smaller for $B_z=0.01$T since, the spin polarization driving
the oscillations is weak already at the start. The $\langle x\rangle(t)$ oscillations pertain when the spin polarization is averaged out to zero due to the initial deflection of the wave packets. On the other hand  for weak spin-orbit coupling $\alpha=5$ meVnm oscillations of the z-component are visible for the entire time scale presented in Fig. \ref{lastguwno}.
For $\alpha=5$ meVnm the value of 0.1T is applied for which the initial z-component of the spin is $0.9926 
\times \frac{\hbar }{2}$ .
In Fig. \ref{lastguwno} the electron starts from zero momentum.  
The amplitude of $\langle x \rangle$ is small at the start of the simulation. The ''resonant'' momentum
is reached for $t\simeq \frac{\hbar \Delta E}{2\alpha eF}$, i.e. 1.73ps and 17.3ps for $\alpha=50$ and 25 meVnm, respectively.

 \begin{figure}[!t]
\begin{tabular}{l}
\includegraphics[width=0.6\columnwidth]{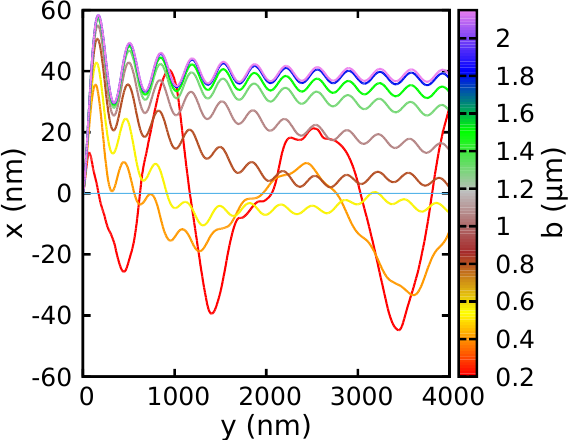} \put(-45,22){(a)}\\
\includegraphics[width=0.6\columnwidth]{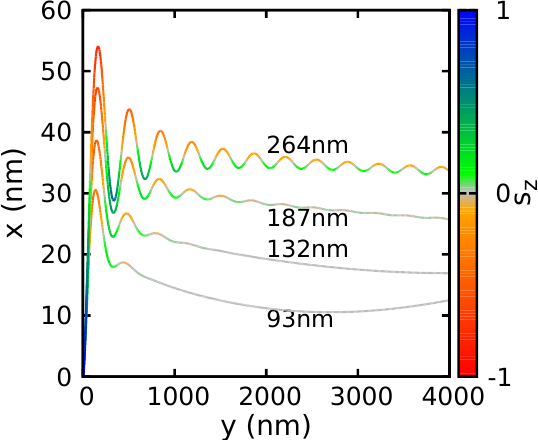} \put(-45,22){(b)}\\
\end{tabular}
\caption{ Average value of $x$ as a function of an average value of $y$ for the initial condition set 
as Gaussian $N\exp(-(x^2+y^2)/2\delta^2)$ spin-polarized in the $+z$ direction. The electric field of
the amplitude of $eF=0.01$ meV/nm is oriented along the $y$ direction and the channel width in the $x$ direction is $b$. In (a) we set $\delta=264$ nm and consider $b=0.2$, 0.25, 0.4, 0.55, 0.8, 1.05, 1.3, 1.5, 1.8 and 2.15 $\mu$m.
In (b) we set $b=1.95\mu$m and show the results for varied values of $\delta$ given in the figure. 
The color of the line corresponds to the average $z$ component of the spin in $\hbar/2$ units.  }
\label{sd}
\end{figure}

\subsection{Side jump due to the spin-orbit coupling}
The results of Section III.C -- for the electron spin initially oriented along the $z$ direction --
and strong spin-orbit coupling -- indicate a presence of a side-jump of the mean electron position
with respect to the channel axis (see Fig. \ref{gat20}(a,b)). At later times the mean position oscillates
near the axis to channel. 
Ref. \cite{Schli2} studied electron wave
packet in initially Gaussian form in and unconfined space (infinite channel width) with
a polarized spin and in-plane homogenous electric field.
A similar side-jump was reported at the initial stage of the motion. For larger packet widths 
the oscillations of the electron position occur off the starting position in the direction perpendicular to the electric field -- in contrast to the results
to our results of  Fig. \ref{gat20}(a,b) and Fig.\ref{insb14} -- where the oscillations occur along
the axis of the channel. 
In order to clarify this qualitative difference we performed calculations for the initial wave function in form of a Gaussian of Eq. (\ref{gauszian}) with the width $\delta=\sqrt\frac{\hbar}{m^*\omega_x}$.
The relation between $\delta$ and variance of the electron position in the Gaussian packet is 
 $\langle x^2\rangle^{1/2}=\delta/\sqrt{2}$. 
 We set the spin $+z$ polarized Gaussian in the center of channel of width $b$ and start the simulation 
 with the electric field $eF=0.01$ meV/nm. The initial wave function is set to zero at the edges of the channel, so for low $b$ -- comparable to $\delta$ -- the initial condition is not exactly Gaussian.
 The electron trajectory i.e. the average $x$ as a function of the average $y$ is shown in Fig. \ref{sd}(a)
 for $\delta=264$ nm for varied $b$. The initial side-jump is observed for all $b$ and the side jump seems to saturate for $d\simeq 2$ $\mu$m, where the electron wave packet no longer reaches the edge of the channel.
The trajectory than oscillates at a constant level of $x\simeq 40$ nm, qualitatively similar to the results
of Ref. \cite{Schli2}. For lower values of $b$ the oscillations level drops with $y$. For $b=0.4\mu$m and below
negative values of average $x$ are observed and for $b=0.2\mu$m we no longer observe the ''fast'' oscillation
with the period of $\simeq 333$ nm. Note, that in the main results of the paper we consider channels thinner
than 200 nm. 

Fig. \ref{sd}(b) shows the results for $b=1.95\mu$m as a function of $\delta$. 
For $\delta=264$ nm we see how the oscillations of $\langle s_z\rangle$ induce the oscillations 
of the mean position of the packet. The initial jump is due to the fully polarized initial state.
The electron spin oscillates but never reaches the initial value hence stabilization of the center of the electron packet above the axis of the channel. Both the spin and the spatial amplitude of the oscillations 
decreases in time (in $y$) the faster the stronger is the initial localization of the packet. 

For the results of the present model given in Fig. \ref{insb14}: the initial $\sqrt{\langle x^2\rangle}$  and ${\sqrt{\langle y^2\rangle}}$ 
are equal to 31.9 and 130.1 nm (or in terms of parameter $\delta$ the corresponding values are 45 nm and 184 nm)
and $b=175$ nm. The initial side jump is not observed in Fig. \ref{insb14} since the initial values of $\langle s_z\rangle$ is zero.
The packet oscillates not due to the spin oscillations but due to scattering by the edges of a relatively thin channel. 

\subsection{In-plane magnetic field effects}
\begin{figure}[!t]
\begin{tabular}{l}
\includegraphics[width=0.8\columnwidth]{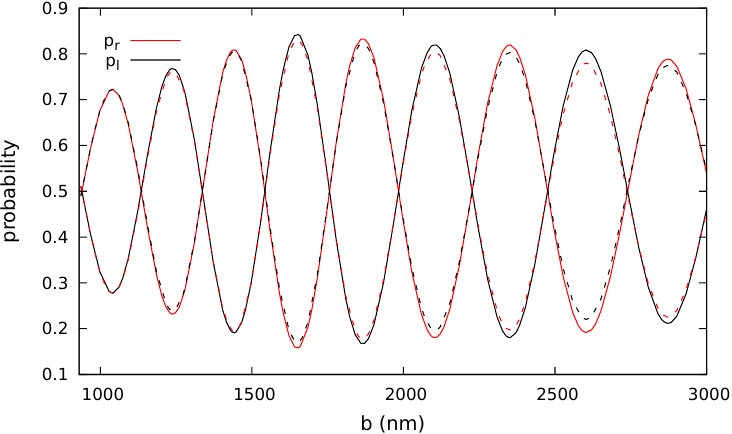}\put(-25,25){(a)}\\
\includegraphics[width=0.8\columnwidth]{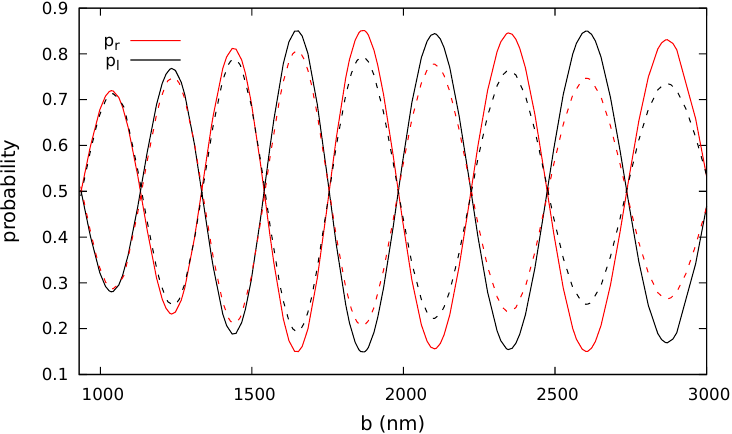}\put(-25,25){(b)}\\
\includegraphics[width=0.8\columnwidth]{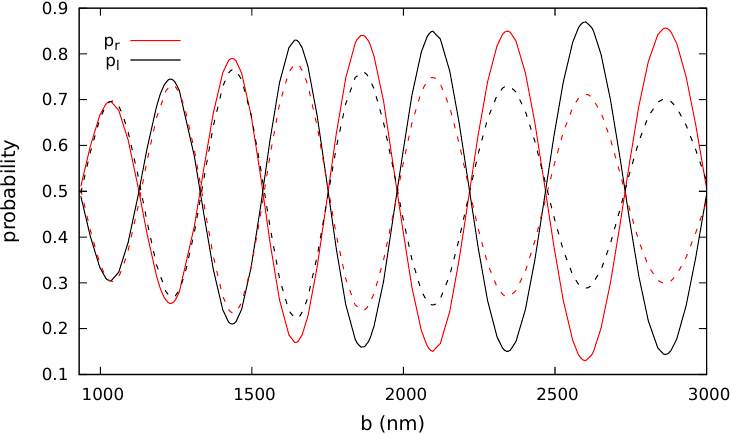}\put(-25,25){(c)}
\end{tabular}
\caption{The transfer probability of the wave packet released from the QD to the left and right output channels as a function of $b$ for the channel
of width $d=100$~nm and the initial state set as the ground state (solid lines) and the higher state of the Kramers doublet (dashed linbes) 
for $B_y=10$ mT (a), 50 mT (b) and 100 mT(c). The results for $B_y=10\mu$T were given in Fig. \ref{wfub}(d). }
\label{wpolu}
\end{figure}

For the purpose of our calculations in Fig. 3-10 we applied the magnetic field parallel to the $y$ channel of a very small value of $10\mu$ T.
The purpose of this near zero field is to lift the QD-confined ground-state degeneracy. This field has no visible influence on the precession
of the spin that occurs solely in $B_{SO}$ field, moreover the spin texture of the ground-state and the higher energy state of the Kramers doublet
as well as the average values of the spin are exactly opposite to one another (see Fig. 4). In consequence the probability of the electron transfer 
to the left or right channel were nearly exactly inverted for the ground-state and the excited state. 
However, the InSb ions carry the nuclear spin that gives rise to the hyperfine interactions that for a typical III-V quantum-dot produces 
the effective Overhauser field of the order of 1 mT \cite{Hanson2007}. 
One needs a stronger magnetic field in order to dominate  possible effects of this field for the initial states of the evolution and next for a possible orbital
effects of the spin precession. The results for $B_y=10$ mT, 50 mT and 100 mT are given in Fig. \ref{wpolu}.
For the selected values of the magnetic field the average $s_y$ in $\hbar/2$ units for the lower and higher states of the Kramers doublet 
are (cf. Fig. \ref{spectrum}(c,d)):   $0.1144$ and $-0.027$ for $B_y=$ 10mT, $0.1144$ and $-0.027$ for $B_y=$ 10mT;
$0.294$ and $0.135$ for $B_y=$ 50mT, and $0.512$ and $0.316$ for $B_y=$ 50mT. The results for the transfer probability that are shown in Fig. \ref{wpolu} demonstrate
that the $p_l/p_r$ for the two initial states are no longer exactly inversed. The differences grow for longer $b$ which indicate that the 
role of the external field precession is larger than the changes to the spin structure of the initial state.
The differences for $B_y=10$ mT [Fig. \ref{wpolu}(a)] are not very large. 
Note, that based on this result for an oriented magnetic field one can state that the locally random fields of the amplitude of about 1 mT should not significantly alter the motion of the wave packet during the
time evolution or due to a modification of the spin structure of the initial state confined in the quantum dot. 

In Fig. \ref{wpolu} we notice  that the amplitude of the $p_l$ and $p_r$ oscillations increase (decrease) forthe ground-state (the excited state).
Note, that the spin polarization in the $y$ direction is strenghented for the ground-state and weakened and next inversed by the $B_y$ field
for the excited state.  At the the value of $b=1650$ nm and $B_y=10$ mT the probability to the electron transfer from the ground state
to the left  channel is 0.8422  while the transfer probabiluty to the right channel  0.83.
For B=50mT ($B=100$ mT) the corresponding values are 0.85 and 0.806  (for 0.83 and 0.77). Concluding, as the in-plane field increases the amplitude of $p_l/p_r$ oscillations as a function of $b$ increases for the ground-state
but decreases for the excited state and for optimal resolution of both the in-plane field should be kept small. For $B_y=10$ mT which
should suffice to dominate the random Overhauser field the resolution for both the states remains very similar.

\subsection{Geometrical asymmetry}
\begin{figure}[!t]
\begin{tabular}{l}
\includegraphics[width=0.8\columnwidth]{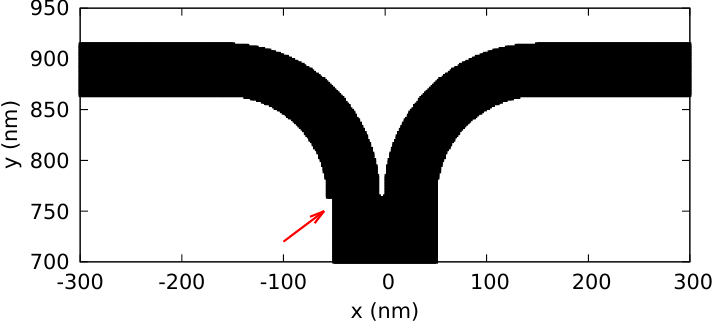}\put(-25,25){(a)}\\
\includegraphics[width=0.8\columnwidth]{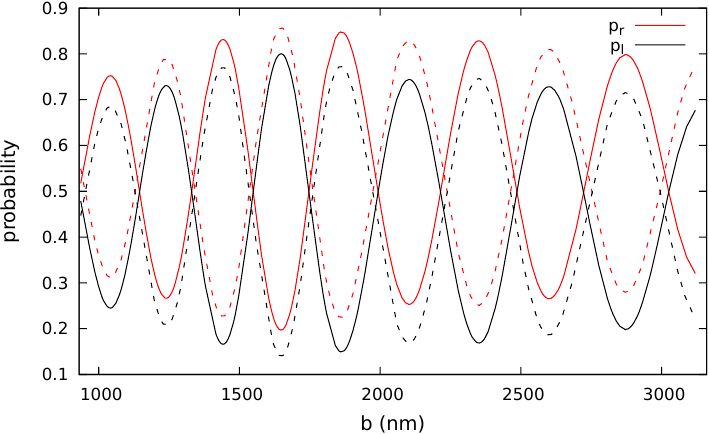}\put(-25,25){(b)}\\
\end{tabular}
\caption{ (a) The split of the channel for the structure with $b=100$ nm and the exit to the left channel shifted by 6.24 nm (the arrow).
(b) The transfer probability to the left and right channel. The values for the ground state (excited state) are given by solid (dashed) lines for $B_y=10\mu$T.
The results for the ideally symmetric system were given in Fig. \ref{wfub}(d). }
\label{defect}
\end{figure}

The discussion of the detection of the initial state assumed that the split of the channel is ideally symmetric with respect to the
axis of the feeding channel. An ideal symmetry in a real strucure is rather excluded. We considered a structure that deviates from the symmetry
by a shift of the left channel to the left by a value of 6.24 nm (see Fig. \ref{defect}(a)).
The transfer probabilities for the field of $B=10\mu$T is given in Fig. \ref{defect}(b). 
Naturally, $p_r$ grows and $p_l$ decreases with respect to the ideally symmetric case of Fig. \ref{wfub}(d). 
For $d=1650$ nm the transfer probability to the left channel from the ground-state is $0.803$ while 
the transfer probability to the right channel from the excited state is 0.856. The transfer characteristics
remains similar to the ideal case, however the resolution of the initial state is deteriorated. Naturally in the limit of strongly deformed structure the exit to the left channel can be effectively closed so that the electrons will go to the other exit for both the initial states. 
A system as close to an ideal symmetry as possible is best suited for the initial state readout.

\section{Conclusions}
In this paper, we have investigated spin-dependent trajectory of electrons released from quantum dots formed in nanowires, determined by the strong spin-orbit interaction present in the host materials. The system we have studied comprises a quantum dot defined by electrostatic gating subject to an electric field along the axis of the nanowire.  We have analyzed the energy spectrum of the quantum dot and the dynamics of the wave packet released from the dot.
Our simulations have shown that the spin state initialized in the quantum dot, and then released into the channel, evolves under the influence of spin-orbit coupling, resulting in spin precession and spin-dependent lateral deflection during propagation. This behavior results in a snake-like trajectory that facilitates spin-to-charge conversion for spin state detection in the T-junction geometry. Our findings have been supported by semiclassical calculations of the electron’s trajectory based on a classical Hamiltonian model coupled to the Bloch equations, as well as the spin-dependent force derived in the Heisenberg picture.
We have demonstrated that the spin dynamics have strongly depended on the structure geometry, with more pronounced spin oscillations and trajectory deviations observed for wider channels. Notably, the snake-like trajectories have persisted even at low external magnetic fields, despite incomplete initial spin polarization and the rapid evolution of the spin polarization towards a state oriented perpendicular to the channel axis shortly after electron release.
We have achieved spin state detection fidelities exceeding 82\% for optimized channel widths and confinement potentials.

\color{black}Finally, the proposed spin-detection effect was analyzed with respect to structural asymmetry, showing that it can survive when only a small asymmetry is introduced into the system. At the same time, we expect that for strong asymmetry the detection of the electron spin becomes impossible. Therefore, a system as close to ideal symmetry as possible is best suited for the initial state readout.

A similar conclusion applies to disorder or imperfections. For strong disorder, one cannot expect high detection efficiency, as spin coherence is lost and/or the spin state may change due to scattering events in a material with the strong spin–orbit interaction required to induce the preferential injection of a given spin into one of the arms.
\color{black}

\section*{Acknowledgements}
We gratefully acknowledge Roberta Citro for valuable discussions.
This work is financed by the Horizon Europe EIC Pathfinder under the grant IQARO number 101115190
titled "Spin-orbitronic quantum bits in reconfigurable 2D-oxides" and partly supported by program „Excellence initiative – research university” for the AGH University. Computing infrastructure PLGrid (HPC Centers: ACK Cyfronet AGH) within computational grant no. PLG/2024/017175 and no. PLG/2025/018197 was used.

\bibliography{ref.bib} 

%apsrev4-2.bst 2019-01-14 (MD) hand-edited version of apsrev4-1.bst
%Control: key (0)
%Control: author (8) initials jnrlst
%Control: editor formatted (1) identically to author
%Control: production of article title (0) allowed
%Control: page (0) single
%Control: year (1) truncated
%Control: production of eprint (0) enabled
\begin{thebibliography}{40}%
\makeatletter
\providecommand \@ifxundefined [1]{%
 \@ifx{#1\undefined}
}%
\providecommand \@ifnum [1]{%
 \ifnum #1\expandafter \@firstoftwo
 \else \expandafter \@secondoftwo
 \fi
}%
\providecommand \@ifx [1]{%
 \ifx #1\expandafter \@firstoftwo
 \else \expandafter \@secondoftwo
 \fi
}%
\providecommand \natexlab [1]{#1}%
\providecommand \enquote  [1]{``#1''}%
\providecommand \bibnamefont  [1]{#1}%
\providecommand \bibfnamefont [1]{#1}%
\providecommand \citenamefont [1]{#1}%
\providecommand \href@noop [0]{\@secondoftwo}%
\providecommand \href [0]{\begingroup \@sanitize@url \@href}%
\providecommand \@href[1]{\@@startlink{#1}\@@href}%
\providecommand \@@href[1]{\endgroup#1\@@endlink}%
\providecommand \@sanitize@url [0]{\catcode `\\12\catcode `\$12\catcode
  `\&12\catcode `\#12\catcode `\^12\catcode `\_12\catcode `\%12\relax}%
\providecommand \@@startlink[1]{}%
\providecommand \@@endlink[0]{}%
\providecommand \url  [0]{\begingroup\@sanitize@url \@url }%
\providecommand \@url [1]{\endgroup\@href {#1}{\urlprefix }}%
\providecommand \urlprefix  [0]{URL }%
\providecommand \Eprint [0]{\href }%
\providecommand \doibase [0]{https://doi.org/}%
\providecommand \selectlanguage [0]{\@gobble}%
\providecommand \bibinfo  [0]{\@secondoftwo}%
\providecommand \bibfield  [0]{\@secondoftwo}%
\providecommand \translation [1]{[#1]}%
\providecommand \BibitemOpen [0]{}%
\providecommand \bibitemStop [0]{}%
\providecommand \bibitemNoStop [0]{.\EOS\space}%
\providecommand \EOS [0]{\spacefactor3000\relax}%
\providecommand \BibitemShut  [1]{\csname bibitem#1\endcsname}%
\let\auto@bib@innerbib\@empty
%</preamble>
\bibitem [{\citenamefont {\ifmmode \check{Z}\else
  \v{Z}\fi{}uti\ifmmode~\acute{c}\else \'{c}\fi{}}\ \emph
  {et~al.}(2004)\citenamefont {\ifmmode \check{Z}\else
  \v{Z}\fi{}uti\ifmmode~\acute{c}\else \'{c}\fi{}}, \citenamefont {Fabian},\
  and\ \citenamefont {Das~Sarma}}]{Fabian2004}%
  \BibitemOpen
  \bibfield  {author} {\bibinfo {author} {\bibfnamefont {I.}~\bibnamefont
  {\ifmmode \check{Z}\else \v{Z}\fi{}uti\ifmmode~\acute{c}\else \'{c}\fi{}}},
  \bibinfo {author} {\bibfnamefont {J.}~\bibnamefont {Fabian}},\ and\ \bibinfo
  {author} {\bibfnamefont {S.}~\bibnamefont {Das~Sarma}},\ }\bibfield  {title}
  {\bibinfo {title} {Spintronics: Fundamentals and applications},\ }\href
  {https://doi.org/10.1103/RevModPhys.76.323} {\bibfield  {journal} {\bibinfo
  {journal} {Rev. Mod. Phys.}\ }\textbf {\bibinfo {volume} {76}},\ \bibinfo
  {pages} {323} (\bibinfo {year} {2004})}\BibitemShut {NoStop}%
\bibitem [{\citenamefont {Ladd}\ \emph {et~al.}(2010)\citenamefont {Ladd},
  \citenamefont {Jelezko}, \citenamefont {Laflamme}, \citenamefont {Nakamura},
  \citenamefont {Monroe},\ and\ \citenamefont {O'Brien}}]{Ladd2010}%
  \BibitemOpen
  \bibfield  {author} {\bibinfo {author} {\bibfnamefont {T.~D.}\ \bibnamefont
  {Ladd}}, \bibinfo {author} {\bibfnamefont {F.}~\bibnamefont {Jelezko}},
  \bibinfo {author} {\bibfnamefont {R.}~\bibnamefont {Laflamme}}, \bibinfo
  {author} {\bibfnamefont {Y.}~\bibnamefont {Nakamura}}, \bibinfo {author}
  {\bibfnamefont {C.}~\bibnamefont {Monroe}},\ and\ \bibinfo {author}
  {\bibfnamefont {J.~L.}\ \bibnamefont {O'Brien}},\ }\bibfield  {title}
  {\bibinfo {title} {Quantum computers},\ }\href@noop {} {\bibfield  {journal}
  {\bibinfo  {journal} {Nature}\ }\textbf {\bibinfo {volume} {464}},\ \bibinfo
  {pages} {45} (\bibinfo {year} {2010})}\BibitemShut {NoStop}%
\bibitem [{\citenamefont {Burkard}\ \emph {et~al.}(2023)\citenamefont
  {Burkard}, \citenamefont {Ladd}, \citenamefont {Pan}, \citenamefont
  {Nichol},\ and\ \citenamefont {Petta}}]{Burkard2023}%
  \BibitemOpen
  \bibfield  {author} {\bibinfo {author} {\bibfnamefont {G.}~\bibnamefont
  {Burkard}}, \bibinfo {author} {\bibfnamefont {T.~D.}\ \bibnamefont {Ladd}},
  \bibinfo {author} {\bibfnamefont {A.}~\bibnamefont {Pan}}, \bibinfo {author}
  {\bibfnamefont {J.~M.}\ \bibnamefont {Nichol}},\ and\ \bibinfo {author}
  {\bibfnamefont {J.~R.}\ \bibnamefont {Petta}},\ }\bibfield  {title} {\bibinfo
  {title} {Semiconductor spin qubits},\ }\href
  {https://doi.org/10.1103/RevModPhys.95.025003} {\bibfield  {journal}
  {\bibinfo  {journal} {Rev. Mod. Phys.}\ }\textbf {\bibinfo {volume} {95}},\
  \bibinfo {pages} {025003} (\bibinfo {year} {2023})}\BibitemShut {NoStop}%
\bibitem [{\citenamefont {Philips}\ \emph {et~al.}(2022)\citenamefont
  {Philips}, \citenamefont {Mądzik}, \citenamefont {Amitonov}, \citenamefont
  {de~Snoo}, \citenamefont {Russ}, \citenamefont {Kalhor}, \citenamefont
  {Volk}, \citenamefont {Lawrie}, \citenamefont {Brousse}, \citenamefont
  {Tryputen}, \citenamefont {Wuetz}, \citenamefont {Sammak}, \citenamefont
  {Veldhorst}, \citenamefont {Scappucci},\ and\ \citenamefont
  {Vandersypen}}]{Philips2022}%
  \BibitemOpen
  \bibfield  {author} {\bibinfo {author} {\bibfnamefont {S.~G.~J.}\
  \bibnamefont {Philips}}, \bibinfo {author} {\bibfnamefont {M.~T.}\
  \bibnamefont {Mądzik}}, \bibinfo {author} {\bibfnamefont {S.~V.}\
  \bibnamefont {Amitonov}}, \bibinfo {author} {\bibfnamefont {S.~L.}\
  \bibnamefont {de~Snoo}}, \bibinfo {author} {\bibfnamefont {M.}~\bibnamefont
  {Russ}}, \bibinfo {author} {\bibfnamefont {N.}~\bibnamefont {Kalhor}},
  \bibinfo {author} {\bibfnamefont {C.}~\bibnamefont {Volk}}, \bibinfo {author}
  {\bibfnamefont {W.~I.~L.}\ \bibnamefont {Lawrie}}, \bibinfo {author}
  {\bibfnamefont {D.}~\bibnamefont {Brousse}}, \bibinfo {author} {\bibfnamefont
  {L.}~\bibnamefont {Tryputen}}, \bibinfo {author} {\bibfnamefont {B.~P.}\
  \bibnamefont {Wuetz}}, \bibinfo {author} {\bibfnamefont {A.}~\bibnamefont
  {Sammak}}, \bibinfo {author} {\bibfnamefont {M.}~\bibnamefont {Veldhorst}},
  \bibinfo {author} {\bibfnamefont {G.}~\bibnamefont {Scappucci}},\ and\
  \bibinfo {author} {\bibfnamefont {L.~M.~K.}\ \bibnamefont {Vandersypen}},\
  }\bibfield  {title} {\bibinfo {title} {Universal control of a six-qubit
  quantum processor in silicon},\ }\href
  {https://doi.org/10.1038/s41586-022-05117-x} {\bibfield  {journal} {\bibinfo
  {journal} {Nature}\ }\textbf {\bibinfo {volume} {609}},\ \bibinfo {pages}
  {919} (\bibinfo {year} {2022})}\BibitemShut {NoStop}%
\bibitem [{\citenamefont {Loss}\ and\ \citenamefont
  {DiVincenzo}(1998)}]{DiVincenzo1998}%
  \BibitemOpen
  \bibfield  {author} {\bibinfo {author} {\bibfnamefont {D.}~\bibnamefont
  {Loss}}\ and\ \bibinfo {author} {\bibfnamefont {D.~P.}\ \bibnamefont
  {DiVincenzo}},\ }\bibfield  {title} {\bibinfo {title} {Quantum computation
  with quantum dots},\ }\href {https://doi.org/10.1103/PhysRevA.57.120}
  {\bibfield  {journal} {\bibinfo  {journal} {Phys. Rev. A}\ }\textbf {\bibinfo
  {volume} {57}},\ \bibinfo {pages} {120} (\bibinfo {year} {1998})}\BibitemShut
  {NoStop}%
\bibitem [{\citenamefont {Hanson}\ \emph {et~al.}(2007)\citenamefont {Hanson},
  \citenamefont {Kouwenhoven}, \citenamefont {Petta}, \citenamefont {Tarucha},\
  and\ \citenamefont {Vandersypen}}]{Hanson2007}%
  \BibitemOpen
  \bibfield  {author} {\bibinfo {author} {\bibfnamefont {R.}~\bibnamefont
  {Hanson}}, \bibinfo {author} {\bibfnamefont {L.~P.}\ \bibnamefont
  {Kouwenhoven}}, \bibinfo {author} {\bibfnamefont {J.~R.}\ \bibnamefont
  {Petta}}, \bibinfo {author} {\bibfnamefont {S.}~\bibnamefont {Tarucha}},\
  and\ \bibinfo {author} {\bibfnamefont {L.~M.~K.}\ \bibnamefont
  {Vandersypen}},\ }\bibfield  {title} {\bibinfo {title} {Spins in few-electron
  quantum dots},\ }\href {https://doi.org/10.1103/RevModPhys.79.1217}
  {\bibfield  {journal} {\bibinfo  {journal} {Rev. Mod. Phys.}\ }\textbf
  {\bibinfo {volume} {79}},\ \bibinfo {pages} {1217} (\bibinfo {year}
  {2007})}\BibitemShut {NoStop}%
\bibitem [{\citenamefont {Zwanenburg}\ \emph {et~al.}(2013)\citenamefont
  {Zwanenburg}, \citenamefont {Dzurak}, \citenamefont {Morello}, \citenamefont
  {Simmons}, \citenamefont {Hollenberg}, \citenamefont {Klimeck}, \citenamefont
  {Rogge}, \citenamefont {Coppersmith},\ and\ \citenamefont
  {Eriksson}}]{Zwanenburg2013}%
  \BibitemOpen
  \bibfield  {author} {\bibinfo {author} {\bibfnamefont {F.~A.}\ \bibnamefont
  {Zwanenburg}}, \bibinfo {author} {\bibfnamefont {A.~S.}\ \bibnamefont
  {Dzurak}}, \bibinfo {author} {\bibfnamefont {A.}~\bibnamefont {Morello}},
  \bibinfo {author} {\bibfnamefont {M.~Y.}\ \bibnamefont {Simmons}}, \bibinfo
  {author} {\bibfnamefont {L.~C.~L.}\ \bibnamefont {Hollenberg}}, \bibinfo
  {author} {\bibfnamefont {G.}~\bibnamefont {Klimeck}}, \bibinfo {author}
  {\bibfnamefont {S.}~\bibnamefont {Rogge}}, \bibinfo {author} {\bibfnamefont
  {S.~N.}\ \bibnamefont {Coppersmith}},\ and\ \bibinfo {author} {\bibfnamefont
  {M.~A.}\ \bibnamefont {Eriksson}},\ }\bibfield  {title} {\bibinfo {title}
  {Silicon quantum electronics},\ }\href
  {https://doi.org/10.1103/RevModPhys.85.961} {\bibfield  {journal} {\bibinfo
  {journal} {Rev. Mod. Phys.}\ }\textbf {\bibinfo {volume} {85}},\ \bibinfo
  {pages} {961} (\bibinfo {year} {2013})}\BibitemShut {NoStop}%
\bibitem [{\citenamefont {Wang}\ \emph {et~al.}(2024)\citenamefont {Wang},
  \citenamefont {John}, \citenamefont {Tidjani}, \citenamefont {Yu},
  \citenamefont {Ivlev}, \citenamefont {Déprez}, \citenamefont {van
  Riggelen-Doelman}, \citenamefont {Woods}, \citenamefont {Hendrickx},
  \citenamefont {Lawrie}, \citenamefont {Stehouwer}, \citenamefont
  {Oosterhout}, \citenamefont {Sammak}, \citenamefont {Friesen}, \citenamefont
  {Scappucci}, \citenamefont {de~Snoo}, \citenamefont {Rimbach-Russ},
  \citenamefont {Borsoi},\ and\ \citenamefont {Veldhorst}}]{Wang2024}%
  \BibitemOpen
  \bibfield  {author} {\bibinfo {author} {\bibfnamefont {C.-A.}\ \bibnamefont
  {Wang}}, \bibinfo {author} {\bibfnamefont {V.}~\bibnamefont {John}}, \bibinfo
  {author} {\bibfnamefont {H.}~\bibnamefont {Tidjani}}, \bibinfo {author}
  {\bibfnamefont {C.~X.}\ \bibnamefont {Yu}}, \bibinfo {author} {\bibfnamefont
  {A.~S.}\ \bibnamefont {Ivlev}}, \bibinfo {author} {\bibfnamefont
  {C.}~\bibnamefont {Déprez}}, \bibinfo {author} {\bibfnamefont
  {F.}~\bibnamefont {van Riggelen-Doelman}}, \bibinfo {author} {\bibfnamefont
  {B.~D.}\ \bibnamefont {Woods}}, \bibinfo {author} {\bibfnamefont {N.~W.}\
  \bibnamefont {Hendrickx}}, \bibinfo {author} {\bibfnamefont {W.~I.~L.}\
  \bibnamefont {Lawrie}}, \bibinfo {author} {\bibfnamefont {L.~E.~A.}\
  \bibnamefont {Stehouwer}}, \bibinfo {author} {\bibfnamefont {S.~D.}\
  \bibnamefont {Oosterhout}}, \bibinfo {author} {\bibfnamefont
  {A.}~\bibnamefont {Sammak}}, \bibinfo {author} {\bibfnamefont
  {M.}~\bibnamefont {Friesen}}, \bibinfo {author} {\bibfnamefont
  {G.}~\bibnamefont {Scappucci}}, \bibinfo {author} {\bibfnamefont {S.~L.}\
  \bibnamefont {de~Snoo}}, \bibinfo {author} {\bibfnamefont {M.}~\bibnamefont
  {Rimbach-Russ}}, \bibinfo {author} {\bibfnamefont {F.}~\bibnamefont
  {Borsoi}},\ and\ \bibinfo {author} {\bibfnamefont {M.}~\bibnamefont
  {Veldhorst}},\ }\bibfield  {title} {\bibinfo {title} {Operating semiconductor
  quantum processors with hopping spins},\ }\href@noop {} {\bibfield  {journal}
  {\bibinfo  {journal} {Science}\ }\textbf {\bibinfo {volume} {385}},\ \bibinfo
  {pages} {447} (\bibinfo {year} {2024})}\BibitemShut {NoStop}%
\bibitem [{\citenamefont {Takeda}\ \emph {et~al.}(2024)\citenamefont {Takeda},
  \citenamefont {Noiri}, \citenamefont {Nakajima}, \citenamefont {Camenzind},
  \citenamefont {Kobayashi}, \citenamefont {Sammak}, \citenamefont
  {Scappucci},\ and\ \citenamefont {Tarucha}}]{Takeda2024}%
  \BibitemOpen
  \bibfield  {author} {\bibinfo {author} {\bibfnamefont {K.}~\bibnamefont
  {Takeda}}, \bibinfo {author} {\bibfnamefont {A.}~\bibnamefont {Noiri}},
  \bibinfo {author} {\bibfnamefont {T.}~\bibnamefont {Nakajima}}, \bibinfo
  {author} {\bibfnamefont {L.~C.}\ \bibnamefont {Camenzind}}, \bibinfo {author}
  {\bibfnamefont {T.}~\bibnamefont {Kobayashi}}, \bibinfo {author}
  {\bibfnamefont {A.}~\bibnamefont {Sammak}}, \bibinfo {author} {\bibfnamefont
  {G.}~\bibnamefont {Scappucci}},\ and\ \bibinfo {author} {\bibfnamefont
  {S.}~\bibnamefont {Tarucha}},\ }\bibfield  {title} {\bibinfo {title} {Rapid
  single-shot parity spin readout in a silicon double quantum dot with fidelity
  exceeding 99\%},\ }\href {https://doi.org/10.1038/s41534-024-00813-0}
  {\bibfield  {journal} {\bibinfo  {journal} {npj Quantum Information}\
  }\textbf {\bibinfo {volume} {10}},\ \bibinfo {pages} {22} (\bibinfo {year}
  {2024})}\BibitemShut {NoStop}%
\bibitem [{\citenamefont {Morello}\ \emph {et~al.}(2010)\citenamefont
  {Morello}, \citenamefont {Pla}, \citenamefont {Zwanenburg}, \citenamefont
  {Chan}, \citenamefont {Tan}, \citenamefont {Huebl}, \citenamefont {Mottonen},
  \citenamefont {Nugroho}, \citenamefont {Dzurak},\ and\ \citenamefont
  {Simmons}}]{Morello2010}%
  \BibitemOpen
  \bibfield  {author} {\bibinfo {author} {\bibfnamefont {A.}~\bibnamefont
  {Morello}}, \bibinfo {author} {\bibfnamefont {J.~J.}\ \bibnamefont {Pla}},
  \bibinfo {author} {\bibfnamefont {F.~A.}\ \bibnamefont {Zwanenburg}},
  \bibinfo {author} {\bibfnamefont {K.~Y.}\ \bibnamefont {Chan}}, \bibinfo
  {author} {\bibfnamefont {K.~W.}\ \bibnamefont {Tan}}, \bibinfo {author}
  {\bibfnamefont {H.}~\bibnamefont {Huebl}}, \bibinfo {author} {\bibfnamefont
  {M.}~\bibnamefont {Mottonen}}, \bibinfo {author} {\bibfnamefont {C.~D.}\
  \bibnamefont {Nugroho}}, \bibinfo {author} {\bibfnamefont {A.~S.}\
  \bibnamefont {Dzurak}},\ and\ \bibinfo {author} {\bibfnamefont {M.~Y.}\
  \bibnamefont {Simmons}},\ }\bibfield  {title} {\bibinfo {title} {Single-shot
  readout of an electron spin in silicon},\ }\href
  {https://doi.org/10.1038/nature09460} {\bibfield  {journal} {\bibinfo
  {journal} {Nature}\ }\textbf {\bibinfo {volume} {467}},\ \bibinfo {pages}
  {687} (\bibinfo {year} {2010})}\BibitemShut {NoStop}%
\bibitem [{\citenamefont {Koppens}\ \emph {et~al.}(2006)\citenamefont
  {Koppens}, \citenamefont {Buizert}, \citenamefont {Tielrooij}, \citenamefont
  {Vink}, \citenamefont {Nowack}, \citenamefont {Meunier}, \citenamefont
  {Kouwenhoven},\ and\ \citenamefont {Vandersypen}}]{Koppens2006}%
  \BibitemOpen
  \bibfield  {author} {\bibinfo {author} {\bibfnamefont {F.~H.~L.}\
  \bibnamefont {Koppens}}, \bibinfo {author} {\bibfnamefont {C.}~\bibnamefont
  {Buizert}}, \bibinfo {author} {\bibfnamefont {K.~J.}\ \bibnamefont
  {Tielrooij}}, \bibinfo {author} {\bibfnamefont {I.~T.}\ \bibnamefont {Vink}},
  \bibinfo {author} {\bibfnamefont {K.~C.}\ \bibnamefont {Nowack}}, \bibinfo
  {author} {\bibfnamefont {T.}~\bibnamefont {Meunier}}, \bibinfo {author}
  {\bibfnamefont {L.~P.}\ \bibnamefont {Kouwenhoven}},\ and\ \bibinfo {author}
  {\bibfnamefont {L.~M.~K.}\ \bibnamefont {Vandersypen}},\ }\bibfield  {title}
  {\bibinfo {title} {Driven coherent oscillations of a single electron spin in
  a quantum dot},\ }\href@noop {} {\bibfield  {journal} {\bibinfo  {journal}
  {Nature}\ }\textbf {\bibinfo {volume} {442}},\ \bibinfo {pages} {766}
  (\bibinfo {year} {2006})}\BibitemShut {NoStop}%
\bibitem [{\citenamefont {Koppens}\ \emph {et~al.}(2005)\citenamefont
  {Koppens}, \citenamefont {Folk}, \citenamefont {Elzerman}, \citenamefont
  {Hanson}, \citenamefont {van Beveren}, \citenamefont {Vink}, \citenamefont
  {Tranitz}, \citenamefont {Wegscheider}, \citenamefont {Kouwenhoven},\ and\
  \citenamefont {Vandersypen}}]{Koppens2005}%
  \BibitemOpen
  \bibfield  {author} {\bibinfo {author} {\bibfnamefont {F.~H.~L.}\
  \bibnamefont {Koppens}}, \bibinfo {author} {\bibfnamefont {J.~A.}\
  \bibnamefont {Folk}}, \bibinfo {author} {\bibfnamefont {J.~M.}\ \bibnamefont
  {Elzerman}}, \bibinfo {author} {\bibfnamefont {R.}~\bibnamefont {Hanson}},
  \bibinfo {author} {\bibfnamefont {L.~H.~W.}\ \bibnamefont {van Beveren}},
  \bibinfo {author} {\bibfnamefont {I.~T.}\ \bibnamefont {Vink}}, \bibinfo
  {author} {\bibfnamefont {H.~P.}\ \bibnamefont {Tranitz}}, \bibinfo {author}
  {\bibfnamefont {W.}~\bibnamefont {Wegscheider}}, \bibinfo {author}
  {\bibfnamefont {L.~P.}\ \bibnamefont {Kouwenhoven}},\ and\ \bibinfo {author}
  {\bibfnamefont {L.~M.~K.}\ \bibnamefont {Vandersypen}},\ }\bibfield  {title}
  {\bibinfo {title} {Control and detection of singlet-triplet mixing in a
  random nuclear field},\ }\href@noop {} {\bibfield  {journal} {\bibinfo
  {journal} {Science}\ }\textbf {\bibinfo {volume} {309}},\ \bibinfo {pages}
  {1346} (\bibinfo {year} {2005})}\BibitemShut {NoStop}%
\bibitem [{\citenamefont {Petta}\ \emph {et~al.}(2005)\citenamefont {Petta},
  \citenamefont {Johnson}, \citenamefont {Taylor}, \citenamefont {Laird},
  \citenamefont {Yacoby}, \citenamefont {Lukin}, \citenamefont {Marcus},
  \citenamefont {Hanson},\ and\ \citenamefont {Gossard}}]{Petta2005}%
  \BibitemOpen
  \bibfield  {author} {\bibinfo {author} {\bibfnamefont {J.~R.}\ \bibnamefont
  {Petta}}, \bibinfo {author} {\bibfnamefont {A.~C.}\ \bibnamefont {Johnson}},
  \bibinfo {author} {\bibfnamefont {J.~M.}\ \bibnamefont {Taylor}}, \bibinfo
  {author} {\bibfnamefont {E.}~\bibnamefont {Laird}}, \bibinfo {author}
  {\bibfnamefont {A.}~\bibnamefont {Yacoby}}, \bibinfo {author} {\bibfnamefont
  {M.~D.}\ \bibnamefont {Lukin}}, \bibinfo {author} {\bibfnamefont {C.~M.}\
  \bibnamefont {Marcus}}, \bibinfo {author} {\bibfnamefont {M.~P.}\
  \bibnamefont {Hanson}},\ and\ \bibinfo {author} {\bibfnamefont {A.~C.}\
  \bibnamefont {Gossard}},\ }\bibfield  {title} {\bibinfo {title} {Coherent
  manipulation of coupled electron spins in semiconductor quantum dots},\
  }\href@noop {} {\bibfield  {journal} {\bibinfo  {journal} {Science}\ }\textbf
  {\bibinfo {volume} {309}},\ \bibinfo {pages} {2180} (\bibinfo {year}
  {2005})}\BibitemShut {NoStop}%
\bibitem [{\citenamefont {Nowack}\ \emph {et~al.}(2007)\citenamefont {Nowack},
  \citenamefont {Koppens}, \citenamefont {Nazarov},\ and\ \citenamefont
  {Vandersypen}}]{Nowack2007}%
  \BibitemOpen
  \bibfield  {author} {\bibinfo {author} {\bibfnamefont {K.~C.}\ \bibnamefont
  {Nowack}}, \bibinfo {author} {\bibfnamefont {F.~H.~L.}\ \bibnamefont
  {Koppens}}, \bibinfo {author} {\bibfnamefont {Y.~V.}\ \bibnamefont
  {Nazarov}},\ and\ \bibinfo {author} {\bibfnamefont {L.~M.~K.}\ \bibnamefont
  {Vandersypen}},\ }\bibfield  {title} {\bibinfo {title} {Coherent control of a
  single electron spin with electric fields},\ }\href@noop {} {\bibfield
  {journal} {\bibinfo  {journal} {Science}\ }\textbf {\bibinfo {volume} {31}},\
  \bibinfo {pages} {1430} (\bibinfo {year} {2007})}\BibitemShut {NoStop}%
\bibitem [{\citenamefont {Barthel}\ \emph {et~al.}(2010)\citenamefont
  {Barthel}, \citenamefont {Kj\ae{}rgaard}, \citenamefont {Medford},
  \citenamefont {Stopa}, \citenamefont {Marcus}, \citenamefont {Hanson},\ and\
  \citenamefont {Gossard}}]{Barthel2009}%
  \BibitemOpen
  \bibfield  {author} {\bibinfo {author} {\bibfnamefont {C.}~\bibnamefont
  {Barthel}}, \bibinfo {author} {\bibfnamefont {M.}~\bibnamefont
  {Kj\ae{}rgaard}}, \bibinfo {author} {\bibfnamefont {J.}~\bibnamefont
  {Medford}}, \bibinfo {author} {\bibfnamefont {M.}~\bibnamefont {Stopa}},
  \bibinfo {author} {\bibfnamefont {C.~M.}\ \bibnamefont {Marcus}}, \bibinfo
  {author} {\bibfnamefont {M.~P.}\ \bibnamefont {Hanson}},\ and\ \bibinfo
  {author} {\bibfnamefont {A.~C.}\ \bibnamefont {Gossard}},\ }\bibfield
  {title} {\bibinfo {title} {Fast sensing of double-dot charge arrangement and
  spin state with a radio-frequency sensor quantum dot},\ }\href
  {https://doi.org/10.1103/PhysRevB.81.161308} {\bibfield  {journal} {\bibinfo
  {journal} {Phys. Rev. B}\ }\textbf {\bibinfo {volume} {81}},\ \bibinfo
  {pages} {161308} (\bibinfo {year} {2010})}\BibitemShut {NoStop}%
\bibitem [{\citenamefont {Mills}\ \emph {et~al.}(2022)\citenamefont {Mills},
  \citenamefont {Guinn}, \citenamefont {Gullans}, \citenamefont {Sigilitto},
  \citenamefont {Feldman}, \citenamefont {Nielsen},\ and\ \citenamefont
  {Petta}}]{Petta2022}%
  \BibitemOpen
  \bibfield  {author} {\bibinfo {author} {\bibfnamefont {A.~R.}\ \bibnamefont
  {Mills}}, \bibinfo {author} {\bibfnamefont {C.~R.}\ \bibnamefont {Guinn}},
  \bibinfo {author} {\bibfnamefont {M.~J.}\ \bibnamefont {Gullans}}, \bibinfo
  {author} {\bibfnamefont {A.~J.}\ \bibnamefont {Sigilitto}}, \bibinfo {author}
  {\bibfnamefont {M.~M.}\ \bibnamefont {Feldman}}, \bibinfo {author}
  {\bibfnamefont {E.}~\bibnamefont {Nielsen}},\ and\ \bibinfo {author}
  {\bibfnamefont {J.~R.}\ \bibnamefont {Petta}},\ }\bibfield  {title} {\bibinfo
  {title} {Two-qubit silicon quantum processor with operation fidelity
  exceeding 99\%},\ }\href {https://doi.org/10.1103/PhysRevB.75.035327}
  {\bibfield  {journal} {\bibinfo  {journal} {Sci. Adv.}\ }\textbf {\bibinfo
  {volume} {8}},\ \bibinfo {pages} {5130} (\bibinfo {year} {2022})}\BibitemShut
  {NoStop}%
\bibitem [{\citenamefont {Nadj-Perge}\ \emph {et~al.}(2010)\citenamefont
  {Nadj-Perge}, \citenamefont {Frolov}, \citenamefont {Bakkers},\ and\
  \citenamefont {Kowenhoven}}]{Nadj-Perge2010}%
  \BibitemOpen
  \bibfield  {author} {\bibinfo {author} {\bibfnamefont {S.}~\bibnamefont
  {Nadj-Perge}}, \bibinfo {author} {\bibfnamefont {S.~M.}\ \bibnamefont
  {Frolov}}, \bibinfo {author} {\bibfnamefont {E.~P. A.~M.}\ \bibnamefont
  {Bakkers}},\ and\ \bibinfo {author} {\bibfnamefont {L.~P.}\ \bibnamefont
  {Kowenhoven}},\ }\bibfield  {title} {\bibinfo {title} {Spin–orbit qubit in
  a semiconductor nanowire},\ }\href@noop {} {\bibfield  {journal} {\bibinfo
  {journal} {Nature}\ }\textbf {\bibinfo {volume} {468}},\ \bibinfo {pages}
  {1084} (\bibinfo {year} {2010})}\BibitemShut {NoStop}%
\bibitem [{\citenamefont {Nadj-Perge}\ \emph {et~al.}(2012)\citenamefont
  {Nadj-Perge}, \citenamefont {Pribiag}, \citenamefont {van~den Berg},
  \citenamefont {Zuo}, \citenamefont {Plissard}, \citenamefont {Bakkers},
  \citenamefont {Frolov},\ and\ \citenamefont {Kouwenhoven}}]{Nadj-Perge2012}%
  \BibitemOpen
  \bibfield  {author} {\bibinfo {author} {\bibfnamefont {S.}~\bibnamefont
  {Nadj-Perge}}, \bibinfo {author} {\bibfnamefont {V.~S.}\ \bibnamefont
  {Pribiag}}, \bibinfo {author} {\bibfnamefont {J.~W.~G.}\ \bibnamefont
  {van~den Berg}}, \bibinfo {author} {\bibfnamefont {K.}~\bibnamefont {Zuo}},
  \bibinfo {author} {\bibfnamefont {S.~R.}\ \bibnamefont {Plissard}}, \bibinfo
  {author} {\bibfnamefont {E.~P. A.~M.}\ \bibnamefont {Bakkers}}, \bibinfo
  {author} {\bibfnamefont {S.~M.}\ \bibnamefont {Frolov}},\ and\ \bibinfo
  {author} {\bibfnamefont {L.~P.}\ \bibnamefont {Kouwenhoven}},\ }\bibfield
  {title} {\bibinfo {title} {Spectroscopy of spin-orbit quantum bits in indium
  antimonide nanowires},\ }\href
  {https://doi.org/10.1103/PhysRevLett.108.166801} {\bibfield  {journal}
  {\bibinfo  {journal} {Phys. Rev. Lett.}\ }\textbf {\bibinfo {volume} {108}},\
  \bibinfo {pages} {166801} (\bibinfo {year} {2012})}\BibitemShut {NoStop}%
\bibitem [{\citenamefont {Koppens}\ \emph {et~al.}(2008)\citenamefont
  {Koppens}, \citenamefont {Nowack},\ and\ \citenamefont
  {Vandersypen}}]{Koppens2008}%
  \BibitemOpen
  \bibfield  {author} {\bibinfo {author} {\bibfnamefont {F.~H.~L.}\
  \bibnamefont {Koppens}}, \bibinfo {author} {\bibfnamefont {K.~C.}\
  \bibnamefont {Nowack}},\ and\ \bibinfo {author} {\bibfnamefont {L.~M.~K.}\
  \bibnamefont {Vandersypen}},\ }\bibfield  {title} {\bibinfo {title} {Spin
  echo of a single electron spin in a quantum dot},\ }\href
  {https://doi.org/10.1103/PhysRevLett.100.236802} {\bibfield  {journal}
  {\bibinfo  {journal} {Phys. Rev. Lett.}\ }\textbf {\bibinfo {volume} {100}},\
  \bibinfo {pages} {236802} (\bibinfo {year} {2008})}\BibitemShut {NoStop}%
\bibitem [{\citenamefont {Elzerman}\ \emph {et~al.}(2004)\citenamefont
  {Elzerman}, \citenamefont {Hanson}, \citenamefont {Willems~van Beveren},
  \citenamefont {Witkamp}, \citenamefont {Vandersypen},\ and\ \citenamefont
  {Kouwenhoven}}]{Elzerman2004}%
  \BibitemOpen
  \bibfield  {author} {\bibinfo {author} {\bibfnamefont {J.~M.}\ \bibnamefont
  {Elzerman}}, \bibinfo {author} {\bibfnamefont {R.}~\bibnamefont {Hanson}},
  \bibinfo {author} {\bibfnamefont {L.~H.}\ \bibnamefont {Willems~van
  Beveren}}, \bibinfo {author} {\bibfnamefont {B.}~\bibnamefont {Witkamp}},
  \bibinfo {author} {\bibfnamefont {L.~M.~K.}\ \bibnamefont {Vandersypen}},\
  and\ \bibinfo {author} {\bibfnamefont {L.~P.}\ \bibnamefont {Kouwenhoven}},\
  }\bibfield  {title} {\bibinfo {title} {Single-shot read-out of an individual
  electron spin in a quantum dot},\ }\href
  {https://doi.org/10.1038/nature02713} {\bibfield  {journal} {\bibinfo
  {journal} {Nature}\ }\textbf {\bibinfo {volume} {430}},\ \bibinfo {pages}
  {431} (\bibinfo {year} {2004})}\BibitemShut {NoStop}%
\bibitem [{\citenamefont {Chittock-Wood}\ \emph {et~al.}(2026)\citenamefont
  {Chittock-Wood}, \citenamefont {Leon}, \citenamefont {Fogarty}, \citenamefont
  {Murphy}, \citenamefont {von Horstig}, \citenamefont {Patomäki},
  \citenamefont {Oakes}, \citenamefont {Williams}, \citenamefont {Johnson},
  \citenamefont {Jussot}, \citenamefont {Kubicek}, \citenamefont {Govoreanu},
  \citenamefont {Wise}, \citenamefont {Morton},\ and\ \citenamefont
  {Gonzalez-Zalba}}]{Chittock-Wood2026}%
  \BibitemOpen
  \bibfield  {author} {\bibinfo {author} {\bibfnamefont {J.~F.}\ \bibnamefont
  {Chittock-Wood}}, \bibinfo {author} {\bibfnamefont {R.~C.~C.}\ \bibnamefont
  {Leon}}, \bibinfo {author} {\bibfnamefont {M.~A.}\ \bibnamefont {Fogarty}},
  \bibinfo {author} {\bibfnamefont {T.}~\bibnamefont {Murphy}}, \bibinfo
  {author} {\bibfnamefont {F.-E.}\ \bibnamefont {von Horstig}}, \bibinfo
  {author} {\bibfnamefont {S.~M.}\ \bibnamefont {Patomäki}}, \bibinfo {author}
  {\bibfnamefont {G.~A.}\ \bibnamefont {Oakes}}, \bibinfo {author}
  {\bibfnamefont {J.}~\bibnamefont {Williams}}, \bibinfo {author}
  {\bibfnamefont {N.}~\bibnamefont {Johnson}}, \bibinfo {author} {\bibfnamefont
  {J.}~\bibnamefont {Jussot}}, \bibinfo {author} {\bibfnamefont
  {S.}~\bibnamefont {Kubicek}}, \bibinfo {author} {\bibfnamefont
  {B.}~\bibnamefont {Govoreanu}}, \bibinfo {author} {\bibfnamefont {D.~F.}\
  \bibnamefont {Wise}}, \bibinfo {author} {\bibfnamefont {J.~J.~L.}\
  \bibnamefont {Morton}},\ and\ \bibinfo {author} {\bibfnamefont {M.~F.}\
  \bibnamefont {Gonzalez-Zalba}},\ }\bibfield  {title} {\bibinfo {title}
  {Radiofrequency cascade readout of coupled spin qubits},\ }\href
  {https://doi.org/10.1038/s41928-026-01582-8} {\bibfield  {journal} {\bibinfo
  {journal} {Nature Electronics}\ }\textbf {\bibinfo {volume} {9}},\ \bibinfo
  {pages} {314} (\bibinfo {year} {2026})}\BibitemShut {NoStop}%
\bibitem [{\citenamefont {Colless}\ \emph {et~al.}(2013)\citenamefont
  {Colless}, \citenamefont {Mahoney}, \citenamefont {Hornibrook}, \citenamefont
  {Doherty}, \citenamefont {Lu}, \citenamefont {Gossard},\ and\ \citenamefont
  {Reilly}}]{Colless2013}%
  \BibitemOpen
  \bibfield  {author} {\bibinfo {author} {\bibfnamefont {J.~I.}\ \bibnamefont
  {Colless}}, \bibinfo {author} {\bibfnamefont {A.~C.}\ \bibnamefont
  {Mahoney}}, \bibinfo {author} {\bibfnamefont {J.~M.}\ \bibnamefont
  {Hornibrook}}, \bibinfo {author} {\bibfnamefont {A.~C.}\ \bibnamefont
  {Doherty}}, \bibinfo {author} {\bibfnamefont {H.}~\bibnamefont {Lu}},
  \bibinfo {author} {\bibfnamefont {A.~C.}\ \bibnamefont {Gossard}},\ and\
  \bibinfo {author} {\bibfnamefont {D.~J.}\ \bibnamefont {Reilly}},\ }\bibfield
   {title} {\bibinfo {title} {Dispersive readout of a few-electron double
  quantum dot with fast rf gate sensors},\ }\href
  {https://doi.org/10.1103/PhysRevLett.110.046805} {\bibfield  {journal}
  {\bibinfo  {journal} {Phys. Rev. Lett.}\ }\textbf {\bibinfo {volume} {110}},\
  \bibinfo {pages} {046805} (\bibinfo {year} {2013})}\BibitemShut {NoStop}%
\bibitem [{\citenamefont {Liu}\ \emph {et~al.}(2021)\citenamefont {Liu},
  \citenamefont {Philips}, \citenamefont {Orona}, \citenamefont {Samkharadze},
  \citenamefont {McJunkin}, \citenamefont {MacQuarrie}, \citenamefont
  {Eriksson}, \citenamefont {Vandersypen},\ and\ \citenamefont
  {Yacoby}}]{Liu2021}%
  \BibitemOpen
  \bibfield  {author} {\bibinfo {author} {\bibfnamefont {Y.-Y.}\ \bibnamefont
  {Liu}}, \bibinfo {author} {\bibfnamefont {S.}~\bibnamefont {Philips}},
  \bibinfo {author} {\bibfnamefont {L.}~\bibnamefont {Orona}}, \bibinfo
  {author} {\bibfnamefont {N.}~\bibnamefont {Samkharadze}}, \bibinfo {author}
  {\bibfnamefont {T.}~\bibnamefont {McJunkin}}, \bibinfo {author}
  {\bibfnamefont {E.}~\bibnamefont {MacQuarrie}}, \bibinfo {author}
  {\bibfnamefont {M.}~\bibnamefont {Eriksson}}, \bibinfo {author}
  {\bibfnamefont {L.}~\bibnamefont {Vandersypen}},\ and\ \bibinfo {author}
  {\bibfnamefont {A.}~\bibnamefont {Yacoby}},\ }\bibfield  {title} {\bibinfo
  {title} {Radio-frequency reflectometry in silicon-based quantum dots},\
  }\href {https://doi.org/10.1103/PhysRevApplied.16.014057} {\bibfield
  {journal} {\bibinfo  {journal} {Phys. Rev. Appl.}\ }\textbf {\bibinfo
  {volume} {16}},\ \bibinfo {pages} {014057} (\bibinfo {year}
  {2021})}\BibitemShut {NoStop}%
\bibitem [{\citenamefont {Crippa}\ \emph {et~al.}(2019)\citenamefont {Crippa},
  \citenamefont {Ezzouch}, \citenamefont {Apr{\'{a}}}, \citenamefont {Amisse},
  \citenamefont {Lavi{\'{e}}ville}, \citenamefont {Hutin}, \citenamefont
  {Bertrand}, \citenamefont {Vinet}, \citenamefont {Urdampilleta},
  \citenamefont {Meunier}, \citenamefont {Sanquer}, \citenamefont {Jehl},
  \citenamefont {Maurand},\ and\ \citenamefont {De~Franceschi}}]{Crippa2019}%
  \BibitemOpen
  \bibfield  {author} {\bibinfo {author} {\bibfnamefont {A.}~\bibnamefont
  {Crippa}}, \bibinfo {author} {\bibfnamefont {R.}~\bibnamefont {Ezzouch}},
  \bibinfo {author} {\bibfnamefont {A.}~\bibnamefont {Apr{\'{a}}}}, \bibinfo
  {author} {\bibfnamefont {A.}~\bibnamefont {Amisse}}, \bibinfo {author}
  {\bibfnamefont {R.}~\bibnamefont {Lavi{\'{e}}ville}}, \bibinfo {author}
  {\bibfnamefont {L.}~\bibnamefont {Hutin}}, \bibinfo {author} {\bibfnamefont
  {B.}~\bibnamefont {Bertrand}}, \bibinfo {author} {\bibfnamefont
  {M.}~\bibnamefont {Vinet}}, \bibinfo {author} {\bibfnamefont
  {M.}~\bibnamefont {Urdampilleta}}, \bibinfo {author} {\bibfnamefont
  {T.}~\bibnamefont {Meunier}}, \bibinfo {author} {\bibfnamefont
  {M.}~\bibnamefont {Sanquer}}, \bibinfo {author} {\bibfnamefont
  {X.}~\bibnamefont {Jehl}}, \bibinfo {author} {\bibfnamefont {R.}~\bibnamefont
  {Maurand}},\ and\ \bibinfo {author} {\bibfnamefont {S.}~\bibnamefont
  {De~Franceschi}},\ }\bibfield  {title} {\bibinfo {title} {Gate-reflectometry
  dispersive readout and coherent control of a spin qubit in silicon},\ }\href
  {https://doi.org/10.1038/s41467-019-10848-z} {\bibfield  {journal} {\bibinfo
  {journal} {Nature Communications}\ }\textbf {\bibinfo {volume} {10}},\
  \bibinfo {pages} {2776} (\bibinfo {year} {2019})}\BibitemShut {NoStop}%
\bibitem [{\citenamefont {Urdampilleta}\ \emph {et~al.}(2019)\citenamefont
  {Urdampilleta}, \citenamefont {Niegemann}, \citenamefont {Chanrion},
  \citenamefont {Jadot}, \citenamefont {Spence}, \citenamefont {Mortemousque},
  \citenamefont {Bäuerle}, \citenamefont {Hutin}, \citenamefont {Bertrand},
  \citenamefont {Barraud}, \citenamefont {Maurand}, \citenamefont {Sanquer},
  \citenamefont {Jehl}, \citenamefont {De~Franceschi}, \citenamefont {Vinet},\
  and\ \citenamefont {Meunier}}]{Urdampilleta2019}%
  \BibitemOpen
  \bibfield  {author} {\bibinfo {author} {\bibfnamefont {M.}~\bibnamefont
  {Urdampilleta}}, \bibinfo {author} {\bibfnamefont {D.~J.}\ \bibnamefont
  {Niegemann}}, \bibinfo {author} {\bibfnamefont {E.}~\bibnamefont {Chanrion}},
  \bibinfo {author} {\bibfnamefont {B.}~\bibnamefont {Jadot}}, \bibinfo
  {author} {\bibfnamefont {C.}~\bibnamefont {Spence}}, \bibinfo {author}
  {\bibfnamefont {P.-A.}\ \bibnamefont {Mortemousque}}, \bibinfo {author}
  {\bibfnamefont {C.}~\bibnamefont {Bäuerle}}, \bibinfo {author}
  {\bibfnamefont {L.}~\bibnamefont {Hutin}}, \bibinfo {author} {\bibfnamefont
  {B.}~\bibnamefont {Bertrand}}, \bibinfo {author} {\bibfnamefont
  {S.}~\bibnamefont {Barraud}}, \bibinfo {author} {\bibfnamefont
  {R.}~\bibnamefont {Maurand}}, \bibinfo {author} {\bibfnamefont
  {M.}~\bibnamefont {Sanquer}}, \bibinfo {author} {\bibfnamefont
  {X.}~\bibnamefont {Jehl}}, \bibinfo {author} {\bibfnamefont {S.}~\bibnamefont
  {De~Franceschi}}, \bibinfo {author} {\bibfnamefont {M.}~\bibnamefont
  {Vinet}},\ and\ \bibinfo {author} {\bibfnamefont {T.}~\bibnamefont
  {Meunier}},\ }\bibfield  {title} {\bibinfo {title} {Gate-based high fidelity
  spin readout in a cmos device},\ }\href
  {https://doi.org/10.1038/s41565-019-0443-9} {\bibfield  {journal} {\bibinfo
  {journal} {Nature Nanotechnology}\ }\textbf {\bibinfo {volume} {14}},\
  \bibinfo {pages} {737} (\bibinfo {year} {2019})}\BibitemShut {NoStop}%
\bibitem [{\citenamefont {Kam}\ and\ \citenamefont {Hu}(2024)}]{KamHu2024}%
  \BibitemOpen
  \bibfield  {author} {\bibinfo {author} {\bibfnamefont {C.-F.}\ \bibnamefont
  {Kam}}\ and\ \bibinfo {author} {\bibfnamefont {X.}~\bibnamefont {Hu}},\
  }\bibfield  {title} {\bibinfo {title} {Fast and high-fidelity dispersive
  readout of a spin qubit with squeezed microwave and resonator nonlinearity},\
  }\href@noop {} {\bibfield  {journal} {\bibinfo  {journal} {npj Quantum
  Information}\ }\textbf {\bibinfo {volume} {10}} (\bibinfo {year}
  {2024})}\BibitemShut {NoStop}%
\bibitem [{\citenamefont {Ono}\ \emph {et~al.}(2002)\citenamefont {Ono},
  \citenamefont {Austing}, \citenamefont {Tokura},\ and\ \citenamefont
  {Tarucha}}]{Ono2002}%
  \BibitemOpen
  \bibfield  {author} {\bibinfo {author} {\bibfnamefont {K.}~\bibnamefont
  {Ono}}, \bibinfo {author} {\bibfnamefont {D.~G.}\ \bibnamefont {Austing}},
  \bibinfo {author} {\bibfnamefont {Y.}~\bibnamefont {Tokura}},\ and\ \bibinfo
  {author} {\bibfnamefont {S.}~\bibnamefont {Tarucha}},\ }\bibfield  {title}
  {\bibinfo {title} {Current rectification by pauli exclusion in a weakly
  coupled double quantum dot system},\ }\href
  {https://doi.org/10.1126/science.1071872} {\bibfield  {journal} {\bibinfo
  {journal} {Science}\ }\textbf {\bibinfo {volume} {297}},\ \bibinfo {pages}
  {1313} (\bibinfo {year} {2002})}\BibitemShut {NoStop}%
\bibitem [{\citenamefont {Johnson}\ \emph {et~al.}(2005)\citenamefont
  {Johnson}, \citenamefont {Petta}, \citenamefont {Taylor}, \citenamefont
  {Yacoby}, \citenamefont {Lukin}, \citenamefont {Marcus}, \citenamefont
  {Hanson},\ and\ \citenamefont {Gossard}}]{Johnson2005}%
  \BibitemOpen
  \bibfield  {author} {\bibinfo {author} {\bibfnamefont {A.}~\bibnamefont
  {Johnson}}, \bibinfo {author} {\bibfnamefont {J.}~\bibnamefont {Petta}},
  \bibinfo {author} {\bibfnamefont {J.}~\bibnamefont {Taylor}}, \bibinfo
  {author} {\bibfnamefont {A.}~\bibnamefont {Yacoby}}, \bibinfo {author}
  {\bibfnamefont {M.}~\bibnamefont {Lukin}}, \bibinfo {author} {\bibfnamefont
  {C.}~\bibnamefont {Marcus}}, \bibinfo {author} {\bibfnamefont
  {M.}~\bibnamefont {Hanson}},\ and\ \bibinfo {author} {\bibfnamefont
  {A.}~\bibnamefont {Gossard}},\ }\bibfield  {title} {\bibinfo {title}
  {Triplet–singlet spin relaxation via nuclei in a double quantum dot},\
  }\href {https://doi.org/10.1038/nature03815} {\bibfield  {journal} {\bibinfo
  {journal} {Nature}\ }\textbf {\bibinfo {volume} {435}},\ \bibinfo {pages}
  {925} (\bibinfo {year} {2005})}\BibitemShut {NoStop}%
\bibitem [{\citenamefont {Nowak}\ \emph {et~al.}(2012)\citenamefont {Nowak},
  \citenamefont {Szafran},\ and\ \citenamefont {Peeters}}]{Nowak2012}%
  \BibitemOpen
  \bibfield  {author} {\bibinfo {author} {\bibfnamefont {M.~P.}\ \bibnamefont
  {Nowak}}, \bibinfo {author} {\bibfnamefont {B.}~\bibnamefont {Szafran}},\
  and\ \bibinfo {author} {\bibfnamefont {F.~M.}\ \bibnamefont {Peeters}},\
  }\bibfield  {title} {\bibinfo {title} {Resonant harmonic generation and
  collective spin rotations in electrically driven quantum dots},\ }\href
  {https://doi.org/10.1103/PhysRevB.86.125428} {\bibfield  {journal} {\bibinfo
  {journal} {Phys. Rev. B}\ }\textbf {\bibinfo {volume} {86}},\ \bibinfo
  {pages} {125428} (\bibinfo {year} {2012})}\BibitemShut {NoStop}%
\bibitem [{\citenamefont {Dyakonov}\ and\ \citenamefont
  {Perel}(1971)}]{ruskie}%
  \BibitemOpen
  \bibfield  {author} {\bibinfo {author} {\bibfnamefont {M.}~\bibnamefont
  {Dyakonov}}\ and\ \bibinfo {author} {\bibfnamefont {V.}~\bibnamefont
  {Perel}},\ }\bibfield  {title} {\bibinfo {title} {Current-induced spin
  orientation of electrons in semiconductors},\ }\href
  {https://doi.org/10.1016/0375-9601(71)90196-4} {\bibfield  {journal}
  {\bibinfo  {journal} {Phys. Lett. A.}\ ,\ \bibinfo {pages} {459}} (\bibinfo
  {year} {1971})}\BibitemShut {NoStop}%
\bibitem [{\citenamefont {Sinova}\ \emph {et~al.}(2004)\citenamefont {Sinova},
  \citenamefont {Culcer}, \citenamefont {Niu}, \citenamefont {Sinitsyn},
  \citenamefont {Jungwirth},\ and\ \citenamefont {MacDonald}}]{sinova2004}%
  \BibitemOpen
  \bibfield  {author} {\bibinfo {author} {\bibfnamefont {J.}~\bibnamefont
  {Sinova}}, \bibinfo {author} {\bibfnamefont {D.}~\bibnamefont {Culcer}},
  \bibinfo {author} {\bibfnamefont {Q.}~\bibnamefont {Niu}}, \bibinfo {author}
  {\bibfnamefont {N.}~\bibnamefont {Sinitsyn}}, \bibinfo {author}
  {\bibfnamefont {T.}~\bibnamefont {Jungwirth}},\ and\ \bibinfo {author}
  {\bibfnamefont {A.}~\bibnamefont {MacDonald}},\ }\bibfield  {title} {\bibinfo
  {title} {Universal intrinsic spin hall effect},\ }\href
  {https://doi.org/10.1103/PhysRevLett.92.126603} {\bibfield  {journal}
  {\bibinfo  {journal} {Phys. Rev. Lett.}\ }\textbf {\bibinfo {volume} {92}},\
  \bibinfo {pages} {126603} (\bibinfo {year} {2004})}\BibitemShut {NoStop}%
\bibitem [{\citenamefont {Sinova}\ \emph {et~al.}(2015)\citenamefont {Sinova},
  \citenamefont {Valenzuela}, \citenamefont {Wunderlich}, \citenamefont
  {Back},\ and\ \citenamefont {Jungwirth}}]{sinova2015}%
  \BibitemOpen
  \bibfield  {author} {\bibinfo {author} {\bibfnamefont {J.}~\bibnamefont
  {Sinova}}, \bibinfo {author} {\bibfnamefont {S.~O.}\ \bibnamefont
  {Valenzuela}}, \bibinfo {author} {\bibfnamefont {J.}~\bibnamefont
  {Wunderlich}}, \bibinfo {author} {\bibfnamefont {C.~H.}\ \bibnamefont
  {Back}},\ and\ \bibinfo {author} {\bibfnamefont {T.}~\bibnamefont
  {Jungwirth}},\ }\bibfield  {title} {\bibinfo {title} {Spin hall effects},\
  }\href {https://doi.org/10.1103/RevModPhys.87.1213} {\bibfield  {journal}
  {\bibinfo  {journal} {Rev. Mod. Phys.}\ }\textbf {\bibinfo {volume} {87}},\
  \bibinfo {pages} {1213} (\bibinfo {year} {2015})}\BibitemShut {NoStop}%
\bibitem [{\citenamefont {Schliemann}\ \emph {et~al.}(2006)\citenamefont
  {Schliemann}, \citenamefont {Loss},\ and\ \citenamefont
  {Westervelt}}]{Schliemann2006}%
  \BibitemOpen
  \bibfield  {author} {\bibinfo {author} {\bibfnamefont {J.}~\bibnamefont
  {Schliemann}}, \bibinfo {author} {\bibfnamefont {D.}~\bibnamefont {Loss}},\
  and\ \bibinfo {author} {\bibfnamefont {R.~M.}\ \bibnamefont {Westervelt}},\
  }\bibfield  {title} {\bibinfo {title} {Zitterbewegung of electrons and holes
  in iii--v semiconductor quantum wells},\ }\href
  {https://doi.org/10.1103/PhysRevB.73.085323} {\bibfield  {journal} {\bibinfo
  {journal} {Phys. Rev. B}\ }\textbf {\bibinfo {volume} {73}},\ \bibinfo
  {pages} {085323} (\bibinfo {year} {2006})}\BibitemShut {NoStop}%
\bibitem [{\citenamefont {Khomitsky}\ \emph {et~al.}(2020)\citenamefont
  {Khomitsky}, \citenamefont {Lavrukhina},\ and\ \citenamefont
  {Sherman}}]{Shermans}%
  \BibitemOpen
  \bibfield  {author} {\bibinfo {author} {\bibfnamefont {D.}~\bibnamefont
  {Khomitsky}}, \bibinfo {author} {\bibfnamefont {E.}~\bibnamefont
  {Lavrukhina}},\ and\ \bibinfo {author} {\bibfnamefont {E.}~\bibnamefont
  {Sherman}},\ }\bibfield  {title} {\bibinfo {title} {Spin rotation by resonant
  electric field in few-level quantum dots: Floquet dynamics and tunneling},\
  }\href {https://doi.org/10.1103/PhysRevApplied.14.014090} {\bibfield
  {journal} {\bibinfo  {journal} {Phys. Rev. Appl.}\ }\textbf {\bibinfo
  {volume} {14}},\ \bibinfo {pages} {014090} (\bibinfo {year}
  {2020})}\BibitemShut {NoStop}%
\bibitem [{\citenamefont {W\'ojcik}\ \emph {et~al.}(2021)\citenamefont
  {W\'ojcik}, \citenamefont {Nowak},\ and\ \citenamefont
  {Zegrodnik}}]{Wojcik_2021}%
  \BibitemOpen
  \bibfield  {author} {\bibinfo {author} {\bibfnamefont {P.}~\bibnamefont
  {W\'ojcik}}, \bibinfo {author} {\bibfnamefont {M.~P.}\ \bibnamefont
  {Nowak}},\ and\ \bibinfo {author} {\bibfnamefont {M.}~\bibnamefont
  {Zegrodnik}},\ }\bibfield  {title} {\bibinfo {title} {Impact of spin-orbit
  interaction on the phase diagram and anisotropy of the in-plane critical
  magnetic field at the superconducting
  $\mathrm{LaAlO}{}_{3}/\mathrm{SrTiO}{}_{3}$ interface},\ }\href@noop {}
  {\bibfield  {journal} {\bibinfo  {journal} {Phys. Rev. B}\ }\textbf {\bibinfo
  {volume} {104}},\ \bibinfo {pages} {174503} (\bibinfo {year}
  {2021})}\BibitemShut {NoStop}%
\bibitem [{\citenamefont {Rashba}\ and\ \citenamefont
  {Efros}(2003)}]{Rashba2003}%
  \BibitemOpen
  \bibfield  {author} {\bibinfo {author} {\bibfnamefont {E.~I.}\ \bibnamefont
  {Rashba}}\ and\ \bibinfo {author} {\bibfnamefont {A.~L.}\ \bibnamefont
  {Efros}},\ }\bibfield  {title} {\bibinfo {title} {Orbital mechanisms of
  electron-spin manipulation by an electric field},\ }\href
  {https://doi.org/10.1103/PhysRevLett.91.126405} {\bibfield  {journal}
  {\bibinfo  {journal} {Phys. Rev. Lett.}\ }\textbf {\bibinfo {volume} {91}},\
  \bibinfo {pages} {126405} (\bibinfo {year} {2003})}\BibitemShut {NoStop}%
\bibitem [{\citenamefont {Meier}\ \emph {et~al.}(2007)\citenamefont {Meier},
  \citenamefont {Salis}, \citenamefont {Shorubalko}, \citenamefont {Gini},
  \citenamefont {Schön},\ and\ \citenamefont {Ensslin}}]{Meier2007}%
  \BibitemOpen
  \bibfield  {author} {\bibinfo {author} {\bibfnamefont {L.}~\bibnamefont
  {Meier}}, \bibinfo {author} {\bibfnamefont {G.}~\bibnamefont {Salis}},
  \bibinfo {author} {\bibfnamefont {I.}~\bibnamefont {Shorubalko}}, \bibinfo
  {author} {\bibfnamefont {E.}~\bibnamefont {Gini}}, \bibinfo {author}
  {\bibfnamefont {S.}~\bibnamefont {Schön}},\ and\ \bibinfo {author}
  {\bibfnamefont {K.}~\bibnamefont {Ensslin}},\ }\bibfield  {title} {\bibinfo
  {title} {Measurement of rashba and dresselhaus spin–orbit magnetic
  fields},\ }\href {https://doi.org/10.1038/nphys675} {\bibfield  {journal}
  {\bibinfo  {journal} {Nature Physics}\ }\textbf {\bibinfo {volume} {3}},\
  \bibinfo {pages} {650} (\bibinfo {year} {2007})}\BibitemShut {NoStop}%
\bibitem [{\citenamefont {Kirichenko}\ \emph {et~al.}(2020)\citenamefont
  {Kirichenko}, \citenamefont {Stephanovich},\ and\ \citenamefont
  {Sherman}}]{sherman}%
  \BibitemOpen
  \bibfield  {author} {\bibinfo {author} {\bibfnamefont {E.~V.}\ \bibnamefont
  {Kirichenko}}, \bibinfo {author} {\bibfnamefont {V.~A.}\ \bibnamefont
  {Stephanovich}},\ and\ \bibinfo {author} {\bibfnamefont {E.~Y.}\ \bibnamefont
  {Sherman}},\ }\bibfield  {title} {\bibinfo {title} {Chaotic cyclotron and
  hall trajectories due to spin-orbit coupling},\ }\href
  {https://doi.org/10.1002/andp.202000012} {\bibfield  {journal} {\bibinfo
  {journal} {Ann. Phys. (Berlin)}\ }\textbf {\bibinfo {volume} {532}},\
  \bibinfo {pages} {2000012} (\bibinfo {year} {2020})}\BibitemShut {NoStop}%
\bibitem [{\citenamefont {Strambini}\ \emph {et~al.}(2009)\citenamefont
  {Strambini}, \citenamefont {Piazza}, \citenamefont {Biasiol}, \citenamefont
  {Sorba},\ and\ \citenamefont {Beltram}}]{Strambini2009}%
  \BibitemOpen
  \bibfield  {author} {\bibinfo {author} {\bibfnamefont {E.}~\bibnamefont
  {Strambini}}, \bibinfo {author} {\bibfnamefont {V.}~\bibnamefont {Piazza}},
  \bibinfo {author} {\bibfnamefont {G.}~\bibnamefont {Biasiol}}, \bibinfo
  {author} {\bibfnamefont {L.}~\bibnamefont {Sorba}},\ and\ \bibinfo {author}
  {\bibfnamefont {F.}~\bibnamefont {Beltram}},\ }\bibfield  {title} {\bibinfo
  {title} {Impact of classical forces and decoherence in multiterminal
  aharonov-bohm networks},\ }\href {https://doi.org/10.1103/PhysRevB.79.195443}
  {\bibfield  {journal} {\bibinfo  {journal} {Phys. Rev. B}\ }\textbf {\bibinfo
  {volume} {79}},\ \bibinfo {pages} {195443} (\bibinfo {year}
  {2009})}\BibitemShut {NoStop}%
\bibitem [{\citenamefont {Schliemann}(2007)}]{Schli2}%
  \BibitemOpen
  \bibfield  {author} {\bibinfo {author} {\bibfnamefont {J.}~\bibnamefont
  {Schliemann}},\ }\bibfield  {title} {\bibinfo {title} {Ballistic side-jump
  motion of electrons and holes in semiconductor quantum wells},\ }\href
  {https://doi.org/10.1103/PhysRevB.75.045304} {\bibfield  {journal} {\bibinfo
  {journal} {Phys. Rev. B}\ }\textbf {\bibinfo {volume} {75}},\ \bibinfo
  {pages} {045304} (\bibinfo {year} {2007})}\BibitemShut {NoStop}%
\end{thebibliography}%
\end{document}